\documentclass[11pt,russian]{article}
\usepackage[koi8-r]{inputenc}
\usepackage{babel}
\usepackage{rotating}

\begin{document}

\begin{center}
     {\bf S.I. Bityukov, N.V. Krasnikov}
\end{center}

\bigskip

\begin{center}
{\Large \bf The use of statistical methods for the search for new  
physics at the LHC (in Russian)}
\end{center}

\bigskip

\begin{center}
{\bf INR RAS, Moscow}
\end{center}

\begin{center}
{\bf IHEP, Protvino}
\end{center}

\bigskip

{We review~\footnote{Extended text of lectures given by S.I. Bityukov for  
students of MIPT and doctoral students of IHEP.}
statistical methods used for the search for new physics at LHC.

} 

\newpage

\begin{center}
     {\bf Битюков Сергей Иванович, Красников Николай Валериевич}
\end{center}

\bigskip

\begin{center}
{\Large \bf Применение статистических методов для поиска новой физики на 
Большом Адронном Коллайдере}
\end{center}

\bigskip

\begin{center}
{\bf ИЯИ РАН, Москва}
\end{center}

\begin{center}
{\bf ИФВЭ, Протвино}
\end{center}

\bigskip

{В настоящей работе~\footnote{Расширенный текст лекций, прочитанный С.И. Битюковым студентам 
МФТИ и аспирантам ИФВЭ.}  
дан обзор статистических методов, используемых при поиске 
новой физики в экспериментах на Большом адронном коллайдере. Приведены 
многочисленные примеры, полезные для физиков, занимающихся обработкой данных с 
детекторов Большого адронного коллайдера. Обзор предназначен для научных 
работников - как теоретиков, так и экспериментаторов; специалистов в области 
моделирования физических процессов при столкновениях частиц высоких энергий на 
современных коллайдерах; всех, кто интересуется извлечением физических 
результатов из экспериментальных данных.

} 

\newpage

\tableofcontents

\newpage

\section*{Введение}

Данный обзор описывает основные статистические методы, использующиеся при обработке результатов 
экспериментов CMS~\cite{CMS} и ATLAS~\cite{ATLAS} в исследованиях на Большом Адронном 
Коллайдере~\cite{LHC} (БАК). Эти методы реализуются в пакете программ 
RooStats~\cite{RooStats}, созданных физиками, участвующими в экспериментах CMS и 
ATLAS. Пакет программ RooStats базируется на системе ROOT~\cite{ROOT}, 
разработанной в ЦЕРНе (Европейском центре ядерных исследований). 
Основное отличие данного обзора от многочисленных 
обзоров и монографий по применению статистических 
методов в экспериментальной физике и других естественнонаучных 
областях~\cite{James,Frodesen,Classic,DAgostini2003,Kramer} это то, что в обзоре  
приводятся многочисленные примеры, ориентированные исключительно для нужд физиков, занимающихся 
обработкой данных с детекторов Большого Адронного Коллайдера. 
В обзоре почти отсутствуют доказательства каких либо 
абстрактных теорем статистики. В каком-то смысле можно сказать, что это ``курс 
молодого бойца'', жаждущего найти проявления новой физики~\cite{Matveev} 
из экспериментальных данных CMS и ATLAS детекторов. 

Для чтения этого обзора необходимо знать лишь основы высшей математики (производная, интеграл) 
в минимальном объеме. Основные понятия теории вероятности кратко приведены в обзоре 
(Глава 1). Также не требуется каких-либо особо глубоких знаний как в теоретической, так и 
в экспериментальной физике. 

Содержание статьи следующее. В первой главе приведены основные понятия теории вероятностей. 
Вторая глава на конкретных примерах (ограничения снизу и сверху на параметр) обсуждает основные подходы, 
используемые в статистике, а именно, частотный и Байесовский подходы и метод максимального 
правдоподобия. Третья глава описывает методы, которые позволяют извлечь информацию о параметрах 
сигнала из экспериментальных данных. В качестве статистик используются статистики, полученные 
из нормального распределения и распределения Пуассона. В четвертой главе приведены методы, 
позволяющие учесть систематические эффекты, связанные с неточным знанием фонов и сигналов. 
В пятой главе обсуждаются проблемы проверки гипотез. В нашем случае основной гипотезой является 
гипотеза о том, что справедлива Стандартная Модель, а альтернативной гипотезой является гипотеза 
о существовании новой физики вне рамок Стандартной Модели, поиск которой является главной задачей 
экспериментов на Большом Адронном Коллайдере. В шестой главе обсуждаются методы, применяемые для 
комбинирования различных экспериментальных данных, в том числе, для комбинирования данных 
разных экспериментов, например, экспериментов CMS и ATLAS. В седьмой главе описаны основные 
возможности программы обработки экспериментальных данных на базе пакета программ RooStats, 
который в настоящее время является основным программным продуктом, официально рекомендованным 
коллаборациями CMS и ATLAS для обработки данных.

\newpage

\section{Основные понятия теории вероятностей}

\subsection{Введение. Интуитивное понятие вероятности}

Теория вероятности по сути дела возникла из игры в бросание монеты. Если мы произвольно бросаем монету 
$N$ раз, то в $N_1$ случаях выпадет ``орел'', а в $N_2$ случаях выпадет ``решка''. Очевидно, 
что в любом случае выпадет либо ``орел'', либо ``решка'', то есть справедливо равенство 
$N=N_1+N_2$. В случае большого числа бросаний монеты (испытаний) вероятность выпадения ``орла'' 
определяется как $P_1$=$P($''орел''$)=\displaystyle \frac{N_1}{N}$ и ``решки'' как 
$P_2$=$P($''решка''$)=\displaystyle \frac{N_2}{N}$. Здесь неявно предполагается, что 
предел при $N \rightarrow \infty$, который по сути дела и является определением вероятности, 
существует. Очевидны следующие свойства вероятности: 

\begin{eqnarray}
P_1 & \ge & 0, \nonumber \\
P_2& \ge & 0, \nonumber \\ 
P_1 + P_2 & = & 1. 
\label{eq:1}
\end{eqnarray}

Поскольку стороны ``орел'' и ``решка'' практически не отличаются друг от друга, так же 
часто предполагается (хотя это и не обязательно), что 
$P($''орел''$)=P($''решка''$)=\displaystyle \frac{1}{2}$.

В качестве более общего случая мы можем представить, что в ящике находится $n$ шаров, 
на которых нарисованы цифры $1, 2, \dots, n$. Если мы $N \gg n$ раз вытаскиваем 
один за другим шары из ящика (с возвращением шара сразу же обратно в ящик), то вероятность 
вынуть шар с номером $k~~(1\le k \le n)$ определяется как 
$P_k(N)=\displaystyle \frac{n_k}{N}$ (где $n_k$-количество раз, когда мы вынули 
шар с номером $k$). Предполагается, что предел $N \rightarrow \infty$ для всех 
$P_k(N)$ существует. Это и есть определение вероятности вытащить шар $k$ из ящика. 

Очевидны следующие свойства вероятностей $P_k$:

\begin{eqnarray}
P_k & \ge & 0, \nonumber \\
\sum_{k=1}^n P_k & = & 1. 
\label{eq:2}
\end{eqnarray}

Нетрудно обобщить вышеприведенные примеры и дать интуитивные определения вероятности 
на случай чисел $n=1,2, \dots, \infty$ и на случай зависимости 
вероятности (точнее плотности вероятности) от непрерывного параметра. Интуитивное понятие 
вероятности можно формализовать на аксиоматическом уровне, что будет показано в следующем параграфе.

\subsection{Аксиомы Колмогорова}

Понятие вероятности на аксиоматическом уровне можно ввести следующим образом. Пусть у нас 
имеется некоторое множество $\Omega$. Вероятность это $P$-вещественная функция, определенная на классе 
всех подмножеств $\Omega$. Для каждого подмножества $A,B, \dots $ множества $\Omega$ функция 
$P(A)$ удовлетворяет следующим аксиомам (аксиомы Колмогорова~\cite{Kolmog}): 

\begin{itemize}
\item $P(A) \ge 0$.
\item Для непересекающихся множеств (т.е. $A\bigcap B = 0$), \\
$P(A\bigcup B) = P(A) + P(B)$. 
\item $P(\Omega) = 1$.
\end{itemize} 

Множество $\Omega$ называется пространством элементарных событий. Подмножества $A, B, \dots$ 
множества $\Omega$ называются событиями.

\subsection{Условные вероятности. Теорема Байеса}

Условная вероятность $P(A|B)$ наступления события $A$ при условии выполнения события $B$ 
определяется как 

\begin{equation}
P(A|B)=\displaystyle \frac{P(A\bigcap B)}{P(B)}.
\label{eq:3}
\end{equation}

Из соотношения~(\ref{eq:3}) и из того факта, что $A\bigcap B = B\bigcap A$, вытекает теорема 
Байеса~\cite{Bayes}

\begin{equation}
P(A|B)=\displaystyle \frac{P(B|A)P(A)}{P(B)}.
\label{eq:4}
\end{equation}

Также справедливо очевидное равенство 

\begin{equation}
P(B)=\displaystyle \sum_i{P(B|A_i)}{P(A_i)} 
\label{eq:5}
\end{equation}

\noindent
при условии, что $\displaystyle \bigcup_i A_i = \Omega$.

С учетом равенства~(\ref{eq:5}) теорема Байеса (\ref{eq:4}) может быть сформулирована как

\begin{equation}
P(A|B)=\displaystyle \frac{P(B|A)P(A)}{\displaystyle \sum_i{P(B|A_i)P(A_i)}},  
\label{eq:6}
\end{equation}

\noindent
где $\displaystyle \bigcup_i A_i = \Omega$.

Теорема Байеса является ключевой при формулировании статистики в Байесовском подходе, где она 
записывается в виде 

\begin{equation}
P(theory|data) \sim \displaystyle P(data|theory)P(theory), 
\label{eq:7}
\end{equation}

\noindent
где ``theory'' представляет собой некоторую гипотезу, а ``data'' представляет собой 
результат эксперимента. $P(theory)$ является априорной вероятностью для теории, отражающую 
априорную степень доверия к ней. Функция $P(data|theory)$ является вероятностью получения 
данных при заданной теоретической модели ``theory'' и называется функцией максимального 
правдоподобия. 

События $A$ и $B$ называются независимыми, если вероятность их одновременного осуществления 
равна произведению вероятностей каждого из них, то есть 

\begin{equation}
P(A\bigcap B)=\displaystyle P(A)P(B). 
\label{eq:8}
\end{equation}

Вспоминая определение (\ref{eq:3}) условной вероятности, получаем, что для независимых событий 
$A$ и $B$ 

\begin{eqnarray}
P(A|B) & = & P(A), \nonumber \\
P(B|A) & = & P(B). 
\label{eq:9}
\end{eqnarray}

\subsection{Случайные величины. Функции распределения}

Случайной величиной называется функция, определенная на множестве элементарных (событий) исходов. 
Случайная величина как функция элементарного исхода $\omega \in \Omega$ обычно обозначается 
как $x(\omega)$. Обычно аргумент $\omega$ для краткости опускается. 

В случае дискретной 
величины $x_1, x_2, \dots$ функция $P(X=x_k)$ есть вероятность события $X(\omega)=x_k$. 

В случае непересекющихся событий $x_k$ ($x_k \bigcap x_l = \emptyset,~k \ne l$), 
аксиомы Колмогорова сводятся к двум условиям: 

\begin{eqnarray}
P(x_k) & \ge & 0, \nonumber \\
\displaystyle \sum_k P(x_k) & = & 1. 
\label{eq:10}
\end{eqnarray}

Для непрерывных случайных величин, когда множество всех возможных значений 
случайных величин заполняют непрерывное или континуальное множество (например, отрезок на прямой, 
часть плоскости или пространства) вводят понятие распределение вероятности. 

Пусть случайная величина $x$ задана при $a \le x \le b$. 
Плотность вероятности $f(x)$ это неотрицательная функция от $x$, 
такая, что вероятность наступления события с $x_1 \le x \le x_2$ равна 
$\displaystyle \int_{x_1}^{x_2}{f(x)dx} = P(x_1 \le x \le x_2)$. 
Имеет место тривиальное условие нормировки  $\displaystyle \int_a^b{f(x)dx} = 1$. 

Обобщение на случай многомерных случайных величин $(x_1,x_2, \dots, x_n)$ очевидно. 

При изменении переменных $y=h(x)$ в силу равенства 

\begin{equation}
g(y)dy = f(x)dx
\label{eq:11}
\end{equation}

\noindent
получаем 

\begin{equation}
g(y) = \displaystyle \frac{f(x)}{|h'(x)|}. 
\label{eq:12}
\end{equation}

Кумулятивной функцией распределения $F(y)$ является вероятность того, что 
$P(x \le y) = F(y)$. Кумулятивная функция выражается через интеграл от плотности вероятности 

\begin{equation}
F(y) = \displaystyle \int_a^y{f(x)dx}.  
\label{eq:13}
\end{equation}

Заметим, что обычно функция $f(x)$ определяется при $-\infty \le x \le \infty$, то есть 
$b = \infty,~a = -\infty$. 

\subsection{Свойства функций распределений}

Пусть $X$ это дискретная случайная величина, принимающая значения $x_1, x_2, \dots$ с вероятностями 
соответственно $p_1, p_2, \dots$. Математическим ожиданием дискретной случайной величины 
называется число 

\begin{equation}
E(X) = \displaystyle x_1p_1+x_2p_2+ \dots .  
\label{eq:14}
\end{equation}

Для непрерывной случайной величины $X$ обобщение~(\ref{eq:14}) имеет вид 

\begin{equation}
E(X) = \displaystyle \int_{-\infty}^{\infty}{xf(x)dx}.  
\label{eq:15}
\end{equation}

Подчеркнем, что для медленно убывающих функций~\footnote{Типичный пример - распределение Коши с 
$f(x) = \displaystyle \frac{1}{\pi}\frac{1}{1+x^2}$.} интеграл~(\ref{eq:15}) может не 
существовать.

Одним из важных свойств случайной величины является линейность: если $a$ и $b$ два числа, а 
$X$ и $Y$ две случайные величины, то 

\begin{equation}
E(aX+bY) = aE(X)+bE(Y).  
\label{eq:16}
\end{equation}

Свойство~(\ref{eq:16}) непосредственно вытекает из определений 
(\ref{eq:14},\ref{eq:15}) математических 
ожиданий для дискретной и непрерывной случайных величин.

Важной характеристикой разброса случайных величин является дисперсия, которая определяется 
как~\footnote{В случае медленно убывающих функций, например, 
$f(x)\sim\displaystyle \frac{1}{1+|x|^3}$, дисперсия не существует, точнее дисперсия 
равна бесконечности.} 

\begin{equation}
D(X) = E(X-E(X))^2.  
\label{eq:17}
\end{equation}

Иными словами, дисперсию можно определить как среднее значение квадрата отклонения случайной 
величины от своего математического ожидания. Формула (\ref{eq:17}) эквивалентна формуле 

\begin{equation}
D(X) = E(X^2)-(E(X))^2.  
\label{eq:18}
\end{equation}

Справедливо следующее свойство дисперсии: 

\begin{equation}
D(cX) = c^2D(X),    
\label{eq:19}
\end{equation}

\noindent
где с - некоторое постоянное число (константа).

Часто используют понятие среднеквадратичного отклонения, определяемого как квадратный корень 
из дисперсии

\begin{equation}
\sigma = \sqrt{D(X)}.   
\label{eq:20}
\end{equation}

Для математического ожидания $E(X)$ справедливо неравенство Маркова 

\begin{equation}
P(X \ge x) \le \displaystyle \frac{E(X)}{x}.   
\label{eq:21}
\end{equation}

Для дисперсии $D(X)$ справедливо неравенство Чебышева 

\begin{equation}
P(|X-E(X)| \ge x) \le \displaystyle \frac{D(X)}{x^2}.   
\label{eq:22}
\end{equation}

Неравенства~(\ref{eq:21},\ref{eq:22}) следуют из определений $E(X)$, $D(X)$ и 
свойства неотрицательности плотности распределения вероятностей $f(x)$.

Напомним, что два события $A$ и $B$ являются независимыми, если 
$P(A\bigcap B)=P(A)\cdot P(B)$. Это понятие независимости двух событий можно 
распространить и на случайные величины. 

\underline{Определение:} Случайные величины  $X$ и $Y$ называются независимыми, если 
при любых $x$ и $y$ независимы события $\{\omega: X(\omega) < x\}$ и 
$\{\omega: Y(\omega) < y\}$.

Можно показать, что если случайные величины независимы, то 

\begin{equation}
E(XY) = E(X)E(Y).   
\label{eq:23}
\end{equation}

На языке плотностей вероятностей независимость событий $X$ и $Y$ означает, что плотность вероятности 
$f(x,y)$ факторизуется: 

\begin{equation}
f(x,y) = f_1(x)f_2(y).   
\label{eq:24}
\end{equation}

Свойство~(\ref{eq:24}) можно рассматривать как определение двух независимых случайных величин. 
Обобщение на случай $n$ независимых величин $x_1, x_2, \dots, x_n$ очевидно, 
а именно, плотность вероятности $f(x_1, x_2, \dots, x_n)$ должна быть 
представлена в виде 

\begin{equation}
f(x_1,x_2,\dots,x_n) = \displaystyle \prod_{k=1}^n{f_k(x_k)}.   
\label{eq:25}
\end{equation}

Степень зависимости случайных величин характеризуется понятием ковариации 
случайных величин $X$ и $Y$. Ковариация определяется как 

\begin{equation}
cov(X,Y) = E[(X-E(X))(Y-E(Y))].   
\label{eq:26}
\end{equation}

Определение~(\ref{eq:26}) можно переписать в виде 
 
\begin{equation}
cov(X,Y) = E(XY)-E(X)E(Y).   
\label{eq:27}
\end{equation}

Если $cov(X,Y) = 0$, то говорят, что случайные величины $X$ и $Y$ являются некоррелированными. 
Легко видеть, что для независимых случайных величин $X$ и $Y$ $cov(X,Y) = 0$. 
Однако обратное утверждение неверно. 

Для дисперсии от суммы двух случайных величин справедлива формула 

\begin{equation}
D(X+Y) = D(X)+D(Y)+2cov(X,Y).   
\label{eq:28}
\end{equation}

\noindent
Отсюда следует, что для независимых величин дисперсия суммы равна сумме дисперсий. 

Часто используют коэффициент корреляции $\rho(x,y)$ двух случайных величин $X$ и $Y$, 
определяемый как 

\begin{equation}
\rho(X,Y) = \displaystyle \frac{cov(X,Y)}{\sqrt{D(X)D(Y)}}.   
\label{eq:29}
\end{equation}

\noindent
Можно показать, что 

\begin{equation}
|\rho(X,Y)| \le 1.   
\label{eq:30}
\end{equation}

\noindent
Более того, если $|\rho(X,Y)| = 1$, то случайные величины $X$ и $Y$ связаны линейной 
зависимостью, то есть существуют числа $a$ и $b$ такие, что  

\begin{equation}
P(Y=aX+b) = 1.   
\label{eq:31}
\end{equation}

При этом если $\rho(X,Y) = 1$, то $a>0$, и, если $\rho(X,Y) = -1$, то $a<0$. 
Можно сказать, что величина коэффициента корреляции характеризует степень линейной 
зависимости.

\subsection{Характеристические функции}

Для случайной переменной $X$ с плотностью вероятности $f(x)$ характеристическая функция 
определяется как 

\begin{equation}
\Phi_X(t) = E(e^{itX}) = \displaystyle \int_{-\infty}^{\infty}{e^{itx}f(x)dx}.    
\label{eq:32}
\end{equation}

\noindent
Для дискретной случайной величины $X$ с конечным числом значений

\begin{equation}
\Phi_X(t) = \displaystyle \sum_{k}{p_ke^{itx_k}}.    
\label{eq:33}
\end{equation}

Характеристическая функция $\Phi_X(t)$ полностью определяет плотность вероятности случайной 
величины, а именно, 

\begin{equation}
f(x) = \displaystyle \frac{1}{2\pi}\int_{-\infty}^{\infty}{\Phi_X(t)e^{-itx}dt}.    
\label{eq:34}
\end{equation}

Для функции $\Phi_X(t)$ справедливы соотношения 

\begin{equation}
\Phi_X(0) = 1,     
\label{eq:35}
\end{equation}

\begin{equation}
|\Phi_X(t)| \le 1,     
\label{eq:36}
\end{equation}

\begin{equation}
\Phi_{aX+b}(t) = e^{ibt}\Phi_X(at).     
\label{eq:37}
\end{equation}

\noindent
Здесь $a$ и $b$ -- константы. 

Если $X$ и $Y$ независимые случайные переменные с характеристическими функциями 
$\Phi_X(t)$ и $\Phi_Y(t)$, то справедливо равенство 

\begin{equation}
\Phi_{X+Y}(t) = \Phi_X(t)\cdot \Phi_Y(t).     
\label{eq:38}
\end{equation}

Обобщение равенства~(\ref{eq:38}) на случай $N$ независимых переменных имеет вид 

\begin{equation}
\Phi_{X_1+\dots+X_N}(t) = \prod_{i=1}^N \Phi_{X_i}(t).     
\label{eq:39}
\end{equation}

\subsection{Основные функции распределения вероятностей}

В этом разделе мы приводим основные функции распределения вероятностей, наиболее часто используемые 
в физике высоких энергий. 

\subsubsection{Биномиальное распределение}

Для биномиального распределения вероятность наступления $r$ событий при $N$ испытаниях 
(r - целое неотрицательное число, $1\le r \le N$) равна 

\begin{equation}
P(r) = \displaystyle \frac{N!}{r!(N-r)!}p^r(1-p)^{N-r},     
\label{eq:40}
\end{equation}

\noindent
где $0 \le p \le 1$ ожидаемая вероятность появления события в одном испытании.
Среднее значение при этом равно 

\begin{equation}
E(r)=Np,     
\label{eq:41}
\end{equation}

\noindent
а дисперсия равна 

\begin{equation}
D(r)=Np(1-p).     
\label{eq:42}
\end{equation}

Биномиальное распределение представляет решение следующей задачи. Пусть имеется общая (или как 
еще говорят генеральная) совокупность, состоящая из $M$ некоторых объектов (например, 
это могут быть события в экспериментальной установке), из которых $R=pM$ 
объектов обладают некоторым признаком $C$, а остальные $M-R=(1-p)M$ не обладают им, так что 
$pM+(1-p)M=M$. Пусть из этой общей совокупности сделана выборка путем независимого случайного 
выбора одного за другим $N$ объектов таким образом, чтобы не нарушилось соотношение между объектами 
с признаком $C$ и без признака $C$. Это возможно в двух случаях: либо общая совокупность бесконечна, 
либо выбранные объекты возвращаются в общую совокупность перед каждым новым выбором. 
В таком случае вероятность того, что из общего числа $N$ объектов выборки $r$ объектов будут обладать 
признаком $C$ будет определяться формулой~(\ref{eq:40}).

Выбор объектов с признаком $C$ или без признака $C$ из общей совокупности называют испытаниями Бернулли. 
При этом выбор объекта с признаком $C$ называют событием или успехом. 

\begin{figure}[htpb]
\begin{center}
   \resizebox{4.1in}{!}{\includegraphics{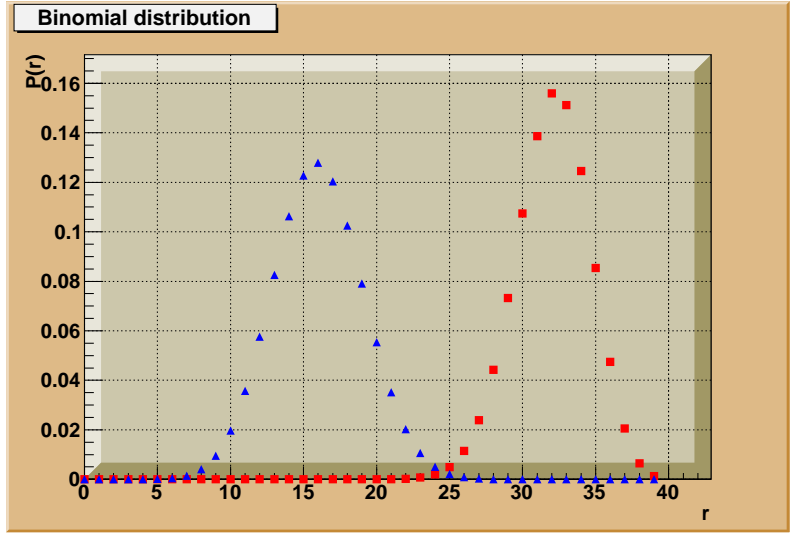}} 
\caption{Распределения вероятностей для биномиального распределения в случае $N=40$ испытаний 
с разными значениями параметра $p$: слева $p=0.4$, справа $p=0.8$.}
    \label{fig:Binom} 
  \end{center}
\end{figure}

На Рис.~\ref{fig:Binom} показаны распределения вероятностей для биномиального распределения  
в случае $N=40$ испытаний с разными значениями параметра $p$: слева $p=0.4$, справа $p=0.8$.

\subsubsection{Распределение Пуассона}

Для распределения Пуассона функция распределения вероятностей имеет вид 

\begin{equation}
P(r|\mu) = \displaystyle \frac{\mu^re^{-\mu}}{r!},     
\label{eq:43}
\end{equation}

\noindent
где $r$ целое неотрицательное число, а параметр $\mu \ge 0$ -- действительное число. 
Среднее значение и дисперсия равны 

\begin{equation}
E(r)=\mu,     
\label{eq:44}
\end{equation}

\begin{equation}
D(r)=\mu.     
\label{eq:45}
\end{equation}

Распределение Пуассона можно рассматривать как предельный случай биномиального распределения 
при выполнении следующего условия. Пусть независимые испытания производятся сериями длины 
$n$ ($n$-я серия, следовательно, состоит из $n$ независимых испытаний), а вероятность 
$p^{(n)}$ некоторого события для всех испытаний $n$-й серии одна и та же, но с 
увеличением $n$ стремится к нулю так, что произведение $np^{(n)}$ сохраняет 
постоянное значение $\mu$:

\begin{equation}
lim_{n\rightarrow\infty}np^{(n)}=\mu,      
\label{eq:45ap}
\end{equation}

\noindent
причем очевидно, что $\mu>0$. 
При этом условии в пределе $n \rightarrow \infty$ 
вероятность $P(r|\mu)$ того, что рассматриваемое событие в бесконечно 
длинной серии испытаний появится ровно $r$ раз описывается формулой~(\ref{eq:43}).

\begin{figure}[htpb]
\begin{center}
   \resizebox{4.1in}{!}{\includegraphics{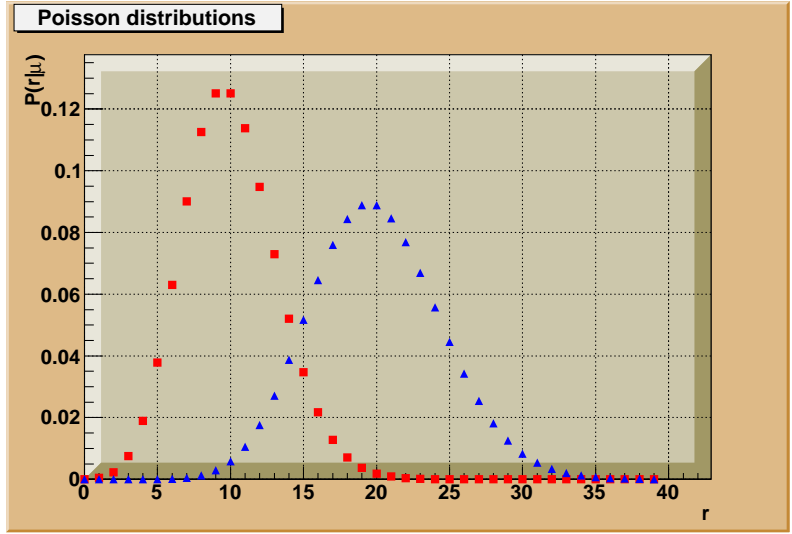}} 
\caption{Распределения вероятностей для распределения Пуассона 
с разными значениями параметров: слева $\mu=10$, справа $\mu=20$.}
    \label{fig:Poiss} 
  \end{center}
\end{figure}

На Рис.~\ref{fig:Poiss} показаны распределения вероятностей для распределения Пуассона 
с разными значениями параметров: слева $\mu=10$, справа $\mu=20$.

\subsubsection{Нормальное одномерное распределение (распределение Гаусса)}

Пусть дана случайная величина $X$, подчиняющаяся нормальному распределению с параметрами 
$\mu$ и $\sigma$. Параметры являются действительными числами: 
$-\infty \le \mu \le \infty$ и $0 \le \sigma \le \infty$.

Плотность вероятности для одномерного нормального распределения равна 

\begin{equation}
f(x|\mu,\sigma^2)=\displaystyle 
\frac{1}{\sigma\sqrt{2\pi}}e^{-\frac{1}{2}\frac{(x-\mu)^2}{\sigma^2}},      
\label{eq:46}
\end{equation}
 
\noindent
где $x$ -- действительное число, $-\infty \le x \le \infty$. 
Среднее значение и дисперсия равны 

\begin{equation}
E(X)=\mu,     
\label{eq:47}
\end{equation}

\begin{equation}
D(X)=\sigma^2.     
\label{eq:48}
\end{equation}

Очень важное свойство нормального распределения состоит в том, что, 
если две случайные величины $X$ и $Y$ подчиняются нормальному распределению 
со средними $\mu_X$ и $\mu_Y$ и дисперсиями $\sigma_X$ и $\sigma_Y$ соответственно, 
то случайная величина $Z=X+Y$ распределена согласно нормальному распределению со 
средним $\mu_Z=\mu_X+\mu_Y$ и дисперсией $\sigma_Z= \sqrt{\sigma_X^2+\sigma_Y^2}$.
Доказательство этого факта заключается во взятии интеграла в формуле 
\begin{center}
$F(z|\mu_X,\mu_Y,\sigma_X,\sigma_Y)=\int{dxdy\delta(z-x-y)N(x|\mu_X,\sigma_X)N(y|\mu_Y,\sigma_Y)}$.
\end{center}
Отсюда, в частности, следует, что если случайные величины $X_1,X_2, \dots, X_N$ подчиняются 
нормальному распределению с универсальными средним $\mu$ и дисперсией $\sigma$, то 
случайная величина $X(N) = \displaystyle \frac{1}{N}\sum_{k=1}^NX_k$ подчиняется 
нормальному распределению со средним $\mu_X=\mu$ и дисперсией 
$\sigma_X^2 = \displaystyle \frac{1}{N}\sigma^2$. 
В пределе $N\rightarrow \infty$ случайная величина $X(N)$ с вероятностью равной $1$ 
принимает значение $\mu$.

Характеристическая функция для нормального распределения равна 

\begin{equation}
\Phi(t) = \displaystyle e^{it\mu-\frac{1}{2}t^2\sigma^2}.    
\label{eq:49}
\end{equation}

Нормальное распределение также можно получить путем предельного перехода из биномиального 
распределения. Так, если $X$ -- число событий (успехов) в $n$ испытаниях Бернулли, а вероятность 
появления события в одном испытании остается той же самой, то какими бы не были действительные 
числа $a$ и $b$ ($a<b$), 

\begin{center}
$\displaystyle P(a\le\frac{X-np}{\sqrt{np(1-p)}}\le b) 
{\displaystyle {\mathop{\longrightarrow}\limits_{n\rightarrow \infty}}} 
\frac{1}{\sqrt{2\pi}}\int_a^b{e^{-\frac{x^2}{2}}dx}$.
\end{center}

\noindent
Это утверждение называется теоремой Муавра-Лапласа.

Весьма важной является также теорема Ляпунова, которая утверждает, что если значения 
независимых случайных величин будут малы в сравнении с их суммой, то, при неограниченном 
возрастании числа этих величин, распределение их суммы становится приближенно нормальным.
Теорема работает при весьма общих условиях для любых распределений. Это позволяет при  
исследовании случайной величины, являющейся суммой многих случайных величин, влияние 
которых на саму сумму незначительно, заранее предполагать, что распределение ошибок при 
измерении этой случайной величины будет нормальным.

\begin{figure}[htpb]
\begin{center}
   \resizebox{4.1in}{!}{\includegraphics{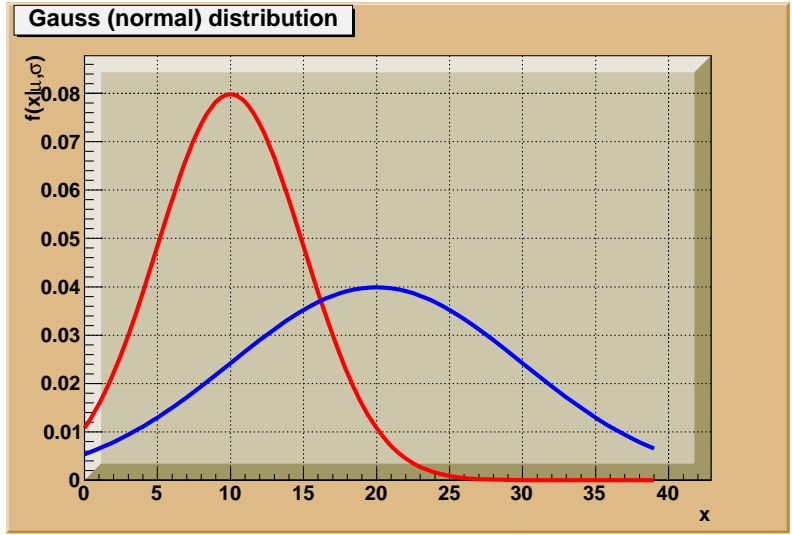}} 
\caption{Распределения плотности вероятности для нормального распределения  
с разными значениями параметров: слева $\mu=10$, $\sigma=5$, справа $\mu=20$, $\sigma=10$.}
    \label{fig:Gauss} 
  \end{center}
\end{figure}

На Рис.~\ref{fig:Gauss} показаны плотности вероятности для нормального распределения  
с разными значениями параметров: слева $\mu=10$, $\sigma=5$, справа $\mu=20$, $\sigma=10$.

\subsubsection{Многомерное нормальное распределение}

Пусть дана $k$-мерная случайная величина $\vec X$, подчиняющаяся многомерному нормальному 
распределению с параметрами $\vec \mu$ ($k$-мерный вектор действительных чисел) и 
$[V]$ (матрица размерности $[k\times k]$ с положительными собственными числами). 

Плотность вероятности равна 

\begin{equation}
f(X|\vec \mu,[V]) = \displaystyle 
\frac{1}{(2\pi)^{\frac{k}{2}}V^{\frac{1}{2}}}
e^{-\frac{1}{2}(\vec x - \vec \mu)^T[V]^{-1}(\vec x - \vec \mu)},   
\label{eq:50}
\end{equation}

\noindent
где переменная $\vec x = (x_1,x_2,\dots,x_k)$ -- $k$-мерный вещественный вектор,   
$V=det[V]$. 

Среднее значение равно 

\begin{equation}
E(\vec X) = \vec \mu.    
\label{eq:51}
\end{equation}

Ковариация определяется как 

\begin{equation}
cov(X_i,X_j) \equiv E(X_iX_j)-E(X_i)E(X_j),~i,j=1,k.    
\label{eq:52}
\end{equation}

Корреляция определяется как 

\begin{equation}
\rho_{ij} = \displaystyle \frac{cov(X_i,X_j)}{\sigma_{X_i}\sigma_{X_j}},    
\label{eq:53}
\end{equation}

\noindent
где $\sigma_{X_i}^2=E((X_i - E(X_i))^2)$. Нетрудно доказать, используя неравенство 
Коши-Буняковского, что $-1 \le \rho_{ij} \le 1$.

Дисперсия и ковариация равны, соответственно,  

\begin{equation}
D(X_i) = V_{ii},~i=1,k,    
\label{eq:54}
\end{equation}

\begin{equation}
cov(X_i,X_j) = V_{ij},~i,j=1,k.    
\label{eq:55}
\end{equation}

Рассмотрим в качестве часто встречающегося примера двумерное нормальное 
распределение~\footnote{Здесь для простоты мы рассматриваем случай 
$\mu_X=\mu_Y=0$. Обобщение для общего случая очевидно.}. 

Путем замены координат в плоскости (X,Y) 

$X' = X~cos \varphi - Y~sin \varphi$

$Y' = X~sin \varphi + Y~cos \varphi$

\noindent
можно выбрать новую систему координат, в которой функция 

$(X,Y)^T [V]^{-1} (X,Y) = \displaystyle \frac{X'^2}{\sigma_{X'}^2}+\frac{Y'^2}{\sigma_{Y'}^2}$

\noindent
диагональна и плотность нормального распределения $N(X',Y'|\dots)$ принимает вид 

$f(x',y')=\displaystyle \frac{1}{2\pi\sigma_{X'}\sigma_{Y'}} 
e^{-\frac{1}{2}(\frac{x'^2}{\sigma_{X'}^2}+\frac{y'^2}{\sigma_{Y'}^2})}.$ 

\noindent
Вследствие того, что вращение сохраняет детерминант матрицы $[V]^{-1}$ имеем равенство 

$\displaystyle \frac{1}{\sigma_{X'}^2\sigma_{Y'}^2} = 
\displaystyle \frac{1}{\sigma_{X}^2\sigma_{Y}^2(1-\rho_{XY}^2)}.$ \\

Матрица $[V]$ и коэффициент корреляции $\rho_{XY}$ в новых  координатах \\

\begin{center}
$[V] =\left(\begin{array}{cc}
 \sigma_X^2                  & \rho_{XY} \sigma_X \sigma_Y  \\
 \rho_{XY}\sigma_X \sigma_Y  & \sigma_Y^2    
\end{array}\right) = $
\end{center}

\begin{center}
$\left(\begin{array}{cc}
 cos^2\varphi\sigma_{X'}^2+sin^2\varphi\sigma_{Y'}^2 & 
               sin\varphi\cos\varphi(\sigma_{Y'}^2-\sigma_{X'}^2) \\
 sin\varphi\cos\varphi(\sigma_{Y'}^2-\sigma_{X'}^2) & 
               sin^2\varphi\sigma_{X'}^2+cos^2\varphi\sigma_{Y'}^2    
\end{array}\right),$
\end{center}

$\rho_{XY} = \displaystyle \frac{cov(X,Y)}{\sigma_X\sigma_Y}= 
\frac{sin2\varphi(\sigma_{Y'}^2-\sigma_{X'}^2)}
{\sqrt{sin2\varphi(\sigma_{Y'}^4-\sigma_{X'}^4)+2\sigma_{Y'}^2\sigma_{X'}^2}},$

$\tan 2\varphi = \displaystyle \frac{2\rho_{XY}\sigma_X\sigma_Y}{\sigma_{Y'}^2-\sigma_{X'}^2}$, 

$f(x,y)=\displaystyle \frac{1}{2\pi\sigma_{X}\sigma_{Y}\sqrt{1-\rho_{XY}^2}} 
e^{-\frac{1}{2(1-\rho_{XY}^2)}(\frac{x^2}{\sigma_{X}^2}+\frac{y^2}{\sigma_{Y}^2}-
\frac{2xy\rho_{XY}}{\sigma_X\sigma_Y})}.$ 

При проецировании функции распределения на оси $X$ и $Y$ получаем 

$f_X(x) \equiv \displaystyle \int_{-\infty}^{\infty}{f(x,y)dy} = 
\frac{1}{\sqrt{2\pi}\sigma_X}e^{-\frac{x^2}{2\sigma_X^2}},$

$f_Y(y) \equiv \displaystyle \int_{-\infty}^{\infty}{f(x,y)dx} = 
\frac{1}{\sqrt{2\pi}\sigma_Y}e^{-\frac{y^2}{2\sigma_Y^2}}.$

Характеристическая функция равна 

\begin{equation}
\Phi(\vec t) = \displaystyle e^{i\vec t\vec \mu-\frac{1}{2}\vec t^T[V]\vec t}.    
\label{eq:56}
\end{equation}

\begin{figure}[htpb]
\begin{center}
   \resizebox{4.1in}{!}{\includegraphics{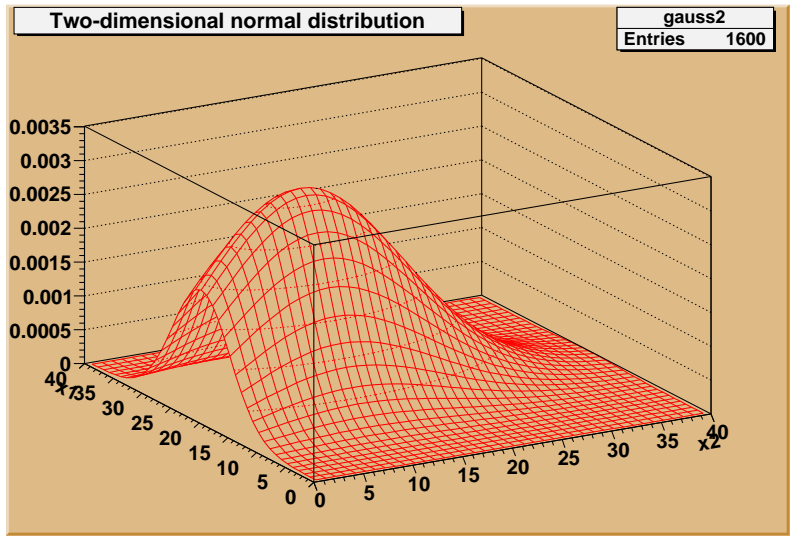}} 
\caption{Распределения плотности вероятности для двумерного нормального распределения  
со следующими значениями параметров: $\mu_1=20$, $\sigma_1=10$, $\mu_2=10$, $\sigma_2=5$, 
$\rho_{12}=0$.}
    \label{fig:Gauss2} 
  \end{center}
\end{figure}

На Рис.~\ref{fig:Gauss2} показана плотность вероятности для двумерного нормального распределения  
со следующими значениями параметров: $\mu_1=20$, $\sigma_1=10$, $\mu_2=10$, $\sigma_2=5$, 
$\rho_{12}=0$.

\subsubsection{$\Gamma$-распределение}

$\Gamma$-распределение относится к числу основных типов распределений. Напомним, что
$\Gamma$-функция 

\begin{equation}
\Gamma(b)=\displaystyle \int_0^{\infty}{x^{b-1}e^{-x}dx} = (b-1)!
\label{eq:84}
\end{equation}

\noindent 
представляет собой интеграл, равный факториалу какого угодно положительного числа $b$.
Этот факт можно реализовать в виде функционального уравнения $\Gamma$-функции, представляющего основное 
свойство $\Gamma$-функции,

$\Gamma(b+1)=b\Gamma(b)$.

Распределение непрерывной случайной величины $X$, зависящей от одного параметра $b$ с плотностью 
вероятности 

$f(x|b) = \displaystyle \frac{x^{b-1}e^{-x}}{\Gamma(b)}$ 

\noindent
называется $\Gamma$-распределением. Кумулятивная функция распределения есть 

$F(x|b) = \displaystyle 
\frac{1}{\Gamma(b)}\int_0^{x}{x'^{b-1}e^{-x'}dx'}$. 

Математическое ожидание и дисперсия случайной величины совпадают и равны 
значению параметра $b$.

Также $\Gamma$-распределением $\Gamma(a,b)$ 
называется распределение с двумя параметрами $a$ и $b$ и c 
плотностью вероятности 

\begin{equation}
f(x|a,b) = \displaystyle \frac{a(ax)^{b-1}e^{-ax}}{\Gamma(b)}\,,
\label{eq:69}
\end{equation}

\noindent
где $a$ и $b$ положительные действительные числа.  В этом случае  
среднее значение и дисперсия равны  

\begin{equation}
E(X)=\displaystyle \frac{b}{a},     
\label{eq:70}
\end{equation}

\begin{equation}
D(X)=\displaystyle \frac{b}{a^2}.     
\label{eq:71}
\end{equation}

Характеристическая функция равна 

\begin{equation}
\Phi(t) = \displaystyle (1-\frac{it}{a})^{-b}.    
\label{eq:72}
\end{equation}

$\Gamma$-распределение тесно связано с нормальным распределением. Параметр $b$ определяет 
форму распределения, а параметр $a$ является параметром масштаба. 

\begin{figure}[htpb]
\begin{center}
   \resizebox{4.1in}{!}{\includegraphics{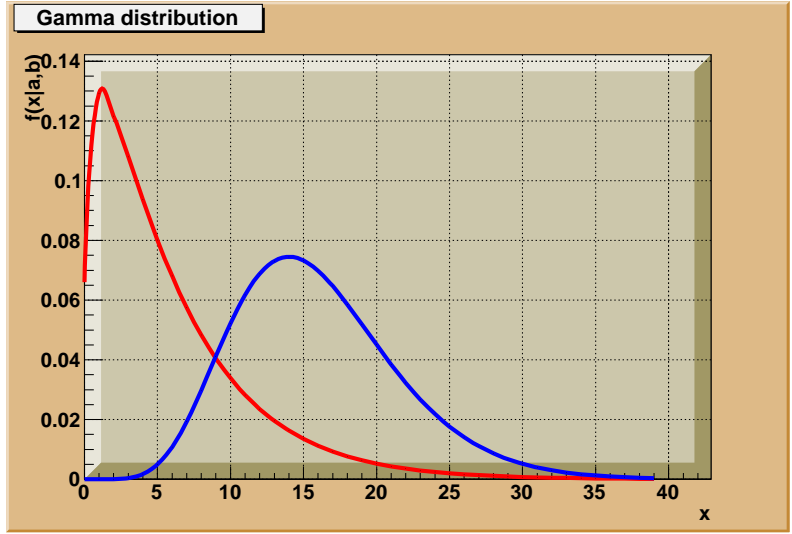}} 
\caption{Распределения плотности вероятности для $\Gamma$-распределения  
с разными значениями параметров: слева $a=1.2$, $b=5$, справа $a=8$, $b=2$.}
    \label{fig:Gamma} 
  \end{center}
\end{figure}

На Рис.~\ref{fig:Gamma} показаны плотности вероятности для  $\Gamma$-распределения  
с разными значениями параметров: слева $a=1.2$, $b=5$, справа $a=8$, $\sigma=2$.

\subsubsection{$\chi^2$-распределение}\label{sec:chi2}

Для $\chi^2$-распределения 
плотность вероятности равна 

\begin{equation}
f(x|N) = \displaystyle 
\frac{\frac{1}{2}(\frac{x}{2})^{\frac{N}{2}-1}}{\Gamma(\frac{N}{2})}e^{-\frac{x}{2}},    
\label{eq:57}
\end{equation}

\noindent
где $0 \le x \le \infty$, а $N$-целое положительное число. 

Среднее значение и дисперсия равны 

\begin{equation}
E(X)=N,     
\label{eq:58}
\end{equation}

\begin{equation}
D(X)=2 N.     
\label{eq:59}
\end{equation}
 
Характеристическая функция равна 

\begin{equation}
\Phi(t) = \displaystyle (1-2it)^{-\frac{N}{2}}.    
\label{eq:60}
\end{equation}

$\chi^2$-распределение связано с нормальным распределением. А именно, пусть 
$X_1, X_2, \dots, X_N$ являются независимыми случайными величинами, подчиняющимися  
стандартному нормальному распределению (то есть $\mu_n=0,~D(X_n)=1, n=1,\dots,N)$. Тогда сумма 
квадратов 

\begin{equation}
\chi^2_{N} = \displaystyle \sum_{i=1}^Nx_i^2    
\label{eq:61}
\end{equation}

\noindent
распределена по $\chi^2$-распределению с $N$ степенями свободы.
$\chi^2$-распределение возникло с развитием статистики для описания поведения функций от наблюденных 
данных и является частным случаем $\Gamma$-распределения 

$\chi^2_{N} =  \displaystyle \Gamma(\frac{1}{2},\frac{N}{2}).$

\begin{figure}[htpb]
\begin{center}
   \resizebox{4.1in}{!}{\includegraphics{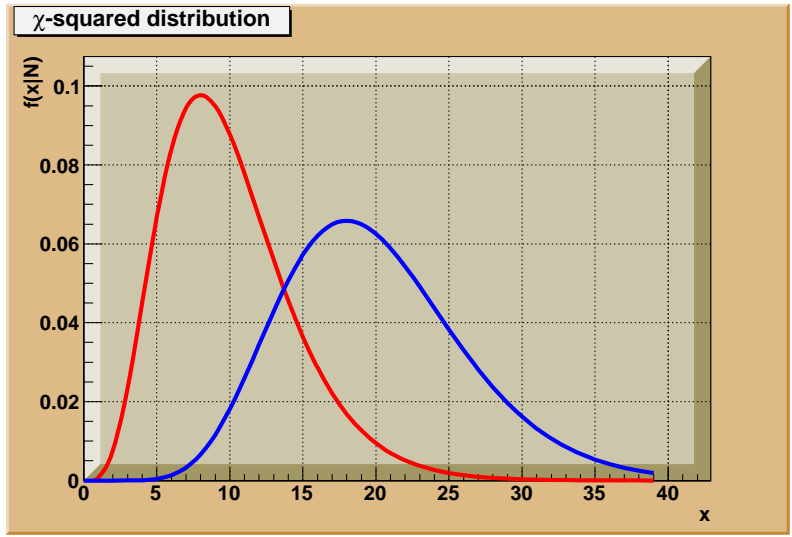}} 
\caption{Распределения плотности вероятности для $\chi^2$-распределения  
с разными значениями параметров: слева $N=10$, справа $N=20$.}
    \label{fig:chi2} 
  \end{center}
\end{figure}

На Рис.~\ref{fig:chi2} показаны плотности вероятности для $\chi^2$-распределения  
с разными значениями параметров: слева $N=10$, справа $N=20$.

Функция $\chi_N=\displaystyle \sqrt{\chi^2_N}$ имеет $\chi(N)$-распределение с функцией плотности 

$f(x|N) = \displaystyle \frac{(\frac{1}{2})^{\frac{N}{2}-1}x^{N-1}}{\Gamma(\frac{N}{2})}
\displaystyle e^{-\frac{1}{2}x^2}.$ 
   
\noindent
В асимптотическом пределе $\chi^2(N)$- и $\chi(N)$-распределения стремятся к нормальному, 
причем эти распределения с хорошей точностью можно считать нормальными при $n>30$. 

Можно также показать, что при больших значениях $N$ величины 
$S_N=\displaystyle \frac{\chi_N^2-N}{\sqrt{2N}}$ и 
$T_N = \displaystyle \sqrt{2\chi^2_N}-\sqrt{2N-1}$ распределены по стандартному 
нормальному закону $N(0,1)$.

\subsubsection{Логнормальное распределение}

Плотность вероятности логарифмически нормального распределения (логнормального) равна 

\begin{equation}
f(x|\mu,\sigma) = \displaystyle 
\frac{1}{\sqrt{2\pi}\sigma}\frac{1}{x}e^{-\frac{1}{2\sigma^2}(ln~x-\mu)^2},    
\label{eq:62}
\end{equation}

\noindent
где $\mu$-действительное число, а $\sigma$-положительное действительное число. 

Среднее значение и дисперсия равны 

\begin{equation}
E(X)=\displaystyle e^{\mu-\frac{1}{2}\sigma^2},     
\label{eq:63}
\end{equation}

\begin{equation}
D(X)=\displaystyle e^{2\mu+\sigma^2}(e^{\sigma^2}-1).     
\label{eq:64}
\end{equation}
 
Логнормальное распределение описывает случайную переменную, логарифм которой подчиняется 
нормальному распределению. 

\begin{figure}[htpb]
\begin{center}
   \resizebox{4.1in}{!}{\includegraphics{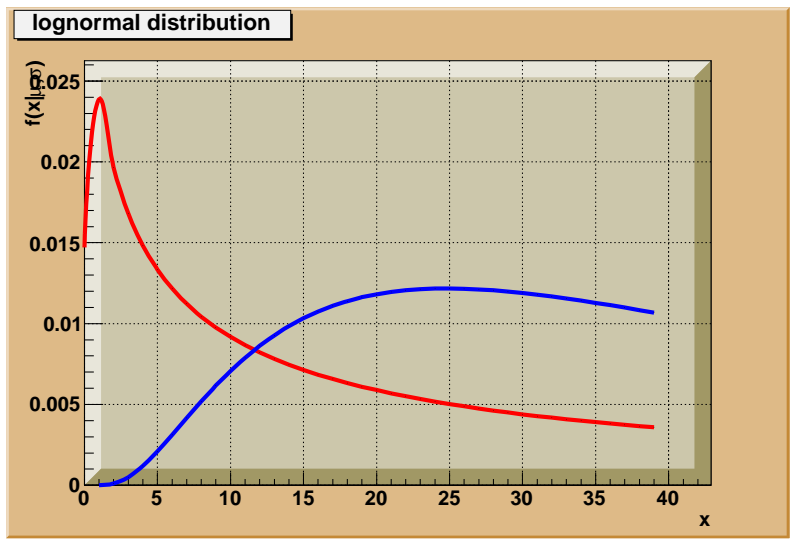}} 
\caption{Распределения плотности вероятности для логнормального распределения  
с разными значениями параметров: слева $\mu=5$, $\sigma=2.5$, справа $\mu=4$, $\sigma=0.9$.}
    \label{fig:loGauss} 
  \end{center}
\end{figure}

На Рис.~\ref{fig:loGauss} показаны плотности вероятности для логнормального распределения  
с разными значениями параметров: слева $\mu=5$, $\sigma=2.5$, справа $\mu=4$, $\sigma=0.9$.

\subsubsection{Равномерное распределение}

Плотность вероятности равна 

\begin{equation}
f(x|a,b) = \displaystyle \frac{1}{b-a}\,, 
\label{eq:65}
\end{equation}

\noindent
где $x$ -- действительная переменная, удовлетворяющая неравенству $a\le x \le b$, 
$a$ и $b$ -- действительные числа. 

Среднее значение и дисперсия равны 

\begin{equation}
E(X)=\displaystyle \frac{a+b}{2},     
\label{eq:66}
\end{equation}

\begin{equation}
D(X)=\displaystyle \frac{(b-a)^2}{12}.     
\label{eq:67}
\end{equation}

Характеристическая функция равна 

\begin{equation}
\Phi(\vec t) = \displaystyle 
\frac{e^{itb}-e^{ita}}{it(b-a)}.    
\label{eq:68}
\end{equation}

\begin{figure}[htpb]
\begin{center}
   \resizebox{4.1in}{!}{\includegraphics{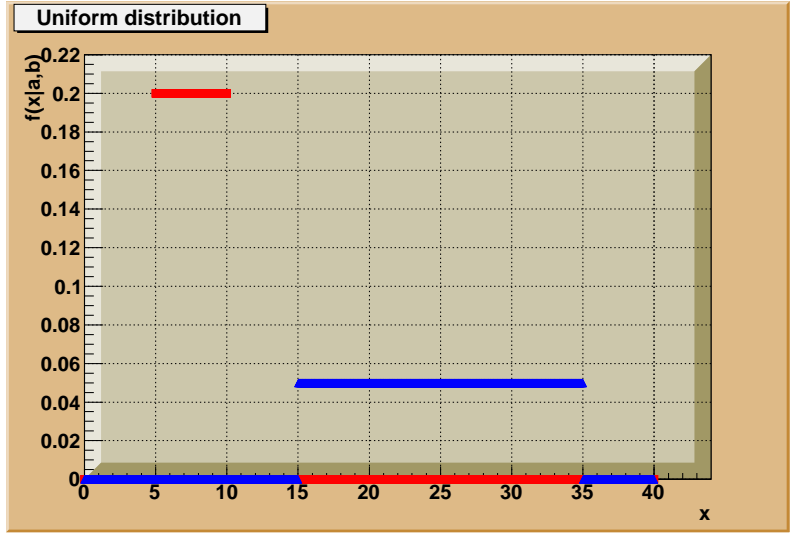}} 
\caption{Распределения плотности вероятности для равномерного распределения  
с разными значениями параметров: слева $a=5$, $b=10$, справа $a=15$, $b=35$.}
    \label{fig:Unif} 
  \end{center}
\end{figure}

На Рис.~\ref{fig:Unif} показаны плотности вероятности для равномерного распределения  
с разными значениями параметров: слева $a=5$, $b=10$, справа $a=15$, $b=35$.

\subsection{Закон больших чисел. Центральная предельная теорема}

\subsubsection{Закон больших чисел}

Пусть $X_1, X_2, \dots, X_N$ - независимые одинаково распределенные 
случайные величины и пусть $E(X)=\mu$. Тогда с вероятностью 1 выполняется соотношение 

\begin{equation}
\displaystyle \mathop{lim}\limits_{N\rightarrow\infty} \frac{1}{N}\sum_{k=1}^N{X_k}=\mu.    
\label{eq:73}
\end{equation}

Закон больших чисел (\ref{eq:73}) следует из неравенства Чебышева (\ref{eq:22}) и 
из того факта, что для независимых величин $X_1, \dots, X_N$ с одинаковым средним значением и 
дисперсией  

\begin{equation}
\displaystyle D(\frac{X_1+\dots+X_N}{N})=\frac{D(X_1)}{N}.    
\label{eq:74}
\end{equation}

\subsubsection{Центральная предельная теорема}

Центральная предельная теорема утверждает, что сумма $N$ независимых случайных величин 
$X_k$ стремится в пределе $N\rightarrow \infty$ к нормальному распределению, а именно

\begin{equation}
\displaystyle 
\frac{\displaystyle \sum_{k=1}^N{X_k}-\sum_{k=1}^N{\mu_k}}
{\sqrt{\displaystyle \sum_{k=1}^N{\sigma_k^2}}} 
\rightarrow N(0,1).
\label{eq:75}
\end{equation}

\noindent
Здесь $\mu_k$ и $\sigma_k^2$ -- средние значения и дисперсии случайных величин $X_k$, 
а $N(0,1)$ стандартное нормальное распределение с плотностью распределения 

\begin{equation}
f(x|\mu=0,\sigma=1) = \displaystyle \frac{1}{\sqrt{2\pi}}e^{-\frac{x^2}{2}}.    
\label{eq:76}
\end{equation}

Доказательство теоремы можно найти, например, в монографии~\cite{Kramer}. 
Центральная предельная теорема по сути дела иллюстрирует тот факт, что нормальное распределение 
явно выделено из всех других распределений.

\subsection{Информация}

Рассмотрим $N$ независимых измерений случайной величины $X$, зависящей от вектора 
параметров $\vec \theta$. Их можно рассматривать как 
случайный вектор c $N$ случайными одинаково распределенными компонентами $X_k,~k=1,2,\dots,N$, 
то есть как $\vec X=(X_1,X_2,\dots,X_N)$. Множество всех допустимых значений $\vec X$ 
обозначим как $\Omega_{\vec \theta}$.

В силу независимости $X_k$ плотность вероятности 
$\vec X$ равна 

\begin{equation}
P(\vec X|\vec \theta) = \displaystyle P(X_1,X_2, \dots, X_n|\vec \theta) = 
\prod_{k=1}^N{f_k(x,\vec \theta)},     
\label{eq:77}
\end{equation}

\noindent
где $f_k(x,\vec \theta)$ -- плотность вероятности случайной величины $X_k$, а $\vec \theta$ -- параметры,
описывающие распределение. В случае, когда переменная $\vec X$ заменяется на наблюдаемые данные 
$\vec x=(x_1,x_2,\dots,x_N)$ функция $P$ уже не является функцией распределения случайных величин, 
а называется функцией правдоподобия, которая зависит от параметров распределения $\vec \theta$

\begin{equation}
L(\vec x|\vec \theta) = \displaystyle P(\vec x|\vec \theta).    
\label{eq:78}
\end{equation}

Любая функция $T=T(x_1,x_2,\dots,x_N)$ от измеренных величин  называется статистикой.

Информация по Фишеру~\cite{Fisher} определяется как 

\begin{equation}
I_x(\theta)=\displaystyle 
E[(\frac{\partial ln L(x|\theta)}{\partial \theta})^2] = \displaystyle 
\int_{\Omega_{\vec \theta}}
{(\frac{\partial ln L(x|\theta)}{\partial \theta})^2L(x|\theta)dx}.    
\label{eq:79}
\end{equation}

В случае, когда $\vec \theta=(\theta_1,\theta_2,\dots,\theta_N)$-$k$-мерный вектор, 
обобщение формулы~(\ref{eq:79}) есть формула 

$[I_x(\vec \theta)]_{ij}=\displaystyle 
E[\frac{\partial ln L(x|\vec \theta)}{\partial \theta_i}\cdot
\frac{\partial ln L(x|\vec \theta)}{\partial \theta_j}]$ 

\begin{equation}
= \displaystyle 
\int_{\Omega_{\vec \theta}}{[\frac{\partial ln L(x|\vec \theta)}{\partial \theta_i}\cdot
\frac{\partial ln L(x|\vec \theta)}{\partial \theta_j}]{L(x|\vec \theta)dx}}.    
\label{eq:80}
\end{equation}

В случае, когда случайная величина $X$ распределена по нормальному закону с неизвестным средним 
$\mu$ и известной дисперсией $\sigma^2$, информация для одного наблюдения $x$ равна 

$I_1(\mu,\sigma)= \displaystyle 
-\int_{-\infty}^{\infty}{\frac{\partial^2}{\partial \mu^2}
(-\frac{x-\mu}{2\sigma^2}-ln~\sigma\sqrt{2\pi})\frac{1}{\sqrt{2\pi}\sigma}
e^{-\frac{1}{2}\frac{(x-\mu)^2}{\sigma^2}}dx}=$

\begin{equation}
\int_{-\infty}^{\infty}{\frac{1}{\sigma^2}\frac{1}{\sqrt{2\pi}\sigma}
e^{-\frac{1}{2}\frac{(x-\mu)^2}{\sigma^2}}dx = \frac{1}{\sigma^2}}.
\label{eq:81}
\end{equation}

В случае $N$ независимых измерений $x_1,x_2,\dots,x_N$ случайной величины $X$, 
подчиняющейся нормальному распределению, информация равна 

$I_N(\mu,\sigma)= \displaystyle 
-\int_{-\infty}^{\infty}\dots\int_{-\infty}^{\infty}{\frac{\partial^2}{\partial \mu^2}
(\sum_{i=1}^N(-\frac{(x_i-\mu)^2}{2\sigma^2})-ln~\sigma\sqrt{2\pi})\times}$

\begin{equation}
\displaystyle
{\prod_{i=1}^N\frac{1}{\sqrt{2\pi}\sigma}e^{-\frac{1}{2}\frac{(x_i-\mu)^2}{\sigma^2}}dx_i}
=\frac{N}{\sigma^2}=NI_1(\mu,\sigma).
\label{eq:82}
\end{equation}

В дискретном случае, когда у нас случайная величина $X$ принимает значения $1,2,\dots,N$ 
с вероятностями $p_1,p_2,\dots,p_N$ соответственно, информация определяется как 

\begin{equation}
H(p_1\dots p_N) = \displaystyle -k\sum_{k=1}^N p_i ln~p_i.    
\label{eq:83}
\end{equation}

\noindent
где $k$-некоторая нормировочная константа, а $\displaystyle \sum_ip_i=1,~p_i\ge 0$. 
Можно показать, что информация~(\ref{eq:83}) максимальна при 
$\displaystyle p_1=p_2=\dots=p_N=\frac{1}{N}$. 



\newpage

\section{Основные задачи и методы математической статистики}

\subsection{Основные задачи статистики}

Если основной задачей теории вероятностей является построение модели случайных 
явлений, позволяющих выразить вероятности более сложных явлений через вероятности 
простых, то задачи математической статистики являются в некотором смысле 
обратными по отношению к теории вероятностей. И состоят они в выработке процедур 
и алгоритмов, позволяющих по экспериментальным данным получить оценки параметров 
функций распределений вероятностей и, тем самым, оценить параметры теоретической 
модели. Итак, графически мы можем проиллюстрировать это следующим образом. 

{\hbox{

\put(25,0){\line(1,0){100}}
\put(25,0){\line(0,1){50}}
\put(25,50){\line(1,0){100}}
\put(125,0){\line(0,1){50}}

\put(40,30){Теоретическая}
\put(50,10){модель}

\put(225,0){\line(1,0){100}}
\put(225,0){\line(0,1){50}}
\put(225,50){\line(1,0){100}}
\put(325,0){\line(0,1){50}}

\put(260,20){Данные}

\put(130,23){\line(1,0){80}}
\put(130,27){\line(1,0){80}}
\put(140,30){Вероятность}

\put(210,35){\line(1,-1){10}}
\put(210,15){\line(1,1){10}}

}}

\bigskip

{\hbox{

\put(25,0){\line(1,0){100}}
\put(25,0){\line(0,1){50}}
\put(25,50){\line(1,0){100}}
\put(125,0){\line(0,1){50}}

\put(40,30){Теоретическая}
\put(50,10){модель}

\put(225,0){\line(1,0){100}}
\put(225,0){\line(0,1){50}}
\put(225,50){\line(1,0){100}}
\put(325,0){\line(0,1){50}}

\put(260,20){Данные}

\put(140,23){\line(1,0){80}}
\put(140,27){\line(1,0){80}}
\put(150,30){Извлечение}
\put(150,13){параметров}
\put(155, 3){модели}

\put(130,25){\line(1,-1){10}}
\put(130,25){\line(1,1){10}}

}}

\noindent
Можно сказать, что основными задачами статистики являются.

\begin{enumerate}
\item Оценка параметров модели из имеющихся экспериментальных данных.
\item Построение интервалов доверия -- нахождение областей параметров модели,
не противоречащих экспериментальным данным.
\item Проверка гипотез -- удовлетворяет ли та или иная гипотеза экспериментальным 
данным и какая из рассматриваемых гипотез удовлетворяет экспериментальным 
данным  лучше.
\item Качество фита -- насколько хорошо модель соответствует экспериментальным данным.
\end{enumerate} 

На сегодняшний день в математической статистике используются в основном три метода, 
позволяющих решить обратную задачу -- извлечение из экспериментальных данных параметров 
распределения: 

\begin{itemize}
\item частотный подход, 
\item метод максимального правдоподобия, 
\item Байесовский метод.
\end{itemize}

Заметим, что частотный подход близок к методу максимального правдоподобия, но не совпадает с ним. 
Численные оценки, полученные с помощью этих трех методов как правило различаются. В следующих 
разделах этой главы мы описываем эти три метода на примере задачи нахождения верхнего предела на 
параметр $\lambda$ распределения Пуассона $P(n|\lambda)$.

\subsection{Частотный подход}

\subsubsection{Доверительный интервал и пределы доверия}

Часто на основании измерений случайной переменной делаются выводы о возможных 
значениях параметров распределений, описывающих поведение этой переменной. 
Так как обычно параметр не является случайной величиной, то мы не можем говорить 
о вероятностных характеристиках параметра. В этом случае можно говорить о степени 
доверия к оценке величины параметра, полученной из результатов измерений. Для 
представления оценок параметров были введены понятия доверительный интервал и 
пределы доверия~\footnote{Существует подход, позволяющий расширить эти понятия 
до понятий доверительная плотность и доверительное распределение. Более подробно с  
данным направлением исследований можно познакомиться в обзоре~\cite{IHEP08}.}. 
Доверительное оценивание в частотной интерпретации часто связывают 
с E. Нейманом~\cite{Neyman0}. 

Доверительный интервал - это интервал, построенный с помощью случайной выборки 
из распределения с неизвестным параметром, такой, что он содержит данный параметр 
с заданной вероятностью (уровнем доверия) $1-\alpha$, где величина 
$\alpha \in [0,1]$ обычно называется уровнем значимости.  

\underline{Определение:} Пусть $\hat x_1, \hat x_2, \dots, \hat x_n$  
есть выборка из распределения $P(\theta)$, где $\theta$ -- неизвестный параметр. 
Пусть также задан уровень доверия $1-\alpha \in [0,1]$. Тогда случайный интервал 
$[L,U]$, где $L=L(\hat x_1, \hat x_2, \dots, \hat x_n)$, 
$U=U(\hat x_1, \hat x_2, \dots, \hat x_n)$ 
есть некоторые статистики имеющейся выборки, такой, что $P(L \le \theta \le U) = 1-\alpha$, 
называется $(1-\alpha)$-доверительным интервалом для параметра $\theta$. 
При одностороннем построении доверительного интервала для параметра в качестве случайной величины 
выступает только одна из границ интервала. В этом случае, если этой границей является 
левая граница интервала $L$, то говорят о нижнем пределе с уровнем доверия $(1-\alpha)$. 
Если же в качестве такой границы выступает правая граница $U$, то говорят об верхнем 
$(1-\alpha)$-пределе.  

\subsubsection{Пределы доверия для параметра распределения Пуассона}

Пусть мы измеряем количество событий в эксперименте, причем события распределены по закону 
Пуассона $P(n|\lambda)=\displaystyle \frac{\lambda^n}{n!}e^{-\lambda}$. 
Предположим мы наблюдаем $n_{obs}$ событий и наша цель получить 
верхний предел $\lambda \le \lambda_{up}$ на параметр $\lambda$ с точностью   
$(1-\alpha)$. Обычно на практике $\alpha$ выбирается равной $0.05$ 
(95\% уровень достоверности). В частотном подходе требуется, чтобы 

\begin{equation}
P(n\le n_{obs}|\lambda)\equiv\displaystyle \sum_{n=0}^{n_{obs}}{P(n|\lambda)}\ge\alpha. 
\label{eq:85}
\end{equation}

\noindent
Иными словами, вероятность наблюдения количества событий меньшей или равной 
$n_{obs}$ должна быть не меньше, чем $\alpha$. Из соотношения~(\ref{eq:85}) можно 
получить ограничение сверху на параметр $\lambda$ в распределении Пуассона 

\begin{equation}
\lambda\le\lambda_{up}(n_{obs},\alpha). 
\label{eq:86}
\end{equation}

Верхний предел зависит как от $n_{obs}$, так и от $\alpha$. Так, например, 

\begin{equation}
\lambda_{up}(n_{obs}=0,\alpha=0.05)=3.0, 
\label{eq:87}
\end{equation}

\begin{equation}
\lambda_{up}(n_{obs}=0,\alpha=0.01)=4.6, 
\label{eq:88}
\end{equation}

\begin{equation}
\lambda_{up}(n_{obs}=2,\alpha=0.05)=6.3, 
\label{eq:89}
\end{equation}

\begin{equation}
\lambda_{up}(n_{obs}=2,\alpha=0.01)=8.4, 
\label{eq:90}
\end{equation}

Задача определения нижнего предела $\lambda_{low}$ на уровне $1-\alpha$ 
достоверности сводится к решению неравенства 

$P(n\ge n_{obs}|\lambda)=\displaystyle 
\sum_{n=n_{obs}}^{\infty}{P(n|\lambda)}\ge\alpha$.

В Таб.~\ref{tab:Main1} представлены верхние $\lambda_{up}$ и нижние $\lambda_{low}$ 
пределы для $n_{obs}\le 10$ и $\alpha=0.1,~~\alpha=0.05,~~\alpha=0.01$~\cite{PDG}. 

\setlength{\tabcolsep}{4pt}
\begin{table}
\begin{center}
\caption{Значения верхних $\lambda_{up}$ и нижних $\lambda_{low}$ 
пределов для $n_{obs}\le 10$ и $\alpha=0.05,~~\alpha=0.1$.}
\begin{tabular}{|c|cc|cc|cc|}
\hline
$n_{obs}$&$\lambda_{low}$& $\lambda_{up}$ & $\lambda_{low}$ & $\lambda_{up}$ &
$\lambda_{low}$ & $\lambda_{up}$\\  
         &$1-\alpha=0.9$ &  & $1-\alpha=0.95$  &  & $1-\alpha=0.99$ \\ 
\hline
  0 &  $-$     &  $2.30$  &  $-$      & $3.00$  & $-$    & $4.60$ \\
  1 &  $0.105$ &  $3.89$  &  $0.051$  & $4.74$  & $0.01$ & $6.63$ \\
  2 &  $0.532$ &  $5.32$  &  $0.355$  & $6.30$  & $0.15$ & $8.40$ \\
  3 &  $1.10$  &  $6.68$  &  $0.818$  & $7.75$  & $0.44$ & $10.04$ \\
  4 &  $1.74$  &  $7.99$  &  $1.37$   & $9.15$  & $0.82$ & $11.60$ \\
  5 &  $2.43$  &  $9.27$  &  $1.97$   & $10.51$ & $1.28$ & $13.10$ \\
  6 &  $3.15$  &  $10.53$ &  $2.61$   & $11.84$ & $1.78$ & $14.57$ \\
  7 &  $3.89$  &  $11.77$ &  $3.29$   & $13.15$ & $2.33$ & $15.99$ \\
  8 &  $4.66$  &  $12.99$ &  $3.98$   & $14.43$ & $2.90$ & $17.40$ \\
  9 &  $5.43$  &  $14.21$ &  $4.70$   & $15.71$ & $3.51$ & $18.78$ \\
 10 &  $6.22$  &  $15.41$ &  $5.43$   & $16.96$ & $4.13$ & $20.14$ \\
\hline
\end{tabular}
\label{tab:Main1}
\end{center}
\end{table}

Подчеркнем, что иногда этот подход к определению верхнего предела критикуется на том основании, 
что в формуле~(\ref{eq:85}) содержится суммирование по ненаблюдаемым вероятностям 
$P(n_{obs}-1,\lambda),~P(n_{obs}-2,\lambda), \dots$.

\subsection{Байесовский подход}

Для распределения Пуассона применение формулы Байеса~(\ref{eq:6}) приводит к формуле 
распределения плотности вероятности для параметра $\lambda$ 

\begin{equation}
P(\lambda|n)=\pi(\lambda)P(n|\lambda), 
\label{eq:91}
\end{equation}

\noindent
где $\pi(\lambda)$ - априорное распределение (приор)~\cite{Prior}, 
удовлетворяющее тривиальному условию нормировки

\begin{equation}
\displaystyle \int_0^{\infty}{P(\lambda|n)d\lambda}=
\int_0^{\infty}{\pi(\lambda)P(n|\lambda)d\lambda}=1. 
\label{eq:92}
\end{equation}

Преимуществом Байесовского подхода является то, что формула~(\ref{eq:91}) по сути 
дела сводит задачу статистики к задаче теории вероятностей, поскольку мы знаем плотность 
вероятности параметра $\lambda$ и можем применять методы теории вероятностей для 
ограничения параметра $\lambda$. 

К недостаткам Байесовского подхода относятся:

\begin{enumerate}
\item Параметр $\lambda$ не является случайным параметром в классическом частотном понятии
теории вероятностей. Параметр $\lambda$ имеет вполне определенное значение, просто мы его не 
знаем. Обычно на эту критику адепты Байесовского подхода отвечают, что формула~(\ref{eq:91}) 
определяет степень доверия к нашему пониманию, каким может быть параметр $\lambda$.
\item Результаты расчетов в Байесовском подходе зависят от неизвестной функции $\pi(\lambda)$, 
которую мы должны определить вообще говоря из ``ad hoc'' соображений. 

Часто в литературе используется либо плоский приор $\pi(\lambda)=const$, либо степенной приор 
$\pi(\lambda)=const\cdot\lambda^{-\beta}$, причем особенно популярен приор 
$\pi(\lambda)=\displaystyle \frac{const}{\sqrt{\lambda}}$. Заметим, что для 
распределения Пуассона в силу тождества 
$\displaystyle \int_0^{\infty}{P(n|\lambda)d\lambda}=1$ 
для плоского приора $const=1$.
\end{enumerate}

Итак, в Байесовском подходе задача нахождения верхнего ограничения на параметр $\lambda$ 
на уровне $(1-\alpha)$ достоверности сводится к решению неравенства 

\begin{equation}
\displaystyle \int_0^{\lambda}{P(\lambda'|n)d\lambda'} \le 1-\alpha 
\label{eq:93}
\end{equation}

\noindent
или

\begin{equation}
\displaystyle \int_{\lambda}^{\infty}{P(\lambda'|n)d\lambda'} \ge \alpha. 
\label{eq:94}
\end{equation}

Из неравенства~(\ref{eq:94}) можно получить ограничение сверху на параметр $\lambda$ 

\begin{equation}
\lambda \le \lambda_{up}(n_{obs},\alpha,\pi). 
\label{eq:95}
\end{equation}

\noindent
При этом параметр $\lambda_{up}(n_{obs},\alpha,\pi)$ определяется из уравнения 

\begin{equation}
\displaystyle \int_0^{\lambda_{up}}{P(\lambda'|n)d\lambda'} = 1-\alpha. 
\label{eq:93eq}
\end{equation}

\noindent
Заметим, что неравенство~(\ref{eq:95}) явно зависит от функции приора $\pi(\lambda)$. 
При выборе плоского приора $\pi(\lambda)=1$ в силу тождеств

\begin{equation}
\displaystyle \int_{\lambda}^{\infty}{P(n_0|\lambda')d\lambda'} = 
\sum_{n=0}^{n_0}{P(n|\lambda)} 
\label{eq:96}
\end{equation}

\noindent
и

\begin{equation}
\displaystyle \int^{\lambda}_0{P(n_0|\lambda')d\lambda'} = 
\sum_{n=n_0+1}^{\infty}{P(n|\lambda)}  
\label{eq:96a}
\end{equation}

\noindent
получаем, что частотное неравенство~(\ref{eq:85}) совпадает с Байесовским 
неравенством~(\ref{eq:94}). Иными словами, при выборе плоского приора $\pi(\lambda)=1$ 
Байесовский подход (численные оценки) для получения верхнего предела на параметр $\lambda$ 
в распределении Пуассона совпадает с частотным подходом, основанном на использовании 
формулы~(\ref{eq:85}).

\begin{figure}[htpb]
\begin{center}
   \resizebox{4.1in}{!}{\includegraphics{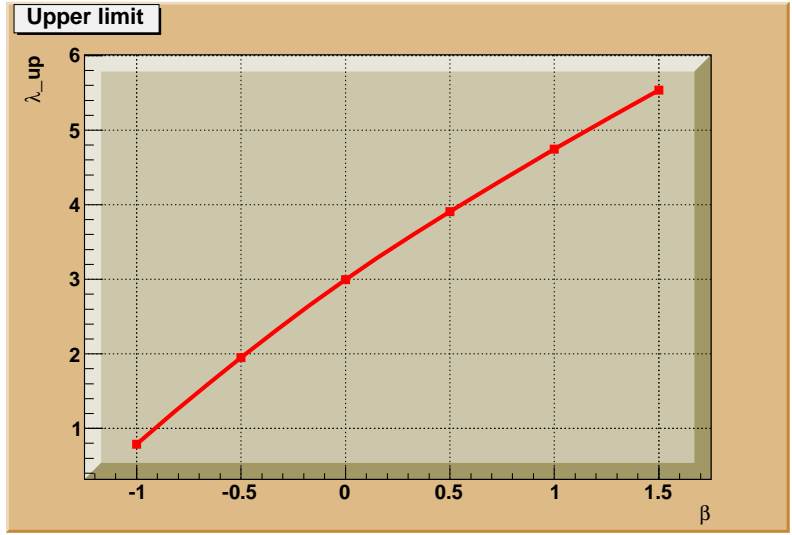}} 
\caption{Зависимость верхнего предела от степени $\beta$ в выражении для степенного приора 
при $n_{obs}=0$ и $\alpha=0.05$.}
    \label{fig:BayLim} 
  \end{center}
\end{figure}

В случае приора $\pi(\lambda)=const\cdot\lambda^{\beta}$ зависимость 
$\lambda_{up}(n_{obs}=0,\alpha=0.05,\beta)$ от параметра $\beta$ показана 
на Рис.~\ref{fig:BayLim}.

Задача об определении ограничения снизу $\lambda_{low}$ на параметр $\lambda$ сводится 
к решению неравенства 

\begin{equation}
\displaystyle \int_{\lambda}^{\infty}{P(\lambda'|n)d\lambda'} \le 1 - \alpha. 
\label{eq:BayLow}
\end{equation}

\noindent
Параметр $\lambda_{low}$ определяется из уравнения 

\begin{equation}
\displaystyle \int_{\lambda_{low}}^{\infty}{P(\lambda'|n)d\lambda'} = 1 - \alpha. 
\label{eq:BayLoweq}
\end{equation}

\noindent
Заметим, что в силу тождества~(\ref{eq:96}) для приора $\displaystyle \frac{const}{\lambda}$ 
результаты Байесовского и частотного подходов для $\lambda_{low}$ совпадают. 

Для часто используемого приора $\pi(\lambda)\sim\displaystyle\frac{1}{\lambda}$~\cite{Prior2} 
в Таб.~\ref{tab:Main2} представлены значения верхних $\lambda_{up}$ и нижних $\lambda_{low}$ 
пределов для $n_{obs}\le 10$ и $\alpha=0.05,~~\alpha=0.1$.

\setlength{\tabcolsep}{4pt}
\begin{table}
\begin{center}
\caption{Значения верхних $\lambda_{up}$ и нижних $\lambda_{low}$ 
пределов для $n_{obs}\le 10$ и $\alpha=0.05,~~\alpha=0.1$. 
Приор $\pi(\lambda)\sim\displaystyle\frac{1}{\lambda}$.}
\begin{tabular}{|c|cc|cc|}
\hline
$n_{obs}$&$\lambda_{low}$& $\lambda_{up}$ & $\lambda_{low}$ & $\lambda_{up}$\\  
          &$1-\alpha=0.9$ &                & $1-\alpha=0.95$  &                 \\ 
\hline
  1       &  $0.105$      &  $2.30$        &  $0.051$        & $3.00$  \\
  2       &  $0.532$      &  $3.89$        &  $0.355$        & $4.74$  \\
  3       &  $1.10$       &  $5.32$        &  $0.818$        & $6.30$  \\
  4       &  $1.74$       &  $6.68$        &  $1.37$         & $7.75$  \\
  5       &  $2.43$       &  $7.99$        &  $1.97$         & $9.15$  \\
  6       &  $3.15$       &  $9.27$        &  $2.61$         & $10.51$  \\
  7       &  $3.89$       &  $10.53$       &  $3.29$         & $11.84$  \\
  8       &  $4.66$       &  $11.77$       &  $3.98$         & $13.15$  \\
  9       &  $5.43$       &  $12.99$       &  $4.70$         & $14.43$  \\
 10       &  $6.22$       &  $14.21$       &  $5.43$         & $15.71$  \\
\hline
\end{tabular}
\label{tab:Main2}
\end{center}
\end{table}

\subsection{Метод максимального правдоподобия}

В методе максимального правдоподобия~\cite{Likelihood} для случая статистики Пуассона функция 
максимального правдоподобия 

\begin{equation}
\displaystyle L(\lambda)=P(n_{obs}|\lambda). 
\label{eq:98lik1}
\end{equation}

Из условия максимума функции правдоподобия 

\begin{equation}
\displaystyle \frac{\partial}{\partial\lambda}L(\lambda)=0
\label{eq:98lik2}
\end{equation}

\noindent
находим наиболее вероятное значение параметра $\lambda$

\begin{equation}
\lambda_{max}=n_{obs}. 
\label{eq:98lik3}
\end{equation}

В методе максимального правдоподобия верхний и нижний пределы на параметр $\lambda$ 
определяются исходя из неравенства~
 
\begin{equation}
\displaystyle \sqrt{2ln\frac{L(\lambda_{max})}{L(\lambda)}}\le s(\alpha). 
\label{eq:98}
\end{equation}

Асимптотически условие~(\ref{eq:98}) сводится к стандартному центральному интервалу с 
$\pm s(\alpha)$ стандартных 
гауссовских отклонений. Для распределения Пуассона неравенство~(\ref{eq:98}) принимает вид 

\begin{equation}
\displaystyle \sqrt{2[n_{obs}ln{\frac{n_{obs}}{\lambda}+\lambda-n_{obs}}]} 
\le s(\alpha). 
\label{eq:99}
\end{equation}

Например, при $n_{obs}=0$ верхний предел на параметр $\lambda$ имеет вид 

\begin{equation}
\lambda \le \lambda_{up}(n_{obs}=0,\alpha=0.05)=1.34.
\label{eq:97}
\end{equation}

Оценка~(\ref{eq:97}) отличается от частотной оценки~(\ref{eq:87}). 

Как видно из Таб.~\ref{tab:Main1}--\ref{tab:Main3} в общем случае как частотный и 
Байесовский подходы, так и метод максимального правдоподобия дают численно разные значения верхнего 
и нижнего пределов $\lambda_{up}$ и $\lambda_{low}$.

\setlength{\tabcolsep}{4pt}
\begin{table}
\begin{center}
\caption{Значения верхних $\lambda_{up}$ и нижних $\lambda_{low}$ 
пределов в методе максимального правдоподобия для $n_{obs}\le 10$ и 
$\alpha=0.05,~~\alpha=0.1$.}
\begin{tabular}{|c|cc|cc|}
\hline
$n_{obs}$&$\lambda_{low}$& $\lambda_{up}$ & $\lambda_{low}$ & $\lambda_{up}$\\  
          &$1-\alpha=0.9$ &                & $1-\alpha=0.95$  &                 \\ 
\hline
  0       &  $-$          &  $0.82$        &  $-$             & $1.34$  \\
  1       &  $0.19$       &  $2.88$        &  $0.11$          & $3.64$  \\
  2       &  $0.69$       &  $4.40$        &  $0.48$          & $5.29$  \\
  3       &  $1.29$       &  $5.80$        &  $0.98$          & $6.80$  \\
  4       &  $1.95$       &  $7.13$        &  $1.55$          & $8.23$  \\
  5       &  $2.65$       &  $8.43$        &  $2.12$          & $9.61$  \\
  6       &  $3.38$       &  $9.71$        &  $2.82$          & $10.96$  \\
  7       &  $4.13$       &  $10.95$       &  $3.51$          & $12.28$  \\
  8       &  $4.90$       &  $12.19$       &  $4.21$          & $13.58$  \\
  9       &  $5.68$       &  $13.41$       &  $4.93$          & $14.86$  \\
 10       &  $6.48$       &  $14.61$       &  $5.67$          & $16.12$  \\
\hline
\end{tabular}
\label{tab:Main3}
\end{center}
\end{table}

\newpage

\section{Оценка параметров}

\subsection{Общая постановка задачи}

Предположим у нас имеется набор экспериментальных данных 
$\vec x = (x_1,x_2,\dots,x_n)$ 
случайной величины $X$, распределенной с плотностью вероятности $f(x|\vec \theta)$, 
где $\vec \theta$-параметры, описывающие распределение. Задача состоит в том, чтобы 
оценить параметры $\vec \theta$ и найти интервалы доверия с доверительной вероятностью 
$1-\alpha$ для полученных оценок. 

В частотном подходе Неймана~\cite{Neyman0} задача решается следующим образом. 
С помощью функции распределения $F(x|\theta)$ мы можем найти такие функции 
$x_1(\theta,\alpha)$ и $x_2(\theta,\alpha)$, что справедливо равенство 

\begin{equation}
P(x_1 < X < x_2|\theta) = 1 - \alpha = \displaystyle 
\int_{x_1}^{x_2}{f(x|\theta)dx}. 
\label{eq:100}
\end{equation}

\begin{figure}[htpb]
\begin{center}
   \resizebox{3.0in}{!}{\includegraphics{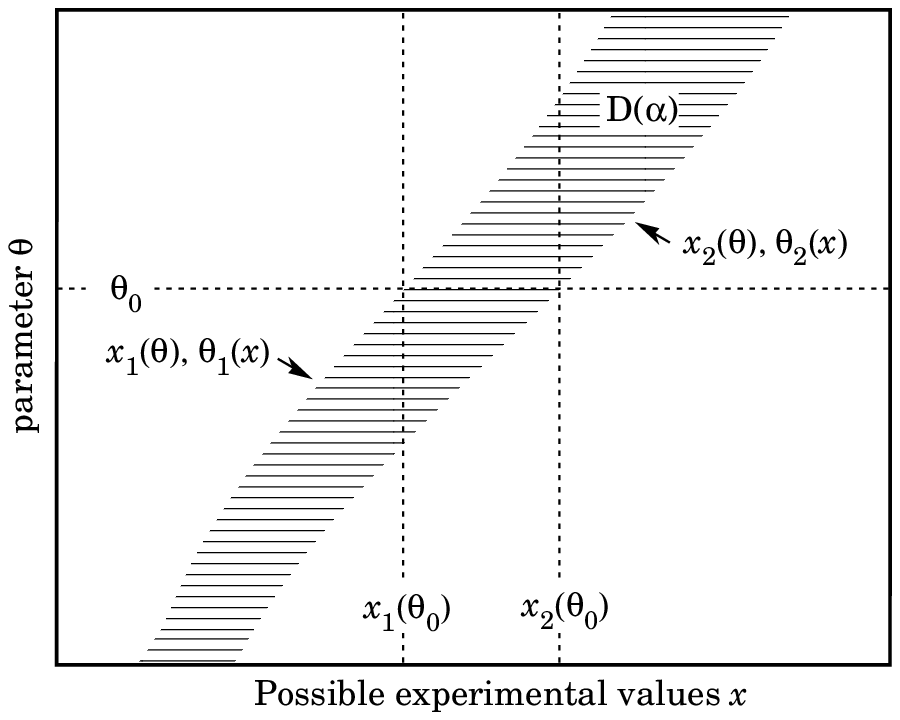}} 
\caption{Построение доверительного интервала.}
    \label{fig:2} 
  \end{center}
\end{figure}

Смысл равенства~(\ref{eq:100}) очевиден и заключается в том, что оно связывает 
вероятность $1-\alpha$ нахождения случайной величины $X$ в пределах 
между $x_1$ и $x_2$ с интегралом от плотности вероятности $f(x|\theta)$. 
Интервал $[x_1(\theta,\alpha),x_2(\theta,\alpha)]$ зависит от 
измеренной величины $\vec x$ и объединение таких интервалов называется доверительным 
интервалом $D(\alpha)$. Для многих распределений (нормальное распределение, 
распределение Пуассона) $x_1(\theta,\alpha)$ и $x_2(\theta,\alpha)$ 
монотонные функции от параметра $\theta$. Смотри  в качестве иллюстрации 
Рис.~\ref{fig:2}. 

В эксперименте мы измеряем случайную величину $X$ и получаем какое-то конкретное 
значение $x_0$. 

Интервал доверия для параметра $\theta$ это набор всех значений $\theta$ для 
которых соответствующая линия сегмента пересекается с вертикальной линией 
(см. Рис.~\ref{fig:2}). Такие интервалы доверия имеют уровень достоверности 
$1-\alpha$. Иными словами, справедливо равенство 

\begin{equation}
1-\alpha = P(x_1(\theta)<X<x_2(\theta))=P(\theta_2(X)<\theta<\theta_1(X)),
\label{eq:101}
\end{equation}

\noindent
где 

\begin{center}
$\theta_2(X) = \displaystyle {max}_{\theta}x_1(\theta,x_0),$
\end{center}
\begin{equation}
\theta_1(X) = \displaystyle {min}_{\theta}x_2(\theta,x_0).
\label{eq:102}
\end{equation} 

В этом вероятностном утверждении границы интервала $\theta_1(X)$ и $\theta_2(X)$ 
являются случайными величинами, в то время как неизвестный параметр $\theta$ 
есть неизвестная константа. Это означает, что если эксперимент повторять много раз,
то интервалы $[\theta_1,\theta_2]$, получаемые в каждом измерении, будут различными, 
но доля интервалов, покрывающих истинное значение параметра $\theta$, будет равна  
значению $1-\alpha$.

Подчеркнем, что условие~(\ref{eq:100}) не определяет однозначно параметры $x_1$, $x_2$. 
Существует бесконечно много значений $(x_1,x_2)$, удовлетворяющих условию~(\ref{eq:100}). 

\subsection{Нормальное распределение}

Рассмотрим в качестве примера нормальное распределение 
\begin{center}
$N(x|\mu,\sigma^2)\sim 
\displaystyle \frac{1}{\sqrt{2\pi}\sigma}e^{-\frac{(x-\mu)^2}{2\sigma^2}}$.  
\end{center}
Существует много способов выбора интервалов параметров 
$x_1$ и $x_2$ в формуле~(\ref{eq:100}).

Перечислим наиболее естественные способы. 

\begin{enumerate}
\item $x_2(\theta)=\infty$. Этот выбор соответствует так называемому верхнему пределу.
\item $x_1(\theta)=-\infty$. Этот выбор соответствует нижнему пределу.
\item $-x_2(\theta)< X-\mu <x_2(\theta)$. Этот выбор соответствует симметричному интервалу.
\end{enumerate}

Заметим, что часто используется интервал $[x_1(\theta),x_2(\theta)]$ такой, что 
плотность распределения $P(x|\theta)$ больше или равна плотности распределения вне интервала
$[x_1(\theta),x_2(\theta)]$~\cite{PDG}~\footnote{При таком выборе справедливо равенство 
$P(x_1|x_1(\theta))=P(x_2|x_2(\theta))$.}. 
В случае распределения Гаусса это 
определение совпадает с предыдущим симметричным определением. В общем случае, например распределение 
Пуассона, симметричное определение и определение, основанное на выборе интервала с максимальной 
плотностью, не совпадают.  

Предположим, что мы знаем дисперсию $\sigma^2$ нормального измерения и хотим определить 
ограничения на параметр $\mu$ нормального распределения. В случае, когда мы хотим найти 
верхний предел на параметр $\mu$ равенство~(\ref{eq:101}) записывается в виде 

\begin{center}
$1-\alpha = P(-\infty<X<x_2)=\displaystyle 
\int_{-\infty}^{x_2}{N(x|\mu,\sigma^2)dx} = $
\end{center}
\begin{equation}
\displaystyle \frac{1}{\sqrt{2\pi}\sigma}
\int_{-\infty}^{x_2}{e^{-\frac{(x-\mu)^2}{2\sigma^2}}dx} = 
\frac{1}{\sqrt{2\pi}}
\int_{-\infty}^{\frac{x_2-\mu}{\sigma}}{e^{-\frac{y^2}{2}}dy}=
1-\frac{1}{\sqrt{2\pi}}
\int^{\infty}_{\frac{x_2-\mu}{\sigma}}{e^{-\frac{y^2}{2}}dy}. 
\label{eq:103}
\end{equation}

Вводим стандартное понятие односторонней значимости как 

\begin{equation}
\alpha = \frac{1}{\sqrt{2\pi}}
\int^{\infty}_s{e^{-\frac{y^2}{2}}dy}.
\label{eq:104}
\end{equation}

\noindent
Заметим, что обычно используемое при установлении пределов в физике высоких энергий 
$\alpha=0.05$ соответствует значению $s=1.64$. 
В Таб.~\ref{tab:1} для ряда значений вероятности $\alpha$ приведены соответствующие им 
значения значимости $s$.

\setlength{\tabcolsep}{4pt}
\begin{table}
\begin{center}
\caption{Соответствие для одностороннего предела между вероятностями $\alpha$ и значимостями $s$.}
\begin{tabular}{|cc||cc|}
\hline
$\alpha$ & $s$ & $\alpha$ & $s$ \\ 
\hline
    $0.1587$             &  $1\sigma$ &  $0.1$  & $1.28\sigma$  \\
    $2.275\cdot 10^{-2}$ &  $2\sigma$ &  $0.05$  & $1.64\sigma$  \\
    $1.35\cdot 10^{-3}$ &  $3\sigma$ &  $0.025$  & $1.96\sigma$  \\
    $3.15\cdot 10^{-5}$ &  $4\sigma$ &  $0.005$  & $2.58\sigma$  \\
    $2.87\cdot 10^{-7}$ &  $5\sigma$ &  $0.0005$  & $3.29\sigma$  \\
    $1.00\cdot 10^{-9}$ &  $6\sigma$ &  $0.5\cdot 10^{-4}$  & $3.89\sigma$  \\
\hline
\end{tabular}
\label{tab:1}
\end{center}
\end{table}

Из уравнения~(\ref{eq:103}) получаем уравнение

\begin{equation}
\displaystyle \frac{x_2-\mu_{up}}{\sigma} = s(\alpha) 
\label{eq:105}
\end{equation}

\noindent
для нахождения параметра $\mu$, решение которого есть $\mu_{up}=x_2+\sigma s(\alpha)$, 
позволяющее определить верхний предел на параметр $\mu$ 

\begin{equation}
\mu \le \mu_{up} = x_2+ \sigma s(\alpha) 
\label{eq:106}
\end{equation}

\noindent
на уровне $1-\alpha$ достоверности. 

В случае, когда мы хотим найти нижний предел на параметр $\mu$, уравнение~$(\ref{eq:103})$ 
запишется в виде 

\begin{center}
$1-\alpha = P(x_1<X<\infty) = \displaystyle 
\int_{x_1}^{\infty}{N(x|\mu,\sigma^2)dx} = $
\end{center}
\begin{equation}
\displaystyle \frac{1}{\sqrt{2\pi}\sigma}
\int_{x_1}^{\infty}{e^{-\frac{(x-\mu)^2}{2\sigma^2}}dx} = 
\frac{1}{\sqrt{2\pi}\sigma}
\int^{\infty}_{\frac{x_1-\mu}{\sigma}}{e^{-\frac{y^2}{2}}dy}=
1-\frac{1}{\sqrt{2\pi}}
\int_{-\infty}^{\frac{x_1-\mu}{\sigma}}{e^{-\frac{y^2}{2}}dy}. 
\label{eq:107}
\end{equation}

\noindent
Отсюда получаем уравнение $\displaystyle \frac{x_1-\mu_{low}}{\sigma} = s(\alpha)$ 
для определения нижнего предела $\mu_{low}=x_1-\sigma s(\alpha)$. Таким образом, 
$\mu \ge \mu_{low}$.

\begin{figure}[htpb]
\begin{center}
   \resizebox{4.0in}{!}{\includegraphics{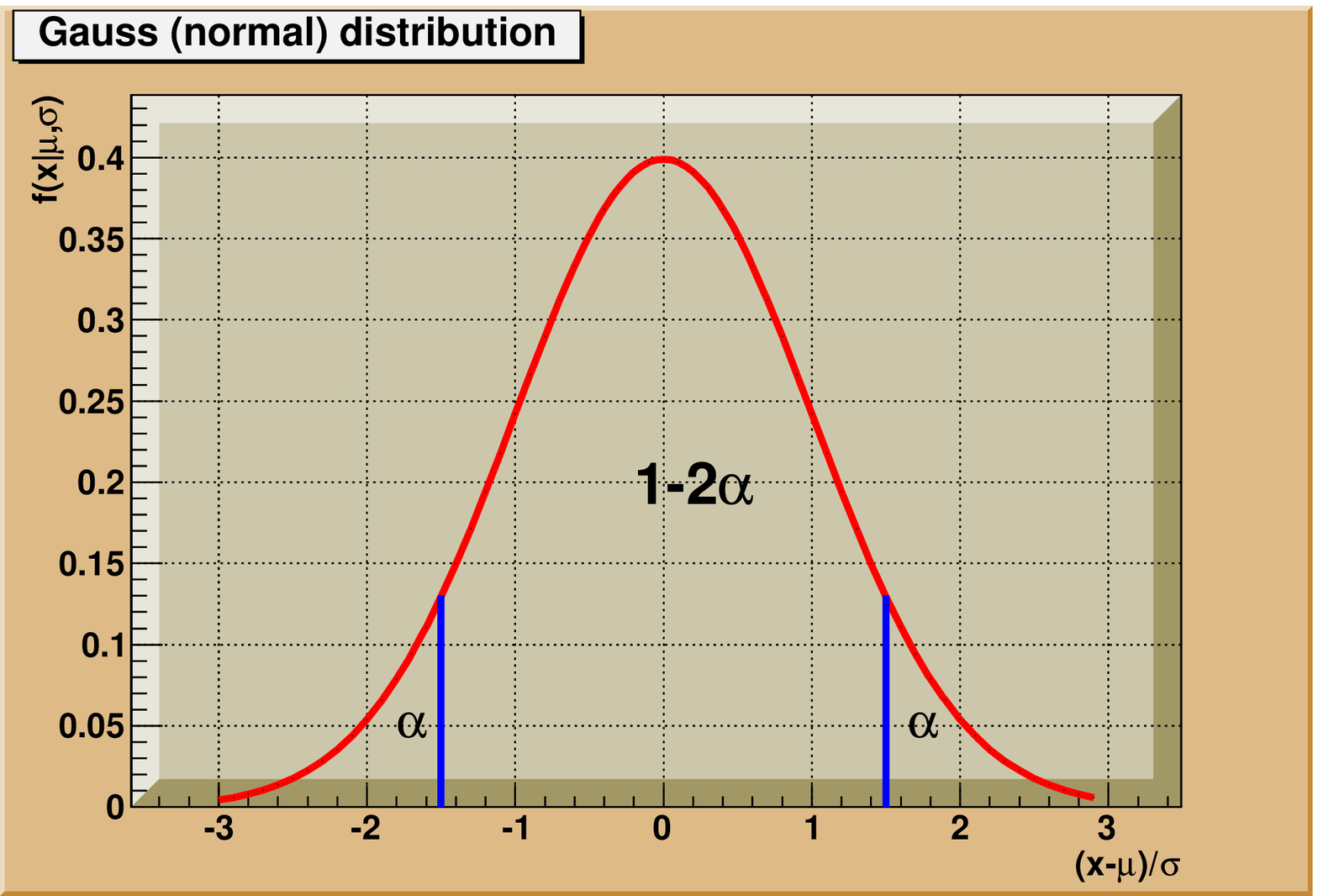}} 
\caption{На рисунке показан симметричный доверительный интервал.}
    \label{fig:3} 
  \end{center}
\end{figure}

В случае, когда мы ищем центральный интервал, который определяется из условия равенства 
$\alpha$ нижнего и верхнего пределов, то есть одновременного выполнения 
равенств (\ref{eq:103}) и (\ref{eq:107}), мы находим (см. Рис~\ref{fig:3}), 
что с вероятностью $1-2\alpha$ 

\begin{equation}
-\sigma s(\alpha)\le \mu-x_1 \le \sigma s(\alpha). 
\label{eq:108}
\end{equation}

В силу симметрии нормального распределения относительно отражения  
$(\mu-x)\rightarrow -(\mu-x)$ верхнее и нижнее значения по модулю $(\mu -x_1)$ 
совпадают, то есть $|x_1-\mu|\le \sigma s(\alpha)$. 

\begin{figure}[htpb]
\begin{center}
   \resizebox{4.0in}{!}{\includegraphics{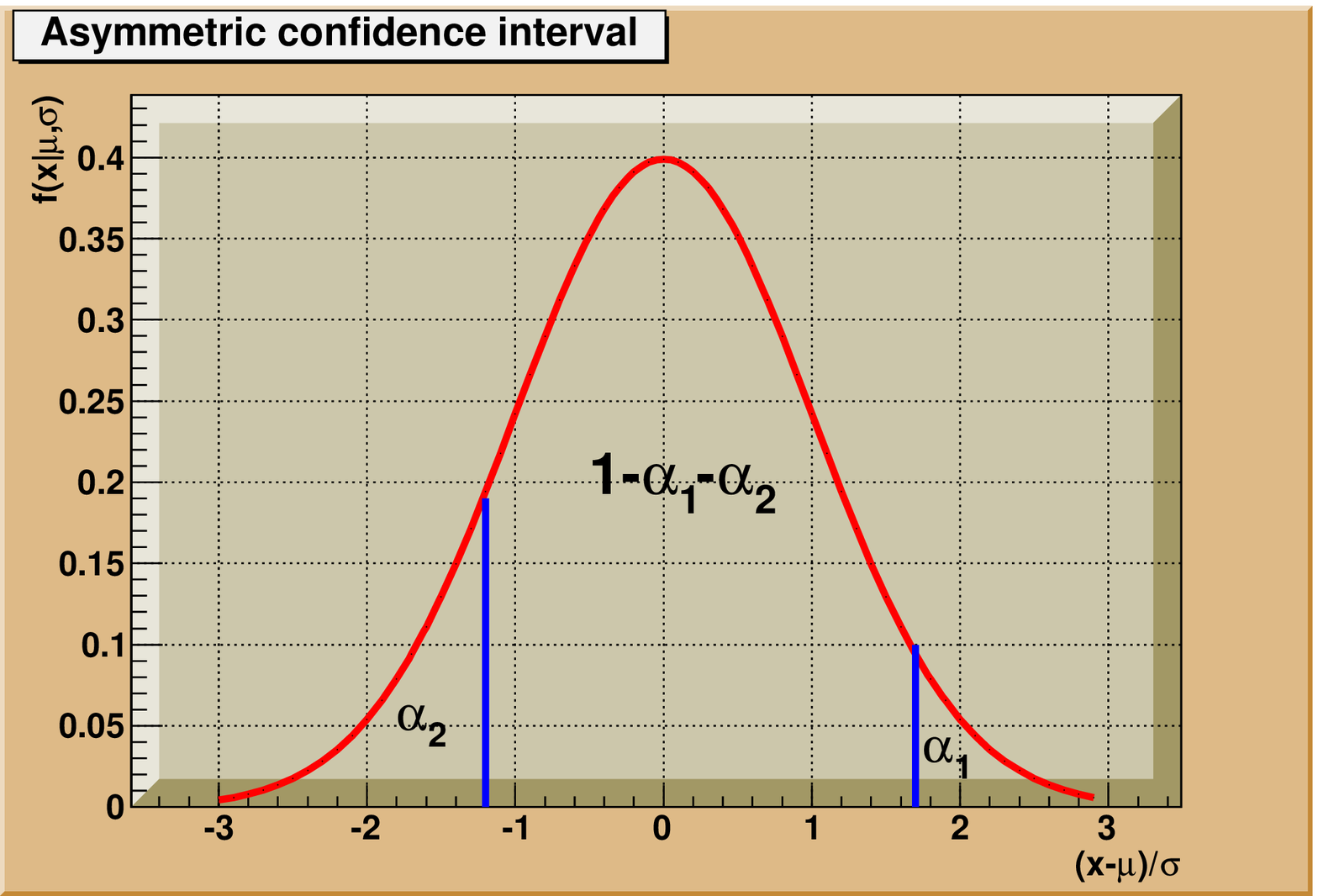}} 
\caption{На рисунке показан асимметричный доверительный интервал.}
    \label{fig:3as} 
  \end{center}
\end{figure}

Заметим, что три вышеперечисленных способа определения доверительных интервалов параметра 
$\mu$ отнюдь не единственны. Например, мы можем требовать справедливость 
уравнения~(\ref{eq:100}) на уровне $(1-\alpha_1)$, уравнения (\ref{eq:103}) 
на уровне $(1-\alpha_2)$ (см.~Рис.\ref{fig:3as}). 
Тогда обобщение симметричного ограничения~(\ref{eq:108}) 
примет вид $-\sigma s(\alpha_2)\le \mu-x_1 \le \sigma s(\alpha_1).$ 

Напомним, что в методе максимального правдоподобия функция правдоподобия зависит от набора $N$ независимых 
измерений $\vec x=(x_1,x_2, \dots, x_N)$

\begin{equation}
L(\vec \theta)=\displaystyle \prod_{i=1}^N{f(x_i|\vec \theta)},
\label{eq:109}
\end{equation}

\noindent
где $f(x_i|\vec \theta)$ плотность вероятности измерения случайной наблюдаемой $X$. 

Параметры $\vec \theta$ функции распределения находятся из условия максимума 
функции правдоподобия 

\begin{center}
$\displaystyle \frac{\partial}{\partial \vec \theta}L(\vec \theta)=0$.
\end{center}

Оценки на параметры $\vec \theta$ находятся из условия 

\begin{equation}
2[ln{L(\vec \theta_{max})} - ln{L(\vec \theta)}]\le s^2, 
\label{eq:110}
\end{equation}

\noindent
где $s$-значимость. Фактически определение~(\ref{eq:110}) выбрано так, чтобы оценка на 
параметры с помощью метода максимального правдоподобия и частотный подход совпадали бы для 
случая нормального распределения. Действительно, для случая нормального распределения с 
известной дисперсией $\sigma^2$ и с неизвестным средним $\mu$ уравнение~(\ref{eq:110}) 
примет вид 

\begin{equation}
\displaystyle \frac{(x_1-\mu)^2}{\sigma^2} \le s^2. 
\label{eq:111}
\end{equation}

\noindent
Решение уравнения~(\ref{eq:111}) 

\begin{equation}
-s\sigma \le x_1-\mu\le s\sigma  
\label{eq:112}
\end{equation}

\noindent
есть ничто иное, как центральный интервал~(\ref{eq:108}) в частотном подходе. В силу четности нормального 
распределения вероятность левого и правого хвостов совпадают.

Рассмотрим задачу определения параметра $\mu$ в Байесовском подходе. При этом мы также 
предполагаем, что дисперсия $\sigma^2$ известна. Плотность распределения вероятности 
в Байесовском подходе имеет вид 

\begin{equation}
P(\mu|x_0,\sigma) = P(\mu)N(x_0|\mu,\sigma), 
\label{eq:113}
\end{equation}

\noindent 
где $P(\mu)$ -- неизвестный приор. При этом единственное условие на приор 
$P(\mu)$ это условие нормировки

\begin{center}
$\displaystyle \int{P(\mu|x_0,\sigma)d\mu} = 1.$ 
\end{center}

\noindent
При заданном приоре $P(\mu)$ формула~(\ref{eq:113}) сводит задачу статистики 
к задаче теории вероятностей и мы можем определять верхние и нижние пределы на 
параметр $\mu$ исходя из уравнений 

\begin{equation}
\displaystyle \int_{\mu_{up}}^{\infty}{P(\mu|x_0,\sigma)d\mu}=\alpha, 
\label{eq:114}
\end{equation}

\begin{equation}
\displaystyle \int^{\mu_{down}}_{-\infty}{P(\mu|x_0,\sigma)d\mu}=\alpha. 
\label{eq:115}
\end{equation}

\noindent
Как следствие уравнений~(\ref{eq:114},\ref{eq:115}) получаем, что для центрального 
интервала справедливо соотношение 

\begin{equation}
\displaystyle \int^{\mu_{up}}_{\mu_{down}}{N(x_0|\mu,\sigma)d\mu}=1-2\alpha. 
\label{eq:116}
\end{equation}

Заметим, что для $P(\mu)=const$ 

\begin{center}
$P(\mu|x_0,\sigma)=
\displaystyle \frac{1}{\sqrt{2\pi}\sigma}e^{-\frac{(\mu-x_0)^2}{2\sigma^2}}.$
\end{center}

\noindent
и, как следствие, для постоянного приора мы получаем точно такие же оценки, как и 
в частотном подходе.

Итак, на примере нормального распределения мы продемонстрировали три подхода к определению 
ограничений на параметр $\mu$. В частотном подходе мы имеем дело с интегралами от 
плотности вероятности, с помощью которых получаются ограничения на параметр $\mu$. 
В методе максимального правдоподобия ограничения выводятся с помощью отношения плотности 
вероятности по отношению к максимальной плотности вероятности (формула~(\ref{eq:110})). 
В Байесовском подходе формула~(\ref{eq:113}) сводит задачу определения ограничений на параметр 
$\mu$ к задаче теории вероятности, причем для приора $P(\mu)=1$. Байесовский подход 
совпадает с частотным. Следует также подчеркнуть, что выбор интервала $\mu_1 \le \mu \le \mu_2$ 
неоднозначен в частотном и в Байесовском подходах, тогда как в методе максимального 
правдоподобия выбор интервала однозначен и он соответствует упорядочению по 
$\displaystyle \frac{\vec \theta}{\vec \theta_{max}}$ (берутся только те 
$\vec \theta$, которые дают максимум этого отношения). Вообще говоря, результаты 
Байесовского подхода могут сильно отличаться от частотного, но в случае постоянного приора 
$P(\mu)=1$ оба подхода совпадают. 

\subsection{Оценка параметра в распределении Пуассона} 

В физике высоких энергий важную роль играет распределение Пуассона. Действительно, на БАКе 
количество событий с той или иной сигнатурой (например количество димюоннов, диджетов, \dots) 
является случайной величиной, распределенной по закону Пуассона $P(n|\lambda)$, где 
$\lambda = L\sigma\epsilon$~\cite{CMSPhys}. Здесь $L$--светимость, $\sigma$--сечение, 
$\epsilon$--эффективность регистрации. Предположим, что в эксперименте 
мы обнаружили $n=n_{obs}$ событий. Встает вопрос, как получить ограничение на параметр 
$\lambda$ из эксперимента?

\subsubsection{Частотный подход}

Для распределения Пуассона $P(n|\lambda)$ задача определения интервалов доверия 
$[\lambda_{low},\lambda_{up}]$ сводится к задачам определения верхнего ограничения $\lambda_{up}$ 
и нижнего ограничения $\lambda_{low}$. Задача получения верхнего ограничения на параметр 
$\lambda \le \lambda_{up}$ на уровне $(1-\alpha)$ достоверности сводится к решению неравенства 

\begin{equation}
P(n\le n_{obs}|\lambda) \equiv \displaystyle 
\sum_{n=0}^{n_{obs}}{P(n|\lambda)}\ge \alpha. 
\label{eq:117}
\end{equation}

\noindent
Из неравенства~(\ref{eq:117}) получаем ограничение сверху 
$\lambda \le \lambda_{up}(n_{obs},\alpha)$ на параметр $\lambda$ на уровне 
$(1-\alpha)$ достоверности. В частности при $n_{obs}=0$ 
$\lambda_{up}\le \displaystyle ln{\frac{1}{\alpha}}$. Для стандартно используемой 
$\alpha=0.05$ $\lambda_{up} \le 3.0$.

Задача же получения нижнего предела на параметр $\lambda$ на уровне $(1-\beta)$ 
достоверности согласно частотному подходу сводится к неравенству  

\begin{equation}
\displaystyle \sum_{n=n_{obs}}^{\infty}{P(n|\lambda)}\ge \beta. 
\label{eq:118}
\end{equation}

\noindent
Из неравенства~(\ref{eq:118}) получаем ограничение снизу на параметр $\lambda$: 
$\lambda \ge \lambda_-(n_{obs},\beta)$  на уровне 
$(1-\beta)$ достоверности. Для $n_{obs}=1$ неравенство~(\ref{eq:118}) примет вид  
$\lambda_{low}\ge \displaystyle ln{\frac{1}{1-\beta}}$ или при  
$\beta=0.05$ $\lambda \ge 0.05$.

%

Верхние и нижние пределы $\lambda_{low}$ и $\lambda_{up}$ для $n_{obs}=0,1,2,\dots,10$ 
при $\alpha=\beta=0.05$ и $\alpha=\beta=0.1$ приведены в Таб.~\ref{tab:Main1}.

\subsubsection{Метод наибольшего правдоподобия}

Для распределения Пуассона функция максимального правдоподобия 

\begin{equation}
L(\lambda)=\displaystyle \frac{1}{n_{obs}!}(\lambda)^{n_{obs}}e^{-\lambda}. 
\label{eq:119}
\end{equation}

\noindent
Решение уравнения 

\begin{equation}
\displaystyle \frac{d}{d\lambda}L(\lambda)=0 
\label{eq:120}
\end{equation}

\noindent
есть 
$\lambda=n_{obs}$. Неравенство~(\ref{eq:110}) в нашем случае запишется в виде 

\begin{equation}
2ln L(\lambda_{max}=n_{obs}) - 2ln L(\lambda) \le s^2  
\label{eq:121}
\end{equation}

\noindent
или 

\begin{equation}
2[(\lambda - n_{obs}) + n_{obs}(ln~{n_{obs}} - ln \lambda)] \le s^2.  
\label{eq:122}
\end{equation}

\noindent
Решение неравенства~(\ref{eq:122}) представимо в виде 

\begin{equation}
\lambda_{low}(n_{obs},s) \le \lambda \le \lambda_{up}(n_{obs},s).  
\label{eq:123}
\end{equation}

Особо следует рассмотреть случай $n_{obs}=0$. В этом случае из неравенства~(\ref{eq:123})  
получаем $\lambda \le \displaystyle \frac{s^2}{2}$. Для 
$\alpha=0.05$ следует $s=1.64$ (см.~Таблица~\ref{tab:1}) и, соответственно, 
$\lambda \le \displaystyle \frac{s^2}{2} = 1.34$. 

В Таб.~(\ref{tab:Main3}) приведены значения $\lambda_{low}$ и $\lambda_{up}$ для 
$n_{obs}=0,1,2,\dots,10$ и для уровней достоверности $1-\alpha=0.95$ и $1-\alpha=0.9$.

Заметим, что уравнение~(\ref{eq:120}) для определения интервала доверия в методе 
максимального правдоподобия дает тот же результат, что и частотный подход только для 
случая нормального распределения. В общем случае результаты могут сильно отличаться. 
Проиллюстрируем это на примере распределения Лапласа 

\begin{equation}
P(x,a) = \frac{1}{2a}e^{-a|x|},~~\-\infty<x<\infty,~~a\ge 0.  
\label{eq:124}
\end{equation}

Метод максимального правдоподобия (уравнение~(\ref{eq:121})) приводит к неравенству 

\begin{equation}
2a|x| \le s^2(\alpha).   
\label{eq:125}
\end{equation}

\noindent
Тогда как частотный подход приводит к неравенству 

\begin{equation}
\displaystyle e^{-a|x|}\ge \alpha   
\label{eq:126}
\end{equation}

\noindent
и, соответственно, $-a|x|\ge ln\alpha,~~\displaystyle a|x|\le ln{\frac{1}{\alpha}}$.  

Численно соотношения сильно различаются. Так, например, при $|x_0|=1$ и $s(\alpha)=3$ 
$(\alpha=2.7\cdot 10^{-3})$ метод максимального правдоподобия приводит к 
$\alpha \le \displaystyle \frac{3}{2}$, тогда как частотный подход дает 
$\alpha \le 5.9$.

\subsubsection{Байесовский метод}

Как уже отмечалось ранее в случае Байесовского подхода задача статистики сводится к задаче теории 
вероятностей. Для распределения Пуассона формула Байеса имеет вид 

\begin{equation}
P(\lambda|n_{obs}) = P(\lambda)P(n_{obs}|\lambda),  
\label{eq:127}
\end{equation}

\noindent
где приор $P(\lambda)$ удовлетворяет условию нормировки 

\begin{equation}
\displaystyle \int_0^{\infty}{P(\lambda|n_{obs})d\lambda} = 1,  
\label{eq:128}
\end{equation}

\noindent
и задача определения параметров $\lambda_{low}$ и $\lambda_{up}$ сводится к решению уравнения 

\begin{equation}
\displaystyle \int_{\lambda_{low}}^{\lambda_{up}}{P(\lambda|n_{obs})d\lambda} = 1-\alpha-\beta.  
\label{eq:129}
\end{equation}

Выбор интервала $(\lambda_{low},\lambda_{up})$ не единственный. По аналогии с частотным случаем его 
можно выбрать, потребовав, чтобы верхние и нижние пределы на $\lambda$ определялись бы из уравнений 

\begin{equation}
\displaystyle \int_{\lambda_{up}}^{\infty}{P(\lambda|n_{obs})d\lambda} = \alpha,   
\label{eq:130}
\end{equation}

\begin{equation}
\displaystyle \int^{\lambda_{low}}_0{P(\lambda|n_{obs})d\lambda} = \beta.  
\label{eq:131}
\end{equation}

В силу тождеств 

\begin{equation}
\displaystyle \int_{\lambda_{up}}^{\infty}{P(n_{obs}|\lambda)d\lambda} = 
\sum_{n=0}^{n_{obs}}{P(n|\lambda_{up})},  
\label{eq:132}
\end{equation}

\begin{equation}
\displaystyle \int^{\lambda_{low}}_0{P(n_{obs}|\lambda)d\lambda} = 
\sum_{n=n_{obs}+1}^{\infty}{P(n|\lambda_{low})}.  
\label{eq:133}
\end{equation}

\noindent
верхний предел $\lambda_{up}$ в Байесовском подходе совпадает с частотным верхним пределом при 
$P(\lambda)=const$, а нижний предел $\lambda_{low}$ совпадает с частотным нижним пределом 
при $P(\lambda)=\displaystyle \frac{const}{\lambda}$.

Итак, в Байесовском подходе уравнения~(\ref{eq:129},\ref{eq:130},\ref{eq:131}) определяют 
допустимую область параметра $\lambda$ 

\begin{equation}
\lambda_{low}\le\lambda\le\lambda_{up}
\label{eq:129a}
\end{equation}

\noindent
на уровне $(1-\alpha-\beta)$ достоверности. Неравенство~(\ref{eq:129a}) в частотном подходе 
для приора $\pi(\lambda)=const$ эквивалентно неравенству 

\begin{equation}
1-\beta \ge \displaystyle \sum_{n=0}^{n_{obs}}{P(n|\lambda)}\ge \alpha.
\label{eq:129b}
\end{equation}

Для приора $\pi(\lambda)=\displaystyle \frac{const}{\lambda}$ неравенство~(\ref{eq:129a}) 
эквивалентно неравенству

\begin{equation}
1-\alpha \ge \displaystyle \sum_{n=n_{obs}}^{\infty}{P(n|\lambda)}\ge \beta.
\label{eq:129c}
\end{equation}

Другой естественный выбор интервала $[\lambda_{low},\lambda_{up}]$ в интеграле~(\ref{eq:129}) 
заключается в принципе упорядочивания, а именно, точки вошедшие в интервал $[\lambda_{low},\lambda_{up}]$
должны обладать большей плотностью распределения по сравнению с точками не вошедшими в 
интервал $[\lambda_{low},\lambda_{up}]$. То есть, 

$P(\lambda \in [\lambda_{low},\lambda_{up}]|n) \ge 
                  P(\lambda \notin [\lambda_{low},\lambda_{up}]|n)$. 

\noindent
Как следствие справедливо равенство 

$P(\lambda_{low}|n)=P(\lambda_{up}|n)$.

\subsection{Оценка сигнала в распределении Пуассона при ненулевом фоне}
 
Очень часто при поиске новых явлений параметр $\lambda$ а распределении Пуассона можно 
представить как сумму двух слагаемых 

\begin{equation}
\lambda=\lambda_b+\lambda_s, 
\label{eq:134}
\end{equation}

\noindent 
где $\lambda_b=L\epsilon_b\sigma_b$ представляет собой вклад от фона и предполагается 
в данном подразделе точно известным, а $\lambda_s=L\sigma_s\epsilon_s$--вклад от сигнала. 
Задача состоит в том, чтобы оценить вклад сигнала, то есть получить ограничение на параметр 
$\lambda_s$. Эта задача очень похожа на рассматриваемую в предыдущем подразделе задачу об 
оценке параметра $\lambda$ исходя из наблюдаемого значения $n_{obs}$ сигнальных событий. 
Разница заключается в том, что в силу соотношений~(\ref{eq:134}) и неравенств 
$\lambda_b \ge 0$, $\lambda_s \ge 0$ ограничение на параметр $\lambda$ имеет вид 
$\lambda \ge \lambda_b$. 

\subsubsection{Частотный подход}

\paragraph{Проблемы с определением интервалов в случае ограничения на 
параметры распределения.}

Проблемы возникают, когда в распределении $P(x,\theta)$ на параметры распределения 
$\theta$ налагаются дополнительные условия, следующие из физических соображений. 

Рассмотрим пример~\cite{Feldman} нормального распределения 

\begin{equation}
N(x|\mu,\sigma=1)=\displaystyle \frac{1}{\sqrt{2\pi}}e^{-\frac{(x-\mu)^2}{2}}, 
\label{eq:135}
\end{equation}

\noindent
причем параметр $\mu$ может принимать только неотрицательные значения $\mu \ge 0$ 
(например, $\mu$ -- масса частицы). Если мы измерили $x=-1.8$, то согласно наивной процедуре 
извлечения параметра $\mu$ на уровне $1\sigma$ мы получаем $\mu=-1.8\pm1$, что 
находится в противоречии с требованием неотрицательности параметра $\mu$. Принцип 
максимального правдоподобия с учетом ограничения $\mu\ge 0$ приводит к наиболее вероятному 
значению для параметра $\mu$ равному 

\begin{equation}
\mu_{best} = \cases{ x_0,~~~x \ge 0 \cr
                      0,~~~~x < 0}~~~~= max(0,x_0).
\label{eq:136}
\end{equation}

Встает также вопрос о построении допустимого интервала на параметр $\mu$. В работе~\cite{Feldman} 
было предложено решение этой проблемы на основе использования упорядочивания с помощью функции 
максимального правдоподобия, которая в нашем случае имеет вид  

\begin{equation}
R(\mu|x) = \displaystyle \frac{N(x|\mu,\sigma=1)}{N(x|\mu_{best},\sigma=1)} =  
\cases{ \displaystyle e^{-\frac{(x-\mu)^2}{2}},~~~x \ge 0 \cr
                      e^{x\mu-\frac{\mu^2}{2}},~~~x < 0}.
\label{eq:137}
\end{equation}

В методе максимального правдоподобия при $x_0\le 0$ неравенство~(\ref{eq:110}) примет вид 

\begin{equation}
(x_0-\mu)^2-x_0^2 \le s^2 
\label{eq:137a}
\end{equation}

\noindent
или  

\begin{equation}
0\le \mu \le x_0+\sqrt{s^2+x_0^2}. 
\label{eq:137b}
\end{equation}

\noindent
В случае $x_0\ge 0$ неравенство~(\ref{eq:110}) имеет стандартный вид 

\begin{equation}
(\mu - x_0)^2 \le s^2 .
\label{eq:137c}
\end{equation}

\noindent
С учетом дополнительного условия $\mu \ge 0$ решение неравенства~(\ref{eq:137c}) есть 

\begin{equation}
max(0,-s+x_0) \le \mu \le x_0+s. 
\label{eq:137d}
\end{equation}

Согласно принципу упорядочивания~\cite{Feldman} функция $R(\mu|x)$ определяет порядок 
добавления точек $x$ в допустимую область при каждом конкретном $\mu$. А именно, 
добавляются точки $x$ с максимальным значением $L(\mu|x)$. Более конкретно это означает,  
что для заданного значения $\mu$ находится интервал $[x_1,x_2]$ такой, что

\begin{center}
$R(\mu|x_1) = R(\mu|x_2)$,
\end{center}

\begin{equation}
\displaystyle \int_{x_1}^{x_2}{N(x|\mu,\sigma=1)dx}=1-\alpha.
\label{eq:138}
\end{equation}

\begin{figure}[htpb]
\begin{center}
   \resizebox{3.3in}{!}{\includegraphics{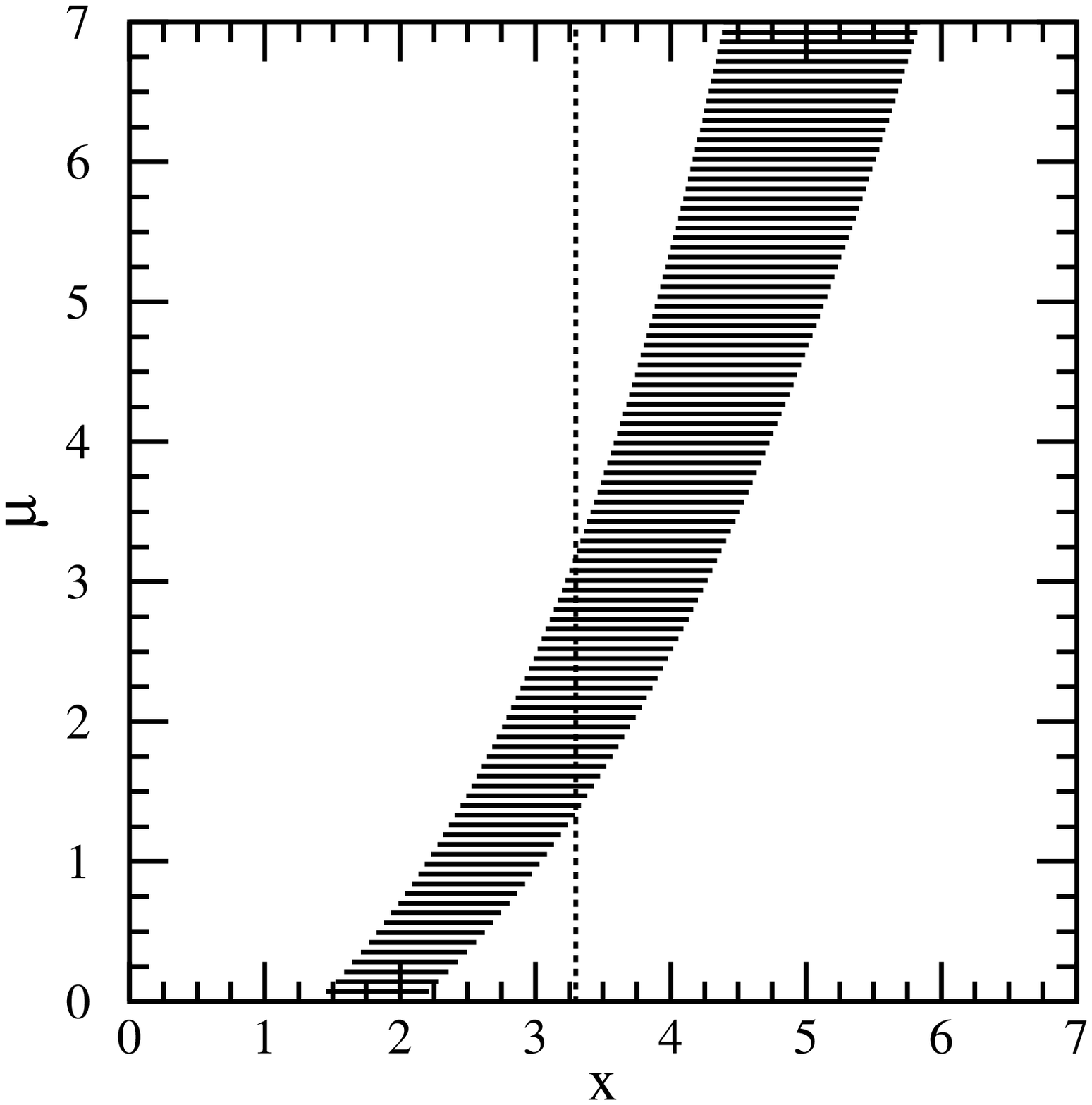}} 
\caption{Построение доверительного коридора и его использование.}
    \label{fig:4}
  \end{center}
\end{figure}

Далее используется стандартная процедура Неймана для определения допустимого интервала на 
значения параметров $[\mu_1,\mu_2]$ (см.~Рис~\ref{fig:4}).

Для измеренного значения $x_0$ параметры $\mu_1,~\mu_2$, определяющие концы интервала 
$[\mu_1,\mu_2]$, соответствуют уравнению~(\ref{eq:138}), когда $x_1(x_2)$ берется 
равным $\mu_2(\mu_1)$.

В таблицах (Таб.~\ref{tab:Main4},\ref{tab:Main5},\ref{tab:Main6}~\cite{Feldman}) 
приведены конкретные значения 
параметров $\mu_1,~\mu_2$ в зависимости от измеренного значения $n_{obs}$ и ожидаемого фона $\lambda_b$
для распределения Пуассона. 

\setlength{\tabcolsep}{4pt}
\begin{table}
\begin{center}
\caption{Значения верхних $\mu_2$ и нижних $\mu_1$ пределов в методе 
Фельдмана-Кузинса для $n_{obs}\le 10$ и $\alpha=0.05,~~\alpha=0.1$. Ожидаемый фон $\lambda_b=0$.}
\begin{tabular}{|c|cc|cc|}
\hline
$n_{obs}$&$\mu_1$& $\mu_2$ & $\mu_1$ & $\mu_2$\\  
          &$1-\alpha=0.9$ &                & $1-\alpha=0.95$  &                 \\ 
\hline
  0       &  $0.0$        &  $2.44$        &  $0.0$           & $3.09$  \\
  1       &  $0.11$       &  $4.36$        &  $0.05$          & $5.14$  \\
  2       &  $0.53$       &  $5.91$        &  $0.36$          & $6.72$  \\
  3       &  $1.10$       &  $7.42$        &  $0.82$          & $8.25$  \\
  4       &  $1.47$       &  $8.60$        &  $1.37$          & $9.76$  \\
  5       &  $1.84$       &  $9.99$        &  $1.84$          & $11.26$  \\
  6       &  $2.21$       &  $11.47$       &  $2.21$          & $12.75$  \\
  7       &  $3.56$       &  $12.53$       &  $2.58$          & $13.81$  \\
  8       &  $3.96$       &  $13.99$       &  $2.94$          & $15.29$  \\
  9       &  $4.36$       &  $15.30$       &  $4.36$          & $16.77$  \\
 10       &  $5.50$       &  $16.50$       &  $4.75$          & $17.82$  \\
\hline
\end{tabular}
\label{tab:Main4}
\end{center}
\end{table}

\setlength{\tabcolsep}{4pt}
\begin{table}
\begin{center}
\caption{Значения верхних $\mu_2$ и нижних $\mu_1$ пределов в методе 
Фельдмана-Кузинса для $n_{obs}\le 10$ и $\alpha=0.05,~~\alpha=0.1$. Ожидаемый фон $\lambda_b=1$.}
\begin{tabular}{|c|cc|cc|}
\hline
$n_{obs}$&$\mu_1$& $\mu_2$ & $\mu_1$ & $\mu_2$\\  
          &$1-\alpha=0.9$ &                & $1-\alpha=0.95$  &                 \\ 
\hline
  0       &  $0.0$        &  $1.61$        &  $0.0$           & $2.33$  \\
  1       &  $0.0$        &  $3.36$        &  $0.00$          & $4.14$  \\
  2       &  $0.0$        &  $4.91$        &  $0.00$          & $5.72$  \\
  3       &  $0.10$       &  $6.42$        &  $0.00$          & $7.25$  \\
  4       &  $0.74$       &  $7.60$        &  $0.37$          & $8.76$  \\
  5       &  $1.25$       &  $8.99$        &  $0.97$          & $10.26$  \\
  6       &  $1.61$       &  $10.47$       &  $1.61$          & $11.75$  \\
  7       &  $2.56$       &  $11.53$       &  $1.97$          & $12.81$  \\
  8       &  $2.96$       &  $12.99$       &  $2.33$          & $14.29$  \\
  9       &  $3.36$       &  $14.30$       &  $3.36$          & $15.77$  \\
 10       &  $4.50$       &  $15.50$       &  $3.75$          & $16.82$  \\
\hline
\end{tabular}
\label{tab:Main5}
\end{center}
\end{table}

\setlength{\tabcolsep}{4pt}
\begin{table}
\begin{center}
\caption{Значения верхних $\mu_2$ и нижних $\mu_1$ пределов в методе 
Фельдмана-Кузинса для $n_{obs}\le 10$ и $\alpha=0.05,~~\alpha=0.1$. Ожидаемый фон $\lambda_b=3$.}
\begin{tabular}{|c|cc|cc|}
\hline
$n_{obs}$&$\mu_1$& $\mu_2$ & $\mu_1$ & $\mu_2$\\  
          &$1-\alpha=0.9$ &                & $1-\alpha=0.95$  &                 \\ 
\hline
  0       &  $0.0$        &  $1.08$        &  $0.0$           & $1.63$  \\
  1       &  $0.0$        &  $1.88$        &  $0.0$           & $2.63$  \\
  2       &  $0.0$        &  $3.04$        &  $0.0$           & $3.84$  \\
  3       &  $0.0$        &  $4.42$        &  $0.0$           & $5.25$  \\
  4       &  $0.0$        &  $5.60$        &  $0.0$           & $6.76$  \\
  5       &  $0.0$        &  $6.99$        &  $0.0$           & $8.26$  \\
  6       &  $0.15$       &  $8.47$        &  $0.0$           & $9.75$  \\
  7       &  $0.89$       &  $9.53$        &  $0.29$          & $10.81$  \\
  8       &  $1.51$       &  $10.99$       &  $0.98$          & $12.29$  \\
  9       &  $1.88$       &  $12.30$       &  $1.62$          & $13.77$  \\
 10       &  $2.63$       &  $13.50$       &  $2.25$          & $14.82$  \\
\hline
\end{tabular}
\label{tab:Main6}
\end{center}
\end{table}

Заметим, что в рамках Байесовского подхода задача состоит в выборе приора 
$P(\mu)$ такого, что $P(\mu)=0$ при $\mu<0$. В частности, 
вместо приора $P(\mu)=1$, часто используемого для нормального распределения, когда параметр $\mu$ 
может принимать произвольные значения, можно использовать приор $const \cdot \theta(\mu)$ 
автоматически учитывающий неотрицательность параметра $\mu$~\footnote{Формула~(\ref{eq:138a}) по 
сути дела есть аналог соответствующей формулы, используемой для распределения Пуассона 
(формула O. Helene~\cite{Helene}).}. При использовании приора $p(\mu)=const \cdot \theta(\mu)$  
нижний предел на параметр $\mu$ определяется с помощью соотношения 

\begin{equation}
\displaystyle \frac{\displaystyle 
\int_{\mu_-}^{\infty}{e^{-\frac{1}{2}(\mu-x_0)^2}d\mu}} 
{\displaystyle \int_{-x_0}^{\infty}{e^{-\frac{1}{2}y^2}}dy} 
\le 1 - \alpha.
\label{eq:138a}
\end{equation}

Заметим, что в классическом частотном подходе в формулах~(\ref{eq:138a}) знаменатель 
предполагается равным $1$. ``Волевым образом'' поделив пределы на вероятность 
выпадения $x_0$, мы получим формулу~(\ref{eq:138a})~\footnote{Из условия нормировки 
$P(\mu|x_0) = \theta(\mu)\cdot const \cdot \displaystyle 
e^{-\frac{(\mu-x_0)^2}{2}}$,~~~$\int{P(\mu|x_0)d\mu=1}$, 
следует, что 
$const = \displaystyle 
\frac{1}{\frac{1}{\sqrt{2\pi}}\int_{-x_0}^{\infty}{e^{-\frac{y^2}{2}}dy}}$. 
Здесь $\theta(\mu) = \cases{ 0,~~~\mu=0 \cr 
                            1,~~~\mu\ge 0}$~~.}. 

\noindent
Задача определения верхнего предела на параметр $\mu$ сводится к решению неравенства  

\begin{equation}
\displaystyle \frac{\displaystyle 
\int_{0}^{\mu_+}{e^{-\frac{1}{2}(\mu-x_0)^2}d\mu}} 
{\displaystyle \int_{-x_0}^{\infty}{e^{-\frac{1}{2}y^2}}dy} 
\le 1 - \alpha.
\label{eq:138b}
\end{equation}

В случае распределение Пуассона $P(n|\lambda_b+\lambda_s)$ с ненулевым известным фоном 
$\lambda_b$ функция правдоподобия 
максимальна при $\lambda_{s,best}=max(0,n-\lambda_b)$.

В подходе Фельдмана-Кузинса выбор индивидуальных слагаемых в сумме 

\begin{equation}
\displaystyle \sum_{n=n_-}^{n_+}{P(n|\lambda_b+\lambda_s)} 
\label{eq:139a}
\end{equation}

\noindent
осуществляется согласно принципу упорядочивания по функции 

$R=\displaystyle \frac{P(n|\lambda_b+\lambda_s)}
{P(n|\lambda_b+\lambda_{s,best})}$,  

\noindent
то есть допустимые члены в сумме~(\ref{eq:139a}) выбираются максимально большими. 

Функция 
$R=\displaystyle \frac{P(n|\lambda_b+\lambda_s)}{P(n|\lambda_b+\lambda_{s,best})}$ 
используется для упорядочивания значений $n$. Далее строится интервал доверия такой, что 
$P(\lambda_s\in[\lambda_1,\lambda_2])\ge 1-\alpha$. Это осуществляется с помощью конструкции 
Неймана, а именно, строится сумма 

\begin{equation}
\displaystyle \sum_{n=n_1}^{n_2}{P(n|\lambda_b+\lambda_s)} \ge 1 - \alpha.
\label{eq:139}
\end{equation}

\noindent
В сумме~(\ref{eq:139}) слагаемые подбираются с помощью упорядочивания по величине $R$. 
В итоге мы имеем полосу Неймана $n_1(\lambda_s),~n_2(\lambda_s)$ (см.~Рис~\ref{fig:4}).
Далее стандартным образом находятся $\lambda_{s_{low}}$ и $\lambda_{s_{up}}$ такие, 
что интервал доверия имеет вид $\lambda_{s_{low}} \le \lambda_s \le \lambda_{s_{up}}$.

%
%
%

Заметим, что в частотном подходе при извлечении верхнего предела на сигнал $s$ из распределения 
Пуассона $P(n|\lambda_b+\lambda_s)$ и $n=n_{obs}$ на уровне $(1-\alpha)$ 
достоверности числа наблюденных событий ограничение на $\lambda_s$ получается из неравенства  

\begin{equation}
P(n \le n_{obs}|\lambda_b+\lambda_s) = 
\displaystyle \sum_{n=0}^{n_{obs}}{P(n|\lambda_b+\lambda_s)} \ge \alpha. 
\label{eq:140}
\end{equation}

%

В модифицированном частотном подходе ($CL_s$ подходе) уравнение~(\ref{eq:140}) 
заменяется на  

\begin{equation}
\displaystyle \frac{P(n \le n_{obs}|\lambda_b+\lambda_s)}{P(n\le n_{obs}|\lambda_b)} 
\ge \alpha,  
\label{eq:142}
\end{equation}

%

\noindent 
что соответствует в Байесовском подходе замене функции приора 
$\theta(\lambda) \rightarrow const\cdot\theta(\lambda-\lambda_b)$. 

\subsubsection{Байесовский подход}

В Байесовском подходе тот факт, что $\lambda \ge \lambda_b$ можно учесть самым простым 
способом, а именно, $P(\lambda) \rightarrow \theta(\lambda-\lambda_b)P(\lambda)$. 

Введение дополнительной функции $\theta(\lambda-\lambda_b)$ изменит условия нормировки. 
В остальном же получение ограничений на параметр $\lambda$ или $\lambda_s$ абсолютно такое 
же, как и предыдущем подразделе. Возьмем постоянный приор, для которого, как было показано, 
верхний Байесовский предел совпадает с частотным верхним пределом благодаря равенству~(\ref{eq:96}). 
С учетом модификации  
$ P(\lambda) \rightarrow \theta(\lambda - \lambda_b)P(\lambda) $
 уравнение для определения верхнего предела для приора 
$\pi(\lambda) = const\cdot \theta(\lambda-\lambda_b)$ запишется в виде 

\begin{equation}
\displaystyle \frac{\int_{\lambda_{up}}^{\infty}{P(n_{obs}|\lambda)d\lambda}}
{\int_{\lambda_b}^{\infty}{P(n_{obs}|\lambda)d\lambda}} \ge \alpha   
\label{eq:144}
\end{equation}

\noindent
или 

\begin{equation}
\displaystyle \frac{\int_{\lambda_{up}-\lambda_b=\lambda_s}^{\infty}
{P(n_{obs}|\lambda_b+\lambda_s')d\lambda_s'}}
{\int_{\lambda_b}^{\infty}{P(n_{obs}|\lambda)d\lambda}} \ge \alpha. 
\label{eq:145}
\end{equation}

\noindent
В силу тождеств~(\ref{eq:96},\ref{eq:96a}) ограничение~(\ref{eq:145}) можно переписать 
в виде  

\begin{equation}
\displaystyle \frac{\sum_{n=0}^{n_{obs}}{P(n_{obs}|\lambda_b+\lambda_s)}}
{\sum_{n=0}^{n_{obs}}{P(n_{obs}|\lambda_b)}} \ge \alpha. 
\label{eq:146}
\end{equation}

\noindent
Для $n_{obs}=0$ неравенство примет вид 

$\displaystyle \frac{e^{-\lambda_b-\lambda_s}}{e^{-\lambda_b}} \ge \alpha,$

\noindent
то есть $\lambda_s \le \displaystyle ln{\frac{1}{\alpha}}$. Без знаменателя 
отличного от единицы в формуле~(\ref{eq:146}) мы имели бы неравенство 
$\lambda_b+\lambda_s \le \displaystyle ln{\frac{1}{\alpha}}$.  

Аналогичное неравенство можно получить и для нижнего предела. Для определения нижнего предела 
для приора $\displaystyle \frac{const}{\lambda}$ неравенство для определения нижнего 
предела примет вид 

\begin{equation}
\displaystyle \frac{\int_{\lambda_b}^{\lambda_b+\lambda_s}{P(n_{obs}-1|\lambda)d\lambda}}
{\int_{\lambda_b}^{\infty}{P(n_{obs}-1|\lambda)d\lambda}} \ge \beta. 
\label{eq:147}
\end{equation}

Для $n_{obs}=1$ получаем 
$\displaystyle \frac{e^{-\lambda_b}-e^{-\lambda_b-\lambda_s}}{e^{-\lambda_b}} = \beta$, 
$\displaystyle 1-e^{-\lambda_s}=\beta$, $\lambda_s=ln~\frac{1}{1-\beta} \approx \beta$. 

При $\beta=0.05$ ограничение имеет вид $\lambda_s > 0.05$.

%
%
%
%
%
%
%
%
%

\subsubsection{Оценка параметров в методе наименьших квадратов}

Часто используют метод наименьших квадратов, основанный на минимизации квадратичной формы. 
Рассмотрим выборку $\vec y=(y_1,y_2,\dots,y_n)$ наблюдений при измерении случайной величины $Y$. 
Пусть математическое ожидание в каждом измерении есть $E(Y_i,\vec \theta)$, а матрица вторых 
моментов соответствующих распределений есть $[V]$. Здесь $\vec \theta$ неизвестные 
параметры, а $V_{ij}(\vec \theta)$ известные функции от $\theta$. В методе наименьших квадратов 
оценками параметров $\vec \theta$ служат такие значения $\vec {\hat \theta}$, которые обращают в 
минимум квадратичную форму 

$Q^2 = \displaystyle \sum_{i=1}^N
{\sum_{j=1}^N{[y_i-E(Y_i,\vec \theta)]([V]^{-1})_{ij}[y_j-E(Y_j,\vec \theta)]}}=$

\begin{equation}
[\vec y-E(\vec y,\vec \theta)]^T[V]^{-1}[\vec y - E(\vec y,\vec \theta)].
\label{eq:147a}
\end{equation}

\noindent
В случае отсутствия корреляций ковариантная матрица $V_{ii}=\sigma^2_i(\vec \theta)$ 
диагональна и выражение~(\ref{eq:147a}) к известной сумме квадратов 

$Q^2=\displaystyle \sum_{i=1}{\sigma_i^{-2}(\vec \theta)[Y_i-E(Y_i,\vec \theta)]^2}$.

В случае нормального $N$-мерного распределения метод максимального правдоподобия и 
метод наименьших квадратов совпадают. 

Метод наименьших квадратов часто используется при фитировании гистограмм. А именно, 
при выборе бинов в которых достаточно большое количество событий, так что приближение 
нормального распределения оправдано и в приближении отсутствует корреляция между бинами, 
необходимо минимизировать величину 

\begin{equation}
Q^2=\displaystyle 
\sum_{i=1}^N{\frac{(N_i-\lambda_i(\vec \theta))^2}{\lambda_i(\vec \theta)}}. 
\label{eq:147b}
\end{equation}

\noindent
Это есть не что иное как обычный $\chi^2$-метод, применяемый для фита гистограмм. 

Для практических приложений вместо минимизации величины~(\ref{eq:147b}) более удобно 
минимизировать величину 

\begin{equation}
Q^2_{mod}=\displaystyle 
\sum_{i=1}^N{\frac{(N_i-\lambda_i(\vec \theta))^2}{N_i}}. 
\label{eq:147c}
\end{equation}

\noindent
Поскольку $E(y_i)=\lambda_i(\vec \theta)$, модифицированное выражение~(\ref{eq:147c}) 
является разумным приближением при минимизации~(\ref{eq:147a}). 

Можно так же в методе максимального правдоподобия максимизировать непосредственно 

$ln~L=\displaystyle \sum_{i=1}^N{N_iln\lambda_i(\vec \theta)}-
\sum_{i=1}^N{\lambda_i(\vec \theta)}$.

\noindent
В приближении $ln~L\approx\displaystyle \sum_{i=1}^N{N_iln\lambda_i(\vec\theta)}~~$ 
$(\sum{\lambda_i(\vec\theta)=const})$ задача сводится к более простой задаче максимизации 
$\displaystyle \sum_{i=1}^N{N_iln\lambda_i(\vec \theta)}$. 

\subsubsection{Интервал доверия в случае нескольких параметров}

Пусть существует оценка $\hat {\vec \theta}$ параметров $\vec \theta$ и эта оценка 
распределена по нормальному многомерному закону со средним $\vec\theta$ и матрицей вторых 
моментов $[V]$. В этом случае  

\begin{equation}
f(\hat {\vec \theta}|\vec \theta) = \displaystyle
\frac{1}{(2\pi)^{\frac{N}{2}(det[V])^{\frac{1}{2}}}}
e^{-\frac{1}{2}(\hat {\vec \theta}-\vec \theta)^T[V]^{-1}(\hat {\vec \theta}-\vec\theta)}.
\label{eq:147d}
\end{equation}

\noindent
Ковариантная форма 
$Q(\hat {\vec \theta},\vec\theta)=
(\hat {\vec \theta}-\vec \theta)^T[V]^{-1}(\hat {\vec \theta}-\vec\theta)$ 
имеет $\chi^2(N)$-распределение (см.раздел~\ref{sec:chi2}).
Это означает в частности, что распределение $Q(\hat {\vec \theta},\vec\theta)$ 
не зависит от $\vec \theta$ и можно написать вероятностное утверждение

\begin{equation}
P(Q(\hat {\vec \theta},\vec\theta)\le K_{\beta}^2)=\beta, 
\label{eq:147e}
\end{equation}

\noindent
где $\beta$-вероятность того, что оценка параметра будет находиться в области доверия,  
определяемой точкой $K_{\beta}^2$ для $\chi^2(N)$-распределения.

Область в пространстве 
$\hat{\vec\theta}$, задаваемая уравнением  
$Q(\hat {\vec \theta},\vec\theta)=K_{\beta}^2$, имеет вид гиперэллипсоида, 
поверхность которого соответствует постоянному значению плотности вероятности для 
функции~(\ref{eq:147d}).  

В двумерном случае матрица $[V]$ это  

\begin{center}
$[V] =\left(\begin{array}{cc}
 \sigma_1^2                  & \rho \sigma_1 \sigma_2  \\
 \rho\sigma_1 \sigma_2  & \sigma_2^2    
\end{array}\right)$.
\end{center}

\noindent
Эллипс содержит параметры, одновременно удовлетворяющие 
неравенству~(\ref{eq:147e}).

\subparagraph{Оценка методом максимального правдоподобия в случае нескольких параметров}

Обобщение одномерного неравенства~(\ref{eq:121}) имеет вид 

$2[ln~L(\vec \theta_{max}) - ln~L(\vec \theta)] \le \chi^2_{\beta}(k)$.

Как показано (например,~\cite{Likelihood}), функция $-2ln~l(\vec \theta)$, где 
$l(\vec \theta)= \displaystyle \frac{L(\vec\theta)}{L(\vec\theta_{max})}$, 
распределена асимптотически как $\chi^2(k)$. Это приводит к неравенству 

$\beta=P[-2ln~l(\vec \theta) \le \chi^2_{\beta}(k)]$, 

\noindent
которое определяет интервал доверия для параметров $\vec \theta$. 

\subsubsection{Асимптотическая нормальность оценки максимального правдоподобия}

Оценка параметра распределения $\hat \theta$ по методу максимального правдоподобия 
асимптотически распределена как $N(\theta,\displaystyle \frac{1}{NI})$, 
где $I$ -- информация о параметре $\theta$ в одном измерении, а $N$-число измерений~\cite{James}. 
Отсюда следует, что интервал доверия для $\hat \theta$ имеет асимптотический вид 
$\hat \theta \pm\displaystyle \frac{\lambda_{\frac{\beta}{2}}}{\sqrt{NI(\theta)}}$, 
где $\displaystyle \lambda_{\frac{\beta}{2}}$ есть $\displaystyle \frac{\beta}{2}$ 
точка стандартного нормального распределения. 

Более точная оценка получается из того факта, что 
$\displaystyle \frac{\partial}{\partial\theta}ln~L$ распределено со средним значением 
равным нулю и вариацией $NI(\theta)$. Распределение 
$\displaystyle \frac{1}{\sqrt{NI(\theta)}}\frac{\partial}{\partial\theta}ln~L$ 
является асимптотически нормальным распределением. Доверительный интервал определяется неравенством 

$\displaystyle 
|\frac{1}{\sqrt{NI(\theta)}}\frac{\partial}{\partial\theta}ln~L| \le 
\lambda_{\frac{\beta}{2}}$. 


\subsection{Ожидаемые пределы} 

В условиях будущего эксперимента мы не знаем количество событий $n$ в распределении Пуассона 
$P(n|\lambda)$. Как правило мы знаем (с какой-то точностью) параметр $\lambda=\lambda_b$ для 
СМ. В случае присутствия новых взаимодействий 

$\lambda=\lambda_b+\lambda_s$,

\noindent
где $\lambda_s$ -- вклад физики вне рамок СМ. Поэтому возникает естественный вопрос -- 
какое ограничение сверху на параметр $\lambda_s$ мы можем получить в будущем эксперименте? 
Вопрос сводится к следующему: какое значение $n_{obs}$ в соответствующих формулах (Глава 3) 
необходимо использовать? Наиболее распространенный рецепт (``Azimov dataset'') это использование 
наиболее вероятного значения $n_{obs}=[\lambda_b]$ для распределения Пуассона $P(n|\lambda)$. 
Так, например, ограничение~(\ref{eq:146})~\footnote{Напомним, что ограничение ~(\ref{eq:146}) 
естественно получается в Байесовском подходе  и часто именуется $CL_S$ методом~\cite{CLs}.}  
примет вид  

$\displaystyle \frac{\sum_{n=0}^{[\lambda_b]}{P([\lambda_b]|\lambda_b+\lambda_s)}}
{\sum_{n=0}^{[\lambda_b]}{P([\lambda_b]|\lambda_b)}} \ge \alpha$. 

\noindent
Смысл анзаца $n_{obs}=[\lambda_b]$ очень простой. Вероятности того, что 
$n_{obs} \le [\lambda_b]$ или $n_{obs} > [\lambda_b]$ приблизительно равны 50\%. 
И, поэтому, с вероятностью $~$50\% в случае реального эксперимента можно получить как лучшее, так 
и худшее ограничение, чем ожидаемое ограничение. 

В случае измерения непрерывной величины, распределенной по нормальному закону $N(x|\mu_b,\sigma^2)$, 
наиболее вероятное значение случайной переменной $x=\mu_b$. Ограничение на параметр сигнала 
$\mu_s$ в предположении, что случайная переменная $X$ распределена в этом случае по нормальному 
закону $N(x|\mu_b+\mu_s,\sigma^2)$ на уровне $1~\sigma$ есть $-\sigma\le\mu_s\le\sigma$. 

В случае нормального распределения $N(x|\mu_b,\sigma^2)$ в 68\% случаев ($1~\sigma$) 
случайная величина $X$ должна лежать в интервале $\mu_b-\sigma\le X\le\mu_b+\sigma$. 
Отсюда мы получаем, что в 68\% случаев ограничение на параметр новой физики $\mu_s$ на уровне 
$1\sigma$ лежит 
в интервале $-2~\sigma\le\mu_s\le2~\sigma$. 

Для случая распределения Пуассона обобщение вышеприведенной процедуры очевидно и состоит 
в выборе таких значений $n$, лежащих вблизи $[\lambda_b]$, что 

$\displaystyle \sum_{n_k}P(n_k|\lambda_b)\ge68$\%.

Далее для каждого $n_k$ находим верхний предел на параметр сигнала $\lambda_s$ 
($\lambda=\lambda_s+\lambda_b$) и для каждого $\lambda_b$ выбираются наибольшие ($n_k^{max}$) 
и наименьшие ($n_k^{min}$) значения $n_k$ и с их помощью определяются соответственно 
наименее ``удачное'' и наиболее ``удачное'' значения верхнего предела на $\lambda_s$.

\newpage

\section{Учет систематических ошибок}

\subsection{Введение}

Типичная задача в физике высоких энергий это извлечение из экспериментальных 
данных информации о возможности существования новой физики вне рамок СМ.  Например, при изучении 
той или иной реакции~\footnote{Например, количество димюонных событий с ограничениями 
по поперечным импульсам мюонов и их инвариантной массе.} с наложением ряда 
кинематических ограничений (обрезаний) вероятность детектирования $n$ событий 
определяется распределением Пуассона $P(n|\lambda)$, где 
$\lambda=\lambda_b+\lambda_s,~~\lambda_b=L\epsilon_b\sigma_b,~~
\lambda_s=L\epsilon_s\sigma_s$. 
Здесь $L$ -- полная интегральная светимость, $\epsilon_b,~\epsilon_s$ -- 
эффективность детектирования для, соответственно, фоновых и сигнальных событий, 
$\sigma_b,~\sigma_s$ сечения фона и сигнала. 

Сечение сигнала $\sigma_s$ зависит от параметров модели, описывающих новое 
явление (массы суперсимметричных частиц, масса $Z'$-бозона, $\dots$). 
Основная задача состоит в нахождении из экспериментальных данных величины или 
ограничения на сигнальное сечение $\sigma_s$ или (что почти то же самое) в 
нахождении параметров (или ограничений на них), определяющих новую физику. 

Проблема заключается в том, что параметры модели $L, \epsilon_b, 
\epsilon_s, \sigma_b, \sigma_s$ определяются с какой-то точностью, это 
называется систематическими эффектами.
Так, например, точность определения интегральной светимости $L$ в процессе работы 
БАКа в 2010 году оценивается равной 11\%~\cite{LHCaccur}.

Параметры эффективности регистрации фона и сигнала зависят от работы детектора и связаны 
главным образом с не 100\% эффективностью регистрации частиц, а также с неточным определением 
импульса частиц. 

При определении $\sigma_b$ и $\sigma_s$ возникают неопределенности, связанные с 
неточным теоретическим знанием сечений. Здесь, правда следует сказать, что иногда величину 
фона $\sigma_b$ можно извлечь из экспериментальных данных в другой кинематической области. 

Обычно систематические неопределенности разделяют на три класса: 
\begin{enumerate}
\item систематические неопределенности, которые можно устранить путем измерения величин
в другой кинематической области. При этом систематические неопределенности такого рода 
определяются статистическими неопределенностями измеряемых величин. Например, при определении 
сечения фона сигнатуры с $n \ge 1~jets + E_T^{miss}$ с адронными струями и ненулевым 
поперечным потерянным импульсом, фон, возникающий от реакции 
$n \ge 1~jets + (Z \rightarrow \nu \bar \nu)$, можно оценить исходя из измерения 
сечения процесса $n \ge 1~jets + (Z \rightarrow l^+l^-)$ и известной величины 
$\displaystyle 
\frac{\Gamma(Z\rightarrow\nu\bar \nu)}{\Gamma(Z\rightarrow l^+l^-)}.$ 
\item Неопределенности, связанные с модельными предположениями при извлечении данных. 
Предположения об эффективности регистрации частиц, точности определения их импульсов, 
мисидентификации (инструментальные фоны)~\footnote{Типичный пример мисидентификации -- 
это неправильное отожествление адронной струи в качестве электрона или, что намного 
реже, в качестве мюона.}. 
\item Неопределенности, связанные с неточным определением теоретических сечений 
$\sigma_b$ и $\sigma_s$. Как уже отмечалось выше в ряде случаев, но не всегда, 
определить сечение фона $\sigma_b$ можно исходя из экспериментальных данных в другой 
кинематической области. 
\end{enumerate}

Существует несколько способов учета систематических эффектов. Они будут рассмотрены в 
последующих трех подразделах. 

\subsection{Оценка параметра фона, исходя из измерений в другой кинематической области}

Проблему можно сформулировать следующим образом: пусть интересующая нас для поиска новой физики 
кинематическая область описывается распределением Пуассона $P(n_1, \lambda_b+\lambda_s)$. 
Связанная с ней кинематическая область, где как мы предполагаем сигнал очень мал, 
описывается распределением Пуассона $P(n_2, \tau\lambda_b)$, где $\tau$ -- известное число. 
Задача состоит в том, чтобы по измеренному значению $n_{2obs}$ определить  
параметр $\lambda_b$  и с помощью измерения числа событий $n_{1obs}$ в интересующей 
нас кинематической области получить ограничение на параметр сигнала $\lambda_s$.

Решение этой задачи наиболее просто в методе максимального правдоподобия. Функция 
правдоподобия равна

\begin{equation}
L(\lambda_b,\lambda_s|n_{1obs},n_{2obs}) =
P(n_{2obs}|\lambda_b+\lambda_s)P(n_{1obs}|\tau\lambda_b). 
\label{eq:149}
\end{equation}

\noindent
Далее из условия максимума 

\begin{equation}
\displaystyle \frac{\partial}{\partial \lambda_b}L(~~)=
\frac{\partial}{\partial \lambda_s}L(~~)=0 
\label{eq:150}
\end{equation}

\noindent
находятся наиболее правдоподобные значения параметров $\lambda_b$ и $\lambda_s$ и 
получаются ограничения на $\lambda_b$ и $\lambda_s$.

Заметим, что в Байесовском подходе, как уже отмечалось выше, справедлива формула 

\begin{equation}
P(\lambda_s|n_{2obs})= \displaystyle 
\int{d\lambda_bP(n_{2obs}|\lambda_b+\lambda_s)\pi(\lambda_b)}. 
\label{eq:151}
\end{equation}

\noindent
А приор $\pi(\lambda_b)$ с учетом измерения $n_{2obs}$ для фона можно определить как  

\begin{equation}
\pi(\lambda_b)= \displaystyle 
\frac{P(n_{1obs}|\tau\lambda_b)\eta(\lambda_b)}
{\int{d\lambda_bP(n_{1obs},\tau\lambda_b)\eta(\lambda_b)}}, 
\label{eq:152}
\end{equation}

\noindent
где $\eta(\lambda_b)$ -- приор, используемый при измерении фонового процесса.

\subsection{Оценка систематических неопределенностей в методе максимального правдоподобия}

Предположим, что мы измеряем количество событий, распределенных по закону Пуассона 
$P(n|\lambda_b+\lambda_s)$, и неопределенность фона $\lambda_b$ 
описывается нормальным распределением 

\begin{equation}
P(\lambda_b|\lambda^0_b,\Delta\lambda^0_b)= \displaystyle 
\frac{1}{\sqrt{2\pi}\Delta\lambda^0_b}
e^{-\frac{1}{2}(\frac{\lambda_b-\lambda^0_b}{\Delta\lambda^0_b})^2}.  
\label{eq:153}
\end{equation}

Функция правдоподобия имеет вид

\begin{equation}
L(\lambda_s,\lambda_b|\lambda_b^0,n)=\displaystyle 
\frac{(\lambda_b+\lambda_s)^n}{n!}e^{-\lambda_b-\lambda_s}
\frac{e^{-\frac{1}{2}(\frac{\lambda_b-\lambda_b^0}{\Delta \lambda_b^0})^2}}
{\sqrt{2\pi}\Delta \lambda_b^0}.
\label{eq:154}
\end{equation}

Оценка параметров для случая отсутствия сигнала $(\lambda_s=0)$ получается путем 
решения уравнения 

\begin{equation}
\displaystyle \frac{\partial}{\partial \lambda_b}
L(\lambda_s=0,\lambda_b|\lambda_b^0,n)=0.
\label{eq:155}
\end{equation}

\noindent
Решение имеет вид

\begin{equation}
\lambda_b(\lambda_b^0,\Delta\lambda_b^0,n)= \displaystyle 
\frac{\lambda_b^0-(\Delta \lambda_b^0)^2}{2} +
\sqrt{(\frac{\lambda_b^0-(\Delta \lambda_b^0)^2}{2})^2+n(\Delta \lambda_b^0)^2}.
\label{eq:156}
\end{equation}

Значение вероятности $P$ осуществления $n_{obs}$ измерений в отсутствие сигнала  
определяется стандартной формулой

$P(\lambda_b^0|n_{obs})=P[N\ge n_{obs}|\lambda_b=\lambda_b(\lambda_b^0,\Delta\lambda_b^0,n)]= $
\begin{equation}
\displaystyle 
\sum_{k=n_{obs}}^{\infty}
{\frac{(\lambda_b(\lambda_b^0,\Delta\lambda_b^0,n))^ke^{-\lambda_b(\lambda_b^0,\Delta\lambda_b^0,b)}}{k!}}. 
\label{eq:157}
\end{equation}

\noindent
В случае ненулевого $\lambda_s$ из условия максимума функции правдоподобия $L(~~)$ по отношению к переменным 
$\lambda_s$ и $\lambda_b$

\begin{equation}
\displaystyle \frac{\partial}{\partial \lambda_s}L(~~) = 
\frac{\partial}{\partial \lambda_b}L(~~) = 0
\label{eq:158}
\end{equation}

\noindent
находится функция максимального правдоподобия 
$L(\hat\lambda_s,\hat\lambda_b|\lambda_b^0,\lambda_s^0, n)$ и сравнивается 
с функцией максимального правдоподобия в отсутствии сигнала. Далее, 
используя методы изложенные в Главе (3), 
можно получить ограничение на параметр сигнала $\lambda_s$. 

\subsection{Метод усреднения Кузинса-Хайланда}
 
Метод Кузинса-Хайланда по существу применяет Байесовский подход к проблеме систематики или 
несущественных параметров.

Предположим, что нам известен Байесовский приор $\pi(\theta')$ для всех несущественных параметров 
$\theta'$ (например, параметров, описывающих неопределенности сечения фона и сигнала). Для функции 
распределения $P(x|\theta,\theta')$, где $x$-случайная величина, $\theta$-существенные параметры, 
проинтегрируем по несущественным параметрам $\theta'$ и получим усредненную функцию распределения 

\begin{equation}
P_{av}(x|\theta)=\int{d\theta'\pi(\theta')P(x|\theta,\theta')d\theta'}.
\label{eq:159}
\end{equation}

Далее, при извлечении из экспериментальных данных информации о существенных величинах $\theta$,   
используем усредненную функцию распределения $P_{av}(x|\theta)$. 
При извлечении из экспериментальных данных на основе усредненной функции распределения 
$P_{av}(x|\theta)$ можно применять как частотный, так и Байесовский подходы, а также 
метод максимального правдоподобия. 

Как уже отмечалось ранее, параметры $\lambda_b$ и $\lambda_s$ представимы в виде
$\lambda_b=L\epsilon_b\sigma_b$, $\lambda_s=L\epsilon_s\sigma_s$.
При оценке систематической неопределенности, например, $\lambda_b$ параметра необходимо знать 
распределения параметров $L,\epsilon_b,\sigma_b,\sigma_s$.

%
%


Обычно предполагается, что параметры $L, \epsilon_b, \epsilon_s, \sigma_b, \sigma_s$ 
независимы и подчиняются нормальному распределению~\footnote{Часто при оценках используют также 
логнормальное распределение.}.

%
%
%
%
%
%
%
%
%
%

Проиллюстрируем метод Кузинса-Хайланда на примере измерения случайной величины $X$, распределенной 
по нормальному закону $N(x|\mu,\sigma^2)$. Пусть мы точно знаем дисперсию, а величина 
$\mu$ распределена по нормальному закону $N(\mu|\mu_0,\sigma_{\mu}^2)$. Здесь 
$\mu_0$ -- теоретическое (среднее) значение величины $\mu$, а $\sigma_{\mu}^2$ -- дисперсия, 
отражающая неточное знание параметра $\mu$. Наша задача состоит в получении ограничений на параметр 
$\mu_0$. Напомним, что согласно методу Кузинса-Хайланда 
распределение случайной величины $X$ определяется путем интегрирования по несущественным параметрам  
\begin{equation}
P_{av}(x|\theta)=\int{d\theta'\pi(\theta')P(x|\theta,\theta')d\theta'}
\label{eq:160}
\end{equation}

\noindent
В нашем случае формула~(\ref{eq:160}) принимает вид 

\begin{equation}
G(x|\mu_0,\sigma^2)=\displaystyle 
\int_{-\infty}^{\infty}{d\mu N(x|\mu,\sigma^2)N(\mu|\mu_0,\sigma_{\mu}^2)}=
N(x|\mu_0,\sigma^2+\sigma_{\mu}^2).
\label{eq:161}
\end{equation}

Таким образом, учет систематики, связанной с неточным знанием величины $\mu$, сводится 
к замене дисперсии 

\begin{equation}
\sigma^2 \rightarrow \sigma^2 + \sigma_{\mu}^2.
\label{eq:162}
\end{equation}

Предположим, что мы измерили величину $X$ равную $x_0$. Тогда, с учетом замены~(\ref{eq:162}), 
значимость нашего измерения определяется стандартной формулой 

\begin{equation}
s=\frac{|x_0-\mu_0|}{\sqrt{\sigma^2+\sigma_{\mu}^2}}.
\label{eq:163}
\end{equation}

Отличие систематической ошибки $\sigma_{\mu}^2$, связанной с неточным знанием величины 
$\mu_0$, от статистической ошибки заключается в том, что она, как правило, пропорциональна 
среднему значению $\mu_0$, то есть 
$\sigma_{\mu}^2\sim\mu_0^2$. 

В случае $N$ измерений случайной величины $X$ среднее значение 
$\bar x = \displaystyle \frac{1}{N}\sum_{i=1}^N{x_i}$ с учетом систематики 
распределена по нормальному закону $\displaystyle N(\bar x| \mu_0, \sigma_{\bar x}^2)$, 
причем  
$\displaystyle \sigma_{\bar x}^2=\frac{\sigma^2}{N}+\sigma_{\mu}^2$.

Дисперсия $\sigma_{\mu}^2$ не зависит от числа измерений и, как следствие, получаем, что в пределе 
бесконечного числа измерений 
$\displaystyle \sigma_{\bar x}^2 \rightarrow \sigma_{\mu}^2$,

\noindent
то есть все равно остается неопределенность в нахождении параметра $\mu_0$. Итак, 
в пределе большого числа измерений систематическая неопределенность доминирует и 
не зависит от числа измерений. 

Для задачи извлечения сигнального сечения по экспериментальным данным на основе использования распределения 
Пуассона $P(n|\lambda)$ в случае $n_{obs}\gg 1$ распределение Пуассона аппроксимируется 
нормальным распределением $N(n|\lambda,\lambda)$. Для модели с 
$\lambda = \lambda_b+\lambda_s,~~\lambda_b=L\epsilon_b\sigma_b,~~
\lambda_s=L\epsilon_s\sigma_s$ независимые неопределенности 
$L,~\epsilon_b,~\epsilon_s,~\sigma_b,~\sigma_s$ удобно представить в виде  
$\displaystyle L=L_0(1\pm\delta_{L}),~
\epsilon_b=\epsilon_b^0(1\pm\delta_{\epsilon_b}),~
\epsilon_s=\epsilon_s^0(1\pm\delta_{\epsilon_s}),~
\sigma_b=\sigma_b^0(1\pm\delta_{\sigma_b}),~
\sigma_s=\sigma_s^0(1\pm\delta_{\sigma_s}).$

\noindent
В случае малых неопределенностей $\delta_{L},~\delta_{\epsilon_b},~\delta_{\epsilon_s}$  
уравнение для определения сигнального сечения $\sigma_s^0$ 
можно записать в виде 

\begin{equation}
\displaystyle \sigma_{s}^0 = 
\frac{n_{obs}-L_0\epsilon_{b}^0\sigma_{b}^0}{L_0\epsilon_{s}^0}\pm x,
\label{eq:164}
\end{equation}

\noindent
причем ошибка измерения сигнального сечения $\sigma_s^0$ равна 

\begin{equation}
\displaystyle x = 
\sqrt{(\frac{\epsilon_{b}^0\sigma_{b}^0}{\epsilon_{s}^0})^2
(\delta_{\epsilon_{s}}^2+\delta_{\epsilon_{b}}^2+\delta_{\sigma_{b}}^2)+
\frac{n_{obs}}{(L_0\epsilon_{s}^0)^2}(1+n_{obs}(\delta_{L}^2 + \delta_{\epsilon_s}))}.
\label{eq:165}
\end{equation}

Таким образом, в приближении нормального распределения комбинированный учет различных 
систематических ошибок сводится к квадрированию ошибок и тривиален. Проблема немного усложняется,  
когда количество $n_{obs}$ невелико и мы не можем использовать приближение нормального 
распределения.

Пусть, например, у нас имеется распределение Пуассона $P(n|\lambda_b+\lambda_s)$,
а параметры $\lambda_b$ и $\lambda_s$ распределены по нормальному  закону

\begin{equation}
\displaystyle N(\lambda_b|\lambda_b^0,\sigma_b) = 
\frac{1}{\sqrt{2\pi}\sigma_b}e^{-\frac{(\lambda_b-\lambda_b^0)^2}{2\sigma_b^2}},
\label{eq:166}
\end{equation}

\begin{equation}
\displaystyle N(\lambda_s|\lambda_s^0,\sigma_s) = 
\frac{1}{\sqrt{2\pi}\sigma_s}e^{-\frac{(\lambda_s-\lambda_s^0)^2}{2\sigma_s^2}}.
\label{eq:167}
\end{equation}

Согласно методу Кузинса-Хайланда мы усредняем распределение Пуассона с функциями 
$N(\lambda_b|\lambda_b^0,\sigma_b)$ и $N(\lambda_s|\lambda_s^0,\sigma_s)$

\begin{equation}
\displaystyle P_{av}(n|\lambda_b^0,\lambda_s^0) = 
\int{d\lambda_b d\lambda_s P(n|\lambda_b+\lambda_s)N(\lambda_b|\lambda_b^0,\sigma_b) 
N(\lambda_s|\lambda_s^0,\sigma_s)}.
\label{eq:168}
\end{equation}

В пределе малых дисперсий $\sigma_b$ и $\sigma_s$ 

\begin{equation}
\displaystyle lim_{{\sigma_b\rightarrow 0}\atop{\sigma_s\rightarrow 0}} 
{P_{av}(n|\lambda_b^0,\lambda_s^0,\sigma_b,\sigma_s) =  
P(n|\lambda_b^0,\lambda_s^0)}.
\label{eq:169}
\end{equation}

Разумеется, очень важный вопрос, как распределения или приоры $P(\lambda_b|\lambda_b^0,\sigma_b)$, 
$P(\lambda_s|\lambda_s^0,\sigma_s)$ выбирать для учета систематики. При маленьких дисперсиях 
$\sigma_b$, $\sigma_s$ это несущественно. При больших дисперсиях результат усреднения 
начинает зависеть от формы распределения. Кроме того, из физических соображений 
$\lambda_b\ge 0$ и $\lambda_s\ge 0$, в то время как нормальное распределение 
приводит к возможным значениям $\lambda_b$ и $\lambda_s$ от 
$-\infty$ до $+\infty$. При небольших дисперсиях эта проблема решается волевым 
путем -- обнулением плотностей вероятностей при отрицательных $\lambda_b$ и $\lambda_s$ 
и изменением нормировки с тем, чтобы полный интеграл по $\lambda_b$, $\lambda_s$ от 
$0$ до $\infty$ равнялся бы $1$. Часто используется логнормальное по 
${\lambda_b}$, ${\lambda_s}$ распределение, для которого переменные 
$ln~{\lambda_b}$, $ln~{\lambda_s}$ автоматически лежат от $-\infty$ до $+\infty$. 

В работе~\cite{Highland} изучалось влияние систематики на извлечение верхнего ограничения на 
параметр $\lambda$ распределения Пуассона $P(n,\lambda)$. В предположении, что 
систематическая 
неопределенность в параметре $\lambda$ распределена по  нормальному   закону  с дисперсией 
$\sigma_{\lambda}$, при малых $\sigma_{\lambda}$ учет систематики приводит к 
ограничению $\lambda\le \lambda_{up}$, где 
$\displaystyle \lambda_{up}= 
\lambda_{up}^0(1+\frac{(\lambda_{up}^0-n_{obs})\sigma_{\lambda}^2}{2})$ 
и $\lambda_{up}^0$ -- верхний предел без учета систематики. При $n_{obs}=0$ и 
90\% уровне достоверности $\displaystyle \lambda_{up}(n_{obs}=0,\sigma_{\lambda}) 
= 2.30(1+2.30\frac{\sigma_{\lambda}^2}{2})$.

Как уже отмечалось выше параметр $\lambda$ в распределении Пуассона $P(n,\lambda)$ 
представим 
в виде $\lambda = L\epsilon \sigma$ и мы должны учитывать 
систематические ошибки, 
связанные с неточным знанием светимости $L$, эффективности регистрации $\epsilon$ 
и сечения $\sigma$. 
Предположим, что ошибки распределены по нормальному закону и они маленькие. Мы можем символически 
записать 
$L=L_0(1\pm\delta_L)$, $\epsilon=\epsilon_0(1\pm\delta_{\epsilon})$, 
$\sigma=\sigma_0(1\pm\delta_{\sigma})$, где 
$\delta_L,~\delta_{\epsilon},~\delta_{\sigma} \ll 1$ -- стандартные отклонения. 
Предполагая, что ошибки $\delta_L,~\delta_{\epsilon},~\delta_{\sigma}$ независимы, 
получаем, что неопределенность параметра $\lambda$ распределена по нормальному закону с дисперсией 
$\delta_{\lambda}=\displaystyle 
\sqrt{\delta_L^2+\delta_{\epsilon}^2+\delta_{\sigma}^2}$, 
то есть $\lambda = \lambda_0(1\pm\delta_{\lambda})$, где 
$\lambda_0=L_0\epsilon_0\sigma_0$.

\setlength{\tabcolsep}{4pt}
\begin{table}
\begin{center}
\caption{Верхние пределы $\lambda_{up}$ при 90\% уровне доверия для значений 
$n_{obs}=~0,~1,~2,~3$ и значений $\delta_{\lambda}=0,~0.1,~0.2$.}
\begin{tabular}{|l|c|c|c|c|}
\hline
$\delta_{\lambda}$ $\backslash$ $n_{obs}$ &    0 &    1 &    2 &    3  \\
\hline
        ~0.0                           & 2.30 & 3.89 & 5.32 & 6.68  \\
        ~0.1                           & 2.33 & 3.95 & 5.41 & 6.80  \\
        ~0.2                           & 2.41 & 4.11 & 5.67 & 7.17  \\
\hline
\end{tabular}
\label{tab:5}
\end{center}
\end{table}

\newpage

В таблице (\ref{tab:5}) показано влияние систематики на значение верхнего 
предела 
на параметр распределения $\lambda$ с помощью частотного метода~\cite{Highland}  
при неопределенности параметра $\lambda$ в 0, 10 и 20\% при $n_{obs}\le 3$.

\newpage

\section{Проверка гипотез}

\subsection{Введение}

Одной из основных задач любого эксперимента является проверка той или иной модели (теории), 
предназначенной для поиска экспериментальных подтверждений, а также поиск новых явлений. В 
настоящее время вся совокупность экспериментальных данных в физике высоких энергий описывается 
Стандартной Моделью~\cite{SM}~\footnote{Единственный экспериментальный факт, лежащий 
вне предсказаний Стандартной Модели -- осцилляции нейтрино~\cite{Neutrino}. Заметим,  
что нейтринные осцилляции легко объясняются путем включения масс нейтрино в 
Стандартную Модель.}. Стандартная Модель~(СМ) предсказывает существование бозона Хиггса -- 
последней неоткрытой частицы СМ. Существуют многочисленные расширения СМ (суперсимметрия, 
дополнительные размерности, $Z'$-бозоны, \dots), предсказывающие существование новой физики.
Основной задачей БАКа является поиск бозона Хиггса и новой физики вне рамок СМ. Это делается 
путем детектирования частиц в конечном состоянии (адронных струй, электронов, мюонов, 
$\tau$-лептонов) с определенными ограничениями на импульсы частиц в конечном состоянии. 
В этом случае у нас есть две гипотезы. 

\begin{enumerate}
\item Гипотеза $H_0$ -- все описывается СМ.
\item Гипотеза $H_1$ -- существует новая физика вне рамок СМ.
\end{enumerate}

{\hbox{

\put(25,0){\line(1,0){100}}
\put(25,0){\line(0,1){50}}
\put(25,50){\line(1,0){100}}
\put(125,0){\line(0,1){50}}

\put(40,30){Гипотеза}
\put(50,10){$H_0$}

}}

{\hbox{

\put(225,0){\line(1,0){100}}
\put(225,0){\line(0,1){50}}
\put(225,50){\line(1,0){100}}
\put(325,0){\line(0,1){50}}

\put(255,20){Данные}

\put(135,48){\line(4,-1){80}}
\put(133,44){\line(4,-1){80}}

\put(131,46){\line(2,1){12}}
\put(131,46){\line(1,-1){10}}

\put(135,2){\line(4,1){80}}
\put(133,6){\line(4,1){80}}

\put(131,4){\line(2,-1){12}}
\put(131,4){\line(1,1){10}}

}}

{\hbox{

\put(25,0){\line(1,0){100}}
\put(25,0){\line(0,1){50}}
\put(25,50){\line(1,0){100}}
\put(125,0){\line(0,1){50}}

\put(40,30){Гипотеза}
\put(50,10){$H_1$}

}}

Гипотеза $H_0$ называется простой гипотезой, если она не имеет свободных параметров. 
Это хорошая идеализация для СМ без учета вкладов от распадов бозона Хиггса, 
поскольку точное значение массы бозона Хиггса не известно. Следует подчеркнуть, 
что в реальной жизни существуют неопределенности, связанные, в частности, с 
неумением точно предсказывать сечения СМ, а также экспериментальные неопределенности, 
связанные с неточным знанием полной светимости, неточным знанием импульсов частиц и 
эффективности их регистрации. Поэтому понятие простой гипотезы является всего лишь 
разумной идеализацией. Предположим, что в качестве альтернативы гипотезе $H_0$ 
мы имеем простую гипотезу $H_1$. Например, гипотезу о существовании суперсимметрии~\cite{SSM} 
с вполне определенными массами суперчастиц. Заметим, что гипотеза 
о существовании новой физики является как правило составной, то есть зависящей от 
неопределенных параметров (масс суперчастиц, параметров дополнительных размерностей, 
масс $Z'$-бозона и т.д.). Поэтому у нас на первом этапе стоит задача проверки 
основной гипотезы $H_0$ -- справедливости СМ.

\subsection{Проверка основной гипотезы}

Предположим мы проверяем СМ в процессах рождения лептонных пар на БАКе, накладывая 
определенные ограничения на импульсы лептонов и их инвариантные массы. СМ при 
заданной светимости предсказывает вполне определенное среднее количество событий 
$\lambda_b$. 

Вероятность наблюдения $n$ событий описывается распределением Пуассона $P(n|\lambda)$. 
Предположим, что мы обнаружили $n_{obs}$ событий и наша задача -- определить насколько 
наблюдаемое количество событий $n_{obs}$ совместимо с нашей основной гипотезой, 
что $\lambda=\lambda_b$ в распределении Пуассона.

Вероятность наблюдения $n\ge n_{obs}$ событий равна~\footnote{Вследствие формулы~(\ref{eq:96a}) 
справедливо равенство 

$P_+(n_{obs}|\lambda_b)=\displaystyle \int_0^{\lambda_b}{P(n_{obs}-1|\lambda')d\lambda'}$, 

\noindent
позволяющее связать частотное определение $p$-вероятности с Байесовским определением вероятности 
параметра $\lambda\le\lambda_b$ для приора $\pi(\lambda)=\displaystyle \frac{const}{\lambda}$.}

\begin{equation}
P_+(n_{obs}|\lambda_b)\equiv P(n\ge n_{obs}|\lambda_b) = 
\displaystyle \sum_{n=n_{obs}}^{\infty}{P(n|\lambda_b)}.
\label{eq:170}
\end{equation}

\noindent
Вероятность наблюдения $n\le n_{obs}$ событий равна 

\begin{equation}
P_-(n_{obs}|\lambda_b)\equiv P(n\le n_{obs}|\lambda_b) = \sum_{n=0}^{n=n_{obs}}
P(n|\lambda_b).
\label{eq:171}
\end{equation}

\noindent
Напомним, что $P(n_{obs}|\lambda_b)$ максимальна~\cite{MaxPnobs} при 
$n_{obs}=[\lambda]$\footnote{Здесь квадратные скобки $[\dots]$ означают 
целую часть от числа.}.
Поэтому, если наблюденное количество событий $n_{obs}$ сильно отличается от $[\lambda_b]$ 
в ту или иную сторону, иными словами либо $P_+$, либо $P_-$ очень малы, то можно 
говорить о том, что СМ с ее предсказанием параметра $\lambda_b$ распределения Пуассона 
не верна, то есть гипотеза $H_0$ противоречит экспериментальным данным. Поскольку в поисках 
новой физики, в основном, предсказывается избыток событий по сравнению с предсказаниями 
СМ, то есть $n_{obs}>\lambda_b$, то мы рассмотрим ситуацию, когда $P_+$ мало.
Обычно значение вероятности $P$ связывается со стандартной значимостью 
$S$ согласно формуле 

\begin{equation}
P = \displaystyle \frac{1}{\sqrt{2\pi}}
\int_{S}^{\infty}{e^{-\frac{x^2}{2}}dx}=
\frac{1}{2}[1-erf(\frac{S}{\sqrt{2}})].
\label{eq:172}
\end{equation}

\noindent
Причем в физике высоких энергий общепринято считать, что при значимости $S\ge 5$ 
мы можем объявлять об открытии нового явления (новой физики вне рамок СМ). 
Значимость $S=5$ соответствует вероятности $P=2.9\cdot 10^{-7}$ того, что явление 
описывается СМ~\footnote{Обычно $S\ge 3$ считается указанием в пользу новой физики.}.  
В Таблице~\ref{tab:6} приведены несколько значений значимостей $S$ и соответствующих 
им вероятностей $P_+$ (см. также Таблицу~\ref{tab:1}).

\noindent 
В  Таблице~\ref{tab:6a} приведены значения $\lambda_b$ в зависимости от $n_{obs}$, соответствующие 
уровню открытия $S=5$.

\setlength{\tabcolsep}{4pt}
\begin{table}
\begin{center}
\caption{Значения значимостей $S$ и соответствующие им значения вероятностей $P_+$.}
\begin{tabular}{|c|c|c|c|c|c|c|}
\hline
$S$   & 0 & 1 & 2 & 3 & 4 & 5  \\ 
\hline
$P_+$ & 0.5 & $0.1587$ & $2.275\cdot 10^{-2}$ & $1.35\cdot 10^{-3}$ &  
$3.15\cdot 10^{-5}$ & $2.87\cdot 10^{-7}$  \\
\hline
\end{tabular}
\label{tab:6}
\end{center}
\end{table}

\setlength{\tabcolsep}{4pt}
\begin{table}
\begin{center}
\caption{Значения $\lambda_b$ для наблюденного числа $n_{obs}$, соответствующие уровню открытия $S=5$.}
\begin{tabular}{|c|c|c|c|c|c|c|c|c|c|c|}
\hline
$n_{obs}$   & 1 & 2 & 3 & 4 & 5 & 6 & 7 & 8 & 9 & 10  \\ 
\hline
$\lambda_b$ & $0.00000028$ & $0.00075$ & $0.012$ & $0.05$ & $0.13$ & $0.25$ & $0.41$ & 
$0.61$ & $0.84$ & $1.11$ \\
\hline
\end{tabular}
\label{tab:6a}
\end{center}
\end{table}

\noindent
Заметим, что в реальной жизни мы знаем параметр СМ $\lambda_b$ с какой-то точностью, 
что связано с нашим неумением точно вычислить параметр $\lambda_b$, точно измерять импульсы частиц 
и их идентифицировать. Как было показано в предыдущей главе, неопределенности можно учесть с помощью 
функции распределения $\displaystyle P(\lambda_b|\lambda_b^0, \sigma)$, 
которая обычно выбирается в виде нормального распределения 

\begin{equation}
P(\lambda_b|\lambda_b^0,\sigma) = \displaystyle 
\frac{1}{\sqrt{2\pi}\sigma}e^{-\frac{(\lambda_b-\lambda_b^0)^2}{2\sigma^2}}.
\label{eq:173}
\end{equation}
 
\noindent
Учет неопределенности, связанной с неточным значением $\lambda_b$ производится путем усреднения 
в правой части формул~(\ref{eq:170},\ref{eq:171}) с $P(\lambda_b|\lambda_b^0,\sigma)$, 
а именно, обобщение  соответствующих  формул имеет вид

\begin{equation}
P_+'(n_{obs}|\lambda_b^0,\sigma)= 
\int{P_+(n_{obs}|\lambda_b,\sigma)P(\lambda_b|\lambda_b^0,\sigma)d\lambda_b},
\label{eq:174}
\end{equation}

\begin{equation}
P_-'(n_{obs}|\lambda_b^0,\sigma)= 
\int{P_-(n_{obs}|\lambda_b,\sigma)P(\lambda_b|\lambda_b^0,\sigma)d\lambda_b}.
\label{eq:175}
\end{equation}

Заметим, что в пределе $\sigma\rightarrow 0$~~$P_+'\rightarrow P_+$, 
$P_-'\rightarrow P_-$. 
При измерении случайной величины $X$, распределенной по нормальному закону 

\begin{equation}
P(x|\mu,\sigma)= \displaystyle 
\frac{1}{\sqrt{2\pi}\sigma}e^{-\frac{(x-\mu)^2}{2\sigma^2}}
\label{eq:176}
\end{equation}

\noindent
вероятность 

\begin{equation}
P(x>\mu+S\sigma) = \displaystyle 
\int_{\mu+S\sigma}^{\infty}{P(x|\mu,\sigma)dx} = 
\frac{1}{\sqrt{2\pi}}\int_S^{\infty}{e^{-\frac{x^2}{2}}dx}
\label{eq:177}
\end{equation}

\noindent
и $S=5$ как раз соответствует порогу открытия.

В общем случае, когда мы собираем данные $X$ для проверки гипотезы $H_0$, соответствующей 
распределению $f(x|\theta)$ задача состоит в нахождении тестовой статистики $T(x)$ такой, 
что большие значения $t_{obs}\equiv T(x_{obs})$ являются свидетельством (доказательством) 
против гипотезы $H_0$. С этой целью вводится  вероятность наблюдения $T\ge t_{obs}$ 

\begin{equation}
P = P(T\ge t_{obs}|H_0).
\label{eq:178}
\end{equation}

Как уже отмечалось выше в физике высоких энергий гипотеза $H_0$ отвергается, если 
$P\le 2.9\cdot 10^{-7}$. 

В методе максимального правдоподобия аналог значимости $S$ определяется 
формулой 

\begin{equation}
S= \displaystyle \sqrt{2ln{\frac{f(x_{obs}|\theta_{max})}{f(x_{obs}|\theta)}}}.
\label{eq:179}
\end{equation}

В случае нормального распределения $N(x|\mu,\sigma^2)$ 
формула~(\ref{eq:179}) совпадает со стандартной формулой 


\begin{equation}
S = \displaystyle \frac{|x_{obs}-\mu|}{\sigma}.
\label{eq:180}
\end{equation}

В  Таблице~\ref{tab:7} приведены значения $S$ для нескольких значений $\lambda_b$ и $n_{obs}$. 
Для распределения Пуассона при малых $n_{obs}$ различие между значением 
значимости~(\ref{eq:172}) и значимости~(\ref{eq:180}) может быть весьма существенным
(Таблица~\ref{tab:7}). 
 
\setlength{\tabcolsep}{4pt}
\begin{table}
\begin{center}
\caption{Значения значимостей $S$ для распределения Пуассона, вычисленные по 
формулам~(\ref{eq:170},\ref{eq:172}) для различных значений  
$n_{obs}$ и $\lambda_b$. В скобках приведено значение значимости, вычисленное 
по формуле~(\ref{eq:180}).}
\begin{tabular}{|c|c|c|c|c|c|}
\hline
$n_{obs}$ & $\lambda_b=1$ &  $\lambda_b=2$ &  $\lambda_b=3$ &  
  $\lambda_b=4$ &  $\lambda_b=5$ \\ 
\hline
    $1$ &  $-0.34$ (0)&  $-1.10$ (0.71)& $-1.65$ (1.16)& $-2.09$ (1.5)& $-2.47$ (1.79)\\
    $2$ &  $ 0.63$ (1)&  $-0.24$ (0)   & $-0.85$ (0.58)& $-1.33$ (1.0)& $-1.75$ (1.34)\\
    $3$ &  $ 1.40$ (2)&  $ 0.46$ (0.71)& $-0.19$ (0)   & $-0.71$ (0.5)& $-1.15$ (0.89)\\
    $4$ &  $ 2.08$ (3)&  $ 1.07$ (1.41)& $ 0.38$ (0.58)& $-0.17$ (0)  & $-0.63$ (0.45)\\
    $5$ &  $ 2.68$ (4)&  $ 1.62$ (2.21)& $ 0.90$ (1.16)& $ 0.33$ (0.5)& $-0.15$ (0)\\
    $6$ &  $ 3.24$ (5)&  $ 2.13$ (2.83)& $ 1.38$ (1.73)& $ 0.79$ (1.0)& $ 0.30$ (0.45)\\
    $7$ &  $ 3.77$ (6)&  $ 2.61$ (3.54)& $ 1.83$ (2.31)& $ 1.22$ (1.5)& $ 0.71$ (0.89)\\
    $8$ &  $ 4.26$ (7)&  $ 3.06$ (4.24)& $ 2.26$ (2.89)& $ 1.63$ (2.0)& $ 1.11$ (1.34)\\
    $9$ &  $ 4.73$ (8)&  $ 3.91$ (4.95)& $ 3.06$ (3.46)& $ 2.03$ (2.5)& $ 1.49$ (1.79)\\
   $10$ &  $ 5.18$ (9)&  $ 4.31$ (5.66)& $ 3.44$ (4.04)& $ 2.40$ (3.0)& $ 1.86$ (2.23)\\
\hline
\end{tabular}
\label{tab:7}
\end{center}
\end{table}

В Байесовском подходе вероятность того, что параметр $\lambda$ распределения Пуассона 
меньше или равен $\lambda_b$ определяется формулой 

$P=\displaystyle \int_0^{\lambda_b}{P(\lambda|n_{obs})d\lambda}$.

\noindent
Малость параметра $P$ означает, что основная гипотеза не верна. 

\subsection{Проверка сложной гипотезы}

Предположим мы хотим решить, какая из гипотез $H_0$ (СМ физика) или 
$H_1$ (физика вне рамок СМ) правильная на основе измерения величины $X$.
Пусть $T(x)$ некоторая функция от наблюдений, которую будем называть тестовой статистикой 
и пусть $W$  будет пространством всех значений $T(x)$. Пространство $W$ обычно 
разделяют на критическую область $\omega$, когда гипотеза $H_0$ отвергается, и 
область $W-\omega$, в которой гипотеза $H_0$ не отвергается.  

Вероятность попадания в область $\omega$, когда гипотеза $H_0$ верна 

\begin{equation}
P(T\in\omega|H_0)=\alpha 
\label{eq:181}
\end{equation}

\noindent
называется ошибкой первого рода. Иными словами $\alpha$ это вероятность того, 
что гипотеза $H_0$ отвергается, когда она верна. 

Важно, чтобы выбранная нами тестовая статистика $T$ позволяла бы различить гипотезы 
$H_0$ и $H_1$. Степень способности статистики $T$ различить гипотезы $H_0$ и 
$H_1$ определяется мощностью теста, определяемого как вероятность выполнения гипотезы $H_1$ 
при условии, что данные лежат в критической области 

\begin{equation}
P(T\in\omega|H_1)=1-\beta. 
\label{eq:182}
\end{equation}

\noindent
Другими словами $\beta$ это вероятность того, что $T$ входит в допустимую область $W-\omega$ 
для гипотезы $H_0$

\begin{equation}
P(T\in W-\omega|H_1)=\beta. 
\label{eq:183}
\end{equation}

Итак, при тестировании двух гипотез вводятся два типа ошибок: 

\begin{enumerate}
\item Ошибка первого рода $\alpha$ -- вероятность отвергнуть основную (нулевую) гипотезу $H_0$, 
когда она верна. 
\item Ошибка второго рода $\beta$ -- вероятность принятия основной (нулевой) гипотезы $H_0$, 
когда она не верна. 
\end{enumerate}

Предположим гипотезы $H_0$ и $H_1$ характеризуются параметрами $\theta_0$ и $\theta_1$ и, 
соответственно, функцией распределения $f(x,\theta)$. 

Ранее была определена мощность теста $P(\theta_1)$, позволяющая различать гипотезы 
$\theta=\theta_0$ и $\theta=\theta_1$, а именно, $P(\theta_1)=1-\beta_1$.

В общем случае составной гипотезы (то есть гипотезы, зависящей от параметра) мощность 
теста определяют как $P(\theta)=1-\beta(\theta)$ для альтернативной гипотезы, определяемой 
параметром $\theta$.

Различные тесты можно сравнивать на основе сравнения функции мощности теста $P(\theta)$.

В случае, если альтернативная гипотеза $H_1$-простая (все параметры распределения заданы) 
тест является максимальным, если $P(\theta_1)$ максимален. Тест, который является наиболее мощным 
при всех значениях $\theta$ называется однородно наиболее мощным. 

Тест называется состоятельным, если при наборе данных вероятность ошибки второго рода стремится к 
нулю, то есть 

$\displaystyle lim_{N\rightarrow \infty}{P(T(x)\in\omega|H_1)}=1$ 

Для распределения Пуассона с гипотезой $H_0:~\lambda=\lambda_b$ и 
альтернативной гипотезой $H_1:~\lambda=\lambda_b+\lambda_s$ в частотном подходе 
параметры $\alpha$ и $1-\beta$ равны соответственно 

$\alpha=\displaystyle \sum_{n=n_{obs}}^{\infty}{P(n|\lambda_b)}$,

$1-\beta=\displaystyle \sum_{n=n_{obs}}^{\infty}{P(n|\lambda_b+\lambda_s)}$.

\noindent
Предположим, что параметры $\lambda_b$ и $\lambda_s$ известны точно. Тогда в случае 
$\alpha\le 2.9\cdot 10^{-7}$ как это общепринято в физике высоких энергий -- 
гипотеза $H_0$ отвергается (СМ физика не в состоянии описать экспериментальные данные). 
В случае, когда $\beta$ меньше определенного значения (например, $\beta \le 0.95$) 
можно сказать, что на 95\% уровне достоверности гипотеза $H_1$ не противоречит 
экспериментальным данным.

\subsection{Тест Неймана-Пирсона}

Проблему нахождения наиболее мощного теста гипотезы $H_0$ относительно гипотезы $H_1$ можно 
сформулировать как нахождение наилучшей критической подобласти в пространстве $x$ возможных 
значений наблюдений. Из определений~(\ref{eq:181},\ref{eq:182},\ref{eq:183}) следует, что 

\begin{equation}
\displaystyle \int_{\omega_{\alpha}}{f(\vec x|\theta_0)d{\vec x}}=\alpha, 
\label{eq:184}
\end{equation}

\begin{equation}
\displaystyle \int_{\omega_{\alpha}}{f(\vec x|\theta_1)d{\vec x}}=1-\beta. 
\label{eq:185}
\end{equation}

\noindent
Уравнения~(\ref{eq:184},\ref{eq:185}) можно представить в виде 

\begin{equation}
1-\beta = \displaystyle 
E_{\omega_{\alpha}}(\frac{f(\vec x|\theta_1)}{f(\vec x|\theta_0)}|_{\theta=\theta_0}). 
\label{eq:186}
\end{equation}

Отсюда следует, что $1-\beta$ будет максимально, если $\omega_{\alpha}$ содержит долю 
$x$-данных таких $x$, что отношение $\displaystyle 
\frac{f(\vec x |\theta_1)}{f(\vec x|\theta_0)}$ максимально. Поэтому наилучшая 
критическая область $\omega_{\alpha}$ состоит из точек, удовлетворяющих неравенству 

\begin{equation}
\displaystyle \frac{f(\vec x,\theta_1)}{f(\vec x,\theta_0)}>c_{\alpha}, 
\label{eq:187}
\end{equation}

\noindent 
причем $c_{\alpha}$ выбирается таким образом, чтобы соотношение~(\ref{eq:184}) выполнялось. 

Итак, если выполняется соотношение~(\ref{eq:187}), выбираем гипотезу $H_1$, в противном случае 
выбираем гипотезу $H_0$. Эта процедура известна как тест Неймана-Пирсона. При этом тестовая 
статистика это отношение функций правдоподобия гипотез $H_1$ и $H_0$. 

В случае проверки двух сложных гипотез $H_0$ и $H_1$, функции распределения которых зависят 
соответственно от параметров $\vec \theta_0$ и $\vec \theta_1$, можно определить 
отношение максимального правдоподобия как 

\begin{equation}
L = \displaystyle 
\frac{max_{\vec \theta_0}L(\vec x, \vec \theta_0)}{max_{\vec \theta_1}L(\vec x, \vec \theta_1)} 
\label{eq:188}
\end{equation}

\noindent
и это отношение позволяет разделять две сложные гипотезы. 

Для случая нормального распределения $N(x|\mu,\sigma^2)$, когда гипотеза $H_0$ 
определяется как $\mu=\mu_b,~\sigma=\sigma_0$, а гипотеза $H_1$ соответствует 
$\mu=\mu_b+\mu_s,~\sigma=\sigma_0$, отношение функций правдоподобия для измеренного значения 
$x=x_{obs}$ есть \\

$\displaystyle 
\frac{N(x_{obs}|\mu_b+\mu_s,\sigma_0)}{N(x_{obs}|\mu_b,\sigma_0)}=
e^{-\frac{1}{2\sigma_0^2}[(x_{obs}-\mu_b-\mu_s)^2-(x_{obs}-\mu_b)^2]}$.

Подчеркнем, что в физике высоких энергий гипотеза $H_1$ как правило составная (параметр 
$\lambda$ зависит от неизвестных масс частиц и параметров взаимодействий) и нам важно 
продемонстрировать на первом этапе исследования, что гипотеза $H_1$ обладает допустимыми 
значениями параметров, при которых она не противоречит экспериментальным данным, что, вообще говоря, 
нетривиально. 

Так, например, если $\mu_s\le\sigma$, а $x_{obs}=\mu_b+5\sigma_0$, то 
основная гипотеза $\mu_s=0$ 
исключена на уровне $5\sigma~(\alpha=2.9\cdot 10^{-7})$, а альтернативная 
составная гипотеза исключена (или имеет согласие с экспериментом на уровне 
$4\sigma~(\alpha=3.2\cdot 10^{-5})$), что также позволяет по сути дела исключить альтернативную гипотезу 
$H_1$. В случае, когда $\mu_s\le 1\sigma$, альтернативная гипотеза $H_1$
имеет согласие с экспериментом на уровне $1\sigma~$($\alpha=0.16$), что 
позволяет рассматривать ее как альтернативную замену основной гипотезе $H_0$.

\subsection{Байесовский подход к проверке гипотез}

Рассмотрим для начала проверку простой гипотезы $H_0$ на примере распределения Пуассона 
$P(n|\lambda)$.  

В силу формулы Байеса  
 
\begin{equation}
P(\lambda|n_{obs})=\pi(\lambda)P(n_{obs}|\lambda), 
\label{eq:189}
\end{equation}

\noindent
вероятность того, что $\lambda\ge \lambda_b$


\begin{equation}
P(\lambda\ge\lambda_b)=
\displaystyle \int_{\lambda_b}^{\infty}{P(\lambda|n_{obs})d\lambda}. 
\label{eq:190}
\end{equation}

\noindent
При плоском приоре $\pi(\lambda)=1$ в силу соотношений~(\ref{eq:93eq},\ref{eq:96}) 
Байесовское определение совпадает с частотным определением~(\ref{eq:86}). 
Таким образом, в Байесовском подходе в случае известного приора $\pi(\lambda)$ 
нам необходимо вычислить интеграл~(\ref{eq:190}). Нулевая гипотеза $H_0$ 
(параметр $\lambda$ в распределении Пуассона равен $\lambda_b$) отвергается, 
если $P(\lambda\ge\lambda_b)\le\alpha$. Здесь мы неявно предполагаем, что реализуется 
ситуация с наблюдением избытка событий. В общем случае нулевая гипотеза 
$H_0:~~\lambda=\lambda_b~~$ отвергается, если $P(\lambda\ge\lambda_b)\le\alpha$, 
либо  $P(\lambda\le\lambda_b)\le\alpha$.

Напомним, что в физике высоких энергий 
общепринято полагать  $\alpha=2.9\cdot 10^{-7}$.

В случае, когда мы должны сделать выбор между простыми гипотезами $H_0$ и $H_1$, 
теорема Байеса приводит к следующим формулам для вероятности: 

\begin{equation}
\pi(H_0|x) =\displaystyle \frac{P(x|H_0)\pi_0}{P(x|H_0)\pi_0+P(x|H_1)\pi_1}, 
\label{eq:191}
\end{equation}

\begin{equation}
\pi(H_1|x) =1 -\pi(H_0|x), 
\label{eq:192}
\end{equation}

\noindent
где $\pi_i$ -- априорные вероятности для гипотез $H_i,~i=1,2$, 
$P(x|H_0),~P(x|H_1)$-плотности вероятности события $X$ при гипотезах $H_0$ и $H_1$ 
соответственно.
 
Если $\pi(H_0|x)<\pi(H_1|x)$, то гипотеза $H_1$ более предпочтительна. 
В случае $\pi(H_0|x)>\pi(H_1|x)$ гипотеза $H_0$ более предпочтительна. 
Если же $\pi(H_0|x)=\pi(H_1|x)$, то невозможно сделать выбор между гипотезами $H_0$ и $H_1$. 
В случае $\displaystyle \pi_0=\pi_1=\frac{1}{2}$ отношение 

\begin{equation}
\displaystyle \frac{\pi(H_0|x)}{\pi(H_1|x)}=\frac{P(x|H_0)}{P(x|H_1)}. 
\label{eq:193}
\end{equation}

\noindent
Отношение~(\ref{eq:193}) есть не что иное как отношение функций правдоподобия.

Часто гипотезы $H_0$ и $H_1$ зависят от неизвестных параметров $\theta$ 
(маргинальных параметров систематики). В этом случае обобщение 
формул~(\ref{eq:191},\ref{eq:192}) имеют вид 

\begin{equation}
\pi(H_0|x) =\displaystyle \frac{\int{P(x|\theta,H_0)\pi(\theta|H_0)\pi_0d\theta}}
{\int{[P(x|\theta,H_0)\pi(\theta|H_0)\pi_0+P(x|\theta,H_1)\pi(\theta|H_1)\pi_1]d\theta}}, 
\label{eq:194}
\end{equation}

\begin{equation}
\pi(H_1|x) =1 -\pi(H_0|x). 
\label{eq:195}
\end{equation}

Для $\pi_0=\pi_1=\displaystyle \frac{1}{2}$ отношение вероятностей равно 

\begin{equation}
\displaystyle \frac{\pi(H_0|x)}{\pi(H_1|x)} =
\frac{\int{P(x|\theta,H_0)\pi(\theta|H_0)\pi_0d\theta}}
{\int{P(x|\theta,H_1)\pi(\theta|H_1)\pi_1d\theta}}. 
\label{eq:196}
\end{equation}

\newpage

\section{Комбинирование результатов}

В случае измерения нескольких независимых величин стоит задача о комбинировании 
этих величин, то есть извлечения дополнительной информации по сравнению с единичными 
измерениями. Существует несколько способов это сделать.

\subsection{Комбинирование двух нормальных распределений}

В случае измерения двух независимых случайных величин $X_1$ и $X_2$, распределенных по 
нормальному закону $X_1 \sim N(\mu_1,\sigma)$ и $X_2 \sim N(\mu_2,\sigma)$ с 
одной и той же дисперсией $\sigma$ и неизвестными параметрами $\mu_1$ и $\mu_2$ 
полезно перейти к новым координатам

\begin{equation}
Y_1 = \displaystyle \frac{X_1\mu_1+X_2\mu_2}{\sqrt{\mu_1^2+\mu_2^2}}, 
\label{eq:237}
\end{equation}

\begin{equation}
Y_2 = \displaystyle \frac{-X_1\mu_2+X_2\mu_1}{\sqrt{\mu_1^2+\mu_2^2}}. 
\label{eq:238}
\end{equation}

\noindent
Нетрудно показать, что для $X1\cdot X2 \sim N(\mu_1,\sigma)\cdot N(\mu_2,\sigma)$  
и $Y1\cdot Y2 \sim N(\sqrt{\mu_1^2+\mu_2^2},\sigma)\cdot N(0,\sigma)$ 
справедливо 

\begin{equation} 
N(\mu_1,\sigma)\cdot N(\mu_2,\sigma) = 
N(\sqrt{\mu_1^2+\mu_2^2},\sigma)\cdot N(0,\sigma). 
\label{eq:239}
\end{equation}

Для модели с $\mu_1=\mu_{b1}+\mu_s$, $\mu_2=\mu_{b2}+\tau\mu_s$, 
где $\tau$, $\mu_{b1}$ и $\mu_{b2}$ -- заданные величины, а $\mu_s$ неизвестно,  
необходимо комбинировать $x_1'=x_{1}-\mu_{b1}$, $x_2'=x_2-\mu_{b2}$. Тогда 

\begin{equation}
Y_1' = \displaystyle \frac{X_1'+\tau X_2'}{\sqrt{1+\tau^2}}, 
\label{eq:240}
\end{equation}

\begin{equation}
Y_2' = \displaystyle \frac{-X_1'\tau+X_2'}{\sqrt{1+\tau^2}}  
\label{eq:241}
\end{equation}

\noindent
и для $X_1'\cdot X_2' \sim N(\mu_1-\mu_{b1},\sigma)\cdot N(\mu_2-\mu_{b2},\sigma)$ 
и $Y_1'\cdot Y_2' \sim N(\mu_s\sqrt{1+\tau^2},\sigma)\cdot N(0,\sigma)$.

Отсюда мы находим наиболее вероятное значение 

\begin{equation}
\mu_s^{max} = \displaystyle \frac{y_1}{\sqrt{1+\tau^2}}=
\frac{(x_1-\mu_{b1})+\tau (x_2-\mu_{b2})}{\sqrt{1+\tau^2}}.   
\label{eq:243}
\end{equation}
 
\noindent
Дисперсия для параметра $\mu_s$ равна $ \frac{\sigma}{\sqrt{1+\tau^2}}$. 
Иными словами, на 67\% уровне достоверности 
$|\mu_s-\mu_s^{max}|\le\displaystyle\frac{\sigma}{\sqrt{1+\tau^2}}$.

Заодно по распределению величины $Y_2 \sim N(0,\sigma)$ мы можем проверить справедливость 
гипотезы о том,  что $\mu_1 = \mu_{b1} + \mu_{s}$ и  $\mu_2 = \mu_{b2} + \tau \mu_{s}$. 

Заметим, что принцип максимального правдоподобия  
$\displaystyle \frac{\partial}{\partial \mu_s}{ln~L}=0$, где  

$L=P(x_1|\mu_{b1}+\mu_s,\sigma)\cdot P(x_2|\mu_{b2}+\tau\mu_s,\sigma)$, 
$P(x|\mu,\sigma)=\displaystyle 
\frac{1}{\sqrt{2\pi}\sigma}e^{-\frac{(x-\mu)^2}{2\sigma^2}}$, 
приводит к тому же значению $\mu_s$~(\ref{eq:243}), что и описанная выше процедура.

\subsection{Нормальное распределение. Общий случай}

Рассмотрим задачу комбинирования $n$ независимых величин $X_k$, $k=1,2,\dots,n$, 
распределенных по нормальному закону $X_k \sim N(\mu_k, \sigma_k)$.  

Для случая, когда все дисперсии равны $\sigma_k=\sigma_l=\sigma,~k\ne l$ и 
$\mu_k=\mu_{bk}+s\cdot c_k$, где $\mu_{bk}$ -- вклад фона, а $s$ -- сигнал,  
причем $\sum{c_k^2}=1$, нетрудно показать, что случайная величина 

$Y_1=\displaystyle \sum_{k=1}^{N}{c_k(x_k-\mu_{bk})}$

\noindent
распределена по нормальному закону $Y_1 \sim N(s,\sigma)$.  
Ортогональные к $Y_1$ комбинации $Y_2,Y_3,\dots,Y_n$ распределены по нормальному 
закону $Y_k \sim N(0,\sigma)~~~(k=2,3,\dots,n)$ со средними значениями 
$\mu_k=0$. Отсюда мы можем извлечь комбинированную оценку на параметр $s$
на уровне $1\sigma$, а именно: 

$s = s_0\pm \sigma$,

$s_0 = \displaystyle \sum_{k=1}^{N}{c_k(x_k^0-\mu_{bk})}$, 

\noindent
где $x_k^0$ -- измеренные значения случайной величины $X_k^0$. Заметим, что использование метода 
максимального правдоподобия приводит к такому же результату. К такому же результату приводит и 
Байесовские вычисления с использованием функции правдоподобия 

$\displaystyle 
\prod_{k=1}^n{\frac{1}{\sqrt{2\pi}\sigma}e^{-\frac{(x_k-\mu_k)^2}{2\sigma^2}}}$ 

\noindent
и плоского приора $\pi(\lambda)=const$.

\subsection{Метод наименьших квадратов}

Предположим мы измеряем $N$ независимых величин $X_i$, распределенных по нормальному закону  
с дисперсиями $\sigma_i^2$ и средними $x_{i,av}$. Пусть имеется модель, предсказывающая 
величины средних $x_{i,av}$ как функции от неизвестных параметров модели 
$\theta_k,~~(k=1,2,\dots,l)$. Для определения степени достоверности нашей модели согласно 
методу наименьших квадратов введем величину 

\begin{equation}
T(\vec \theta)=\displaystyle 
\sum_{i=1}^N{\frac{(x_{i,meas}-x_{i,av}(\vec \theta))^2}{\sigma_i^2}},
\label{eq:202}
\end{equation}

\noindent
где $x_{i,meas}$-измеренные значения случайных величин $X_i$. Параметры 
$\vec \theta = (\theta_1,\theta_2,...\theta_l)$ 
определяются исходя из минимума величины $T(\vec \theta)$~\footnote{Здесь мы предполагаем, что 
дисперсии $\sigma_k$ не зависят от неизвестных параметров $\theta_k,~k=1,\dots,l$.} 

\begin{equation}
\displaystyle \frac{\partial}{\partial \theta_i}{T(\vec \theta)}=0,~~i=1,2,\dots,l.
\label{eq:203}
\end{equation}

Эта конструкция очень полезна при анализе данных, поскольку величина $T(\vec \theta)$ в случае 
совпадения $x_{i,av}(\vec \theta^{min})$ с действительными средними $x_{i,meas}$ 
распределена как $\chi^2(T,N)$ с $N$ степенями свободы. 

Напомним, что при больших $N$ величина 
$Z_N = \displaystyle \frac{\chi^2(N)-N}{\sqrt{2N}}$
распределена по нормальному закону с дисперсией $\sigma=1$ и нулевым средним. 

В более общем случае в методе наименьших квадратов нам необходимо минимизировать билинейную 
форму

\begin{equation}
\chi^2 = \displaystyle 
\sum_{i,j=1}^N(m_i-M_i(\vec \theta))_i\sigma_{ij}^{-1}(m_j-M_j(\vec \theta))_j =
(\vec m-\vec M(\vec \theta))^T C^{-1}(\vec m-\vec M(\vec \theta)). 
\label{eq:204}
\end{equation}

\noindent
Здесь $m_i$-результаты измерений, а $M_i(\vec \theta)$-теоретические предсказания, зависящие 
от $l$ неизвестных параметров $\vec \theta=(\theta_1,\dots,\theta_l)$, по которым и происходит 
минимизация билинейной формы. 

Например, в случае $N=2$ и ненулевого корреляционного коэффициента $\rho \ne 0$ нам 
необходимо минимизировать форму 

\begin{equation}
\chi^2 = \displaystyle 
(m-m_1, m-m_2) \left( \begin{array}{cc}
  \sigma_1^2           &  \rho \sigma_1\sigma_2  \\
  \rho\sigma_1\sigma_2 &  \sigma_2^2            
\end{array} \right)^{-1}
{{m-m_1} \choose {m-m_2}}.
\label{eq:205}
\end{equation}

\noindent
Минимизация формы~(\ref{eq:205}) приводит к 

\begin{equation}
m = \displaystyle 
\frac{m_1(\sigma_2^2-\rho\sigma_1\sigma_2)+m_2(\sigma_1^2-\rho\sigma_1\sigma_2)}
{\sigma_1^2-2\rho\sigma_1\sigma_2+\sigma_2^2},
\label{eq:206}
\end{equation}

\begin{equation}
\sigma_m^2 = \displaystyle \frac{\sigma_1^2\sigma_2^2(1-\rho^2)}
{\sigma_1^2-2\rho\sigma_1\sigma_2+\sigma_2^2}.
\label{eq:207}
\end{equation}


\subsection{Метод максимального правдоподобия}

Метод наименьших квадратов естественно получается в методе максимального правдоподобия. 
В случае $N$ независимых измерений функция правдоподобия есть произведение функций 
правдоподобия 

\begin{equation}
L_{tot}(\vec \theta) = \displaystyle \prod_{k=1}^N{L(\vec \theta)},~~
L(\vec \theta)=F_k(x_k|\vec \theta).
\label{eq:208}
\end{equation}

\noindent
Параметры $\vec\theta = (\theta_1, \dots, \theta_l)$ определяются исходя из условия максимизации 
$L_{tot}(\vec\theta)$, то есть 

\begin{equation}
\displaystyle 
\frac{\partial}{\partial \theta_i}L_{tot}(\vec\theta)=0,~~~i=1,\dots,l.
\label{eq:209}
\end{equation}

В случае нормального распределения мы получаем, как следствие, условие минимума 
функции $\chi^2$~(формула \ref{eq:204}). Таким образом видно, что обобщением 
метода наименьших квадратов является метод максимального правдоподобия.  

В случае измерения $N$ независимых величин $n_k$, распределенных согласно закону 
Пуассона с параметрами  $\lambda_k(\vec\theta)$, зависящими от неизвестных величин  
$\vec\theta$ модели, функция правдоподобия 

\begin{equation}
L_{tot}(\vec \theta) = \displaystyle \prod_{k=1}^N{L_k(\vec\theta)},~~
L_k(\vec\theta)=P(n_k^{obs}|\lambda_k(\vec\theta)) 
\label{eq:210}
\end{equation}

\noindent
и неизвестные параметры $\vec\theta$ определяются стандартным образом как максимумы функции 
${\L_{tot}(\vec\theta)}$.

Предположим, что мы имеем гистограмму, разбитую на $N$ бинов, распределение событий 
в каждом из бинов описывается распределением Пуассона с параметрами 
$\lambda_i(\vec\theta)$, где 
$\vec\theta$-параметры теоретической модели. Тогда в методе 
максимального правдоподобия 

\begin{equation}
-2 ln{L} = \displaystyle 
-2 \ln{(\prod_{i=1}^Ne^{-\lambda_i(\vec\theta)}
\frac{\lambda_i(\vec\theta)^{n_i}}{n_i!})}.
\label{eq:211}
\end{equation}

В этом случае $\chi^2$-функция есть 

\begin{equation}
\chi^2 = \displaystyle 
-2 \ln{\frac{L(\lambda_i)}{L(\lambda^{max}_i)}} =
2\sum_{i=1}^N{[\lambda_i(\vec\theta)-n_i
+n_i~ln{\frac{n_i}{\lambda_i(\vec\theta)}}]}.
\label{eq:212}
\end{equation}

Функцию $\chi^2$-можно представить в виде (в случае аппроксимации распределения Пуассона 
нормальным распределением)

\begin{equation}
\chi^2 = \displaystyle 
\sum_i{\frac{(n_i-f_i(\vec\theta))^2}{n_i}} -2~ln{~L} -n~ln{~2n}-\sum_i{ln{~n_i}}. 
\label{eq:213}
\end{equation}

В случае замены $n_i \rightarrow \mu_i=f_i(\vec\theta)$ мы получаем 
$\chi^2$-распределение Пирсона 

\begin{equation}
\chi^2 = \displaystyle 
\sum{\frac{(n_i-f_i(\vec\theta))^2}{f_i(\vec\theta)}} 
\label{eq:214}
\end{equation}

\noindent
для минимизации. 

С помощью $\chi^2$-распределения можно комбинировать данные. Например, пусть у нас есть 
два измерения с некоррелированными ошибками: 
$\mu_1\pm \sigma_1$ и $\mu_2\pm \sigma_2$. Тогда $\chi^2$-функция равна 

\begin{equation}
\chi^2 = \displaystyle 
\frac{(\mu-\mu_1)^2}{\sigma_1^2}+\frac{(\mu-\mu_2)^2}{\sigma_2^2} 
\label{eq:215}
\end{equation}

\noindent
и минимизация $\displaystyle \frac{\partial}{\partial \mu}\chi^2=0$ 
дает оценку

\begin{equation}
\mu = \displaystyle 
\frac{\frac{\mu_1}{\sigma_1^2}+\frac{\mu_2}{\sigma_2^2}}
{\frac{1}{\sigma_1^2}+\frac{1}{\sigma_2^2}}. 
\label{eq:216}
\end{equation}

\noindent
При этом ошибка равна 

\begin{equation}
\displaystyle \frac{1}{\sigma_m^2}=
-\frac{\partial^2}{\partial \mu^2}ln{~L}=
\frac{1}{2}\frac{\partial^2}{\partial \mu^2}\chi^2 =
\frac{1}{\sigma_1^2}+\frac{1}{\sigma_2^2}. 
\label{eq:217}
\end{equation}

\subsection{Комбинирование пределов}

Как уже было отмечено выше комбинирование результатов (скажем двух экспериментов 
ATLAS и CMS) достигается с помощью объединенной функции правдоподобия, которая есть 
произведение функций правдоподобия экспериментов ATLAS и CMS 

\begin{equation}
L_{comb}(\vec\theta) =  L_{CMS}(\vec\theta)L_{ATLAS}(\vec\theta), 
\label{eq:218}
\end{equation}

\noindent
где $\vec\theta$ -- неизвестные параметры модели.

Предположим в обоих экспериментах мы искали одну и ту же сигнатуру, скажем, количество 
димюонов с двумя дополнительными адронными струями и с инвариантной массой 
$M(jet_1 + jet_2 +\mu^+ + \mu^-) \ge M_0$. Предположим, что мы знаем фоны 
$\lambda_{b,CMS}$ и $\lambda_{b,ATLAS}$ и предположим, что ожидаемое число  
событий ($\lambda_{tot}$ в распределении Пуассона) 

$\lambda_{tot,CMS}=\lambda_{b,CMS}+\lambda_{s,CMS},$

$\lambda_{tot,ATLAS}=\lambda_{b,ATLAS}+\lambda_{s,ATLAS},$

$\lambda_{s,ATLAS} = k \lambda_{s,CMS},$

\noindent
где $k$-известное число, связанное с различиями в детекторах CMS и 
ATLAS~\footnote{В случае тождественности CMS и ATLAS детекторов мы имели бы 
$k=1$ и $\lambda_{b,CMS} = \lambda_{b,ATLAS}$ .}. Предположим мы 
не зафиксировали ни одного события, удовлетворяющего нашим 
условиям. Тогда при раздельной обработке событий на 95\% уровне достоверности 
мы имели бы 

$\lambda_{s,CMS} < 3.0,$

$\lambda_{s,ATLAS} < 3.0.$ 

\noindent
Тогда как метод правдоподобия, как и метод комбинирования различных распределений Пуассона, 
что в данном случае одно и то же, дает 

$\lambda_{s,CMS}+\lambda_{s,ATLAS} < 3.0$ 

\noindent
или 

$\lambda_{s,CMS} < \displaystyle \frac{3.0}{1+k}.$

Рассмотрим комбинирование двух распределений Пуассона $P(n_1,\lambda_1)$ и 
$P(n_2,\tau \lambda_1)$, где $\tau$-фиксированное число. Логарифм функции 
правдоподобия 

\begin{equation}
ln{~L(\lambda_1)}=n_1ln{~\lambda_1}+n_2ln{(\tau\lambda_1)}-\lambda_1-\tau\lambda_1-
ln(n_1!)-ln(n_2!)
\label{eq:219}
\end{equation}

\noindent 
уравнения для определения максимума 
$\displaystyle \frac{\partial}{\partial\lambda_1}(ln~L)=0$ приводит к 

$\lambda_1^{max}+\lambda_1^{max}\tau=n_1+n_2$, 

\begin{equation}
ln{~L(\lambda_1)}-ln{~L(\lambda_1=\lambda_1^{max})}
= \displaystyle (n_1+n_2)ln{\frac{\lambda_1(1+\tau)}{n_1+n_2}}-
\lambda_1(1+\tau)+n_1+n_2.
\label{eq:220}
\end{equation}

\noindent
Отсюда при заданном $\lambda$ мы можем определить значимость или p-вероятность. 

Заметим, что абсолютно такой же результат получается, если учесть, что сумма двух 
процессов Пуассона есть процесс Пуассона с $n=n_1+n_2$ и 
$\lambda=\lambda_1+\lambda_2$ и далее уже использовать метод максимального правдоподобия 
для распределения $P(n_1+n_2,\lambda_1+\lambda_2)$.

В более общем случае, когда мы явно учитываем фоны и 
$\lambda_1=\lambda_{1b}+\lambda$ и $\lambda_2=\lambda_{2b}+\tau\lambda$, 
где $\lambda_{1b},\lambda_{2b},\tau$--фиксированные величины, а $\lambda$--параметр 
(число сигнальных событий для первого процесса), метод максимального правдоподобия и метод 
объединения двух Пуассоновских процессов приводят, вообще говоря, к разным результатам.
Действительно в методе максимального правдоподобия уравнение 
$\displaystyle \frac{\partial}{\partial\lambda}L=0$ имеет решение

\begin{center}
$\lambda = \displaystyle -\lambda_{1b}+\frac{n_1+n_2}{2(1+\tau)} -
\frac{1}{2}(\frac{\lambda_{2b}}{\tau}-\lambda_{1b})+$
\end{center}
\begin{equation}
\displaystyle 
\sqrt{(\frac{n_1+n_2}{2(1+\tau)}-\frac{1}{2}(\frac{\lambda_{2b}}{\tau}-\lambda_{1b}))^2-
\frac{\lambda_{1b}\lambda_{2b}}{\tau} + 
\frac{1}{1+\tau}(\frac{n_1\lambda_{2b}}{\tau}+n_2\lambda_{1b})}.
\label{eq:221}
\end{equation}

\noindent
Тогда как при комбинировании двух распределений Пуассона параметр $\lambda$ равен 

\begin{equation}
\displaystyle \lambda = 
\frac{n_1+n_2-\lambda_{1b}-\lambda_{2b}}{1+\tau}.
\label{eq:222}
\end{equation}

В случае $\displaystyle \frac{\lambda_{2b}}{\tau}=\lambda_{1b}$ два подхода совпадают. 

Заметим, что когда мы измеряем $n_{obs,CMS}$ и $n_{obs,ATLAS}$ число событий 
и находим ограничение сверху на параметры распределений Пуассона $\lambda_{tot,CMS}$, 
$\lambda_{tot,ATLAS}$, где  $\lambda_{tot}=\lambda_b+\lambda_s$, обобщение 
вышеизложенного подхода очевидно.

\subsection{Байесовский подход}

Для $N$ независимых измерений с функциями распределения $F_k(x|\vec\theta)$ функция 
правдоподобия $L(\vec\theta)=\displaystyle \prod_{k=1}^N F_k(x_k|\vec\theta)$.
В этом случае согласно Байесовскому подходу комбинированная плотность вероятности

\begin{equation}
f(\vec\theta) = \displaystyle 
\frac{L(\vec\theta)\pi(\vec\theta)}{\int{d\vec\theta L(\vec\theta)\pi(\vec\theta)}}.
\label{eq:223}
\end{equation}

Остановка за малым -- выбрать правильный приор $\pi(\vec\theta)$. 
 Формула~(\ref{eq:223}) задает способ комбинирования результатов в
Байесовском подходе. 
 
Рассмотрим задачу комбинирования двух результатов, полученных на основе Пуассоновской статистики.
Для случая  $\lambda_1=\lambda,~~\lambda_2=k~\lambda,~~k=const$ имеем 
тождество 

$P(n_1,\lambda_1)P(n_2,\lambda_2) = \displaystyle 
\frac{n_1!n_2!}{(n_1+n_2)!}\frac{k^{n_2}}{(1+k)^{n_1+n_2}}P(n_1+n_2,\lambda_1+\lambda_2)$.

\noindent
Иными словами, произведение двух распределений Пуассона с точностью до численного множителя совпадает 
с распределением Пуассона с числом событий $n_1+n_2$ и параметром 
$\lambda=\lambda_1+\lambda_2$.

При выборе плоского приора $\pi(\lambda)=const$ задача комбинирования свелась 
к задаче извлечения параметров из распределения $P(n_1+n_2,\lambda_1+\lambda_2)$. 
Для модели, описывающей как вклад фона, так и 
вклад сигнала с 

$\lambda_1=\lambda_{b1}+\lambda_{s1}$, 

$\lambda_2=\lambda_{b2}+k\lambda_{s1}$  

\noindent
в случае, когда $\lambda_{b2}=k\lambda_{b1}$, произведение распределений Пуассона так же 
сводится с точностью до численного множителя к распределению Пуассона с $n=n_1+n_2$ и 
$\lambda=\lambda_1+\lambda_2$. В случае, когда $\lambda_{b2} \ne k~\lambda_{b1}$ 
задача не сводится к распределению единичного Пуассона. В любом случае, для извлечения информации 
из двух измерений нам необходимо знать функцию приора $\pi(\lambda)$.

При комбинировании двух процессов Пуассона 
мы можем извлечь информацию, рассмотрев случайную величину равную сумме случайных величин, 
описывающихся распределениями Пуассона $P(n_i,\lambda_i), i=1,2,$ 
и описывающуюся процессом Пуассона с 
параметрами $n=n_1+n_2$, $\lambda=\lambda_1+\lambda_2$. При этом извлекать ограничения на параметр 
$\lambda_s$ можно  из распределения Пуассона 
$P(n_1+n_2,\lambda_{b1}+\lambda_{b2}+(1+k)\lambda_{s1})$.  

Заметим при этом, что учет систематических эффектов, связанных с неточным знанием 
$\lambda_{b1},~\lambda_{b2},~\lambda_{s1},~\lambda_{s2}$ 
принципиально ничем не отличается от соответствующего рассмотрения для
единичного распределения Пуассона.  

Интересно отметить, 
что для распределения Пуассона $P(n_k,\lambda_k)$, где $\lambda_k=c_k\tau$ и 
$c_k$-фиксированные числа, а $\tau$-неизвестный параметр, справедливо равенство

\begin{equation}
\displaystyle \prod_{k=1}^N{P(n_k,\lambda_k)} = c P(\sum{n_k},\sum{\lambda_k}),  
\label{eq:225}
\end{equation}

\noindent 
где коэффициент пропорциональности $c$ зависит от величин $n_k$ и $c_k$.

Равенство~(\ref{eq:225}) является отражением того факта, что сумма процессов Пуассона является 
процессом Пуассона с наблюдаемым числом событий равным сумме наблюдаемых 
$n_{obs}=\sum{n_i}$ и параметром распределения $\lambda$ равным сумме параметров распределения 
$\lambda=\sum{\lambda_i}$. На основании этой формулы задача комбинирования сводится 
к задаче определения параметра $\lambda$ ( параметра $\tau$) для единичного 
процесса Пуассона. 

\subsection{Комбинирование уровней значимости (комбинирование значений вероятности)}

Предположим, что мы хотим проверить основную гипотезу $H_0$ путем измерения $N$ независимых величин.
Предположим также, что у нас в результате измерений $N$ независимых величин известны вероятности $P_k$ 
их реализации или, что то же самое в силу формулы 

\begin{equation}
\displaystyle \frac{1}{\sqrt{2\pi}}\int_{S}^{\infty}{e^{-\frac{x^2}{2}}}= P(S),  
\label{eq:226}
\end{equation}

\noindent
их значимостей $S_k$. 

Встает вопрос: как определить комбинированную значимость или комбинированную вероятность. 
Существует несколько способов решения этой проблемы. В методе Фишера используется переменная 
$\alpha'=\displaystyle -2~ln{\prod_{i=1}^N{P_i}}$. Далее предполагается, что $\alpha'$ 
распределена как $\chi^2(2N)$. Как следствие получаем формулу комбинирования. Для 
комбинирования двух вероятностей метод Фишера приводит к правилу комбинирования 

\begin{equation}
P(P_1,P_2)=P_1P_2[1-ln(P_1P_2)]. 
\label{eq:227}
\end{equation}

В методе Стоуфера~\cite{Stouffer} был предложен другой способ комбинирования значимостей. 
Поясним этот метод на примере распределения Пуассона. Пусть мы измеряем $N$ независимых 
величин, распределенных по закону Пуассона $P(n_k|\lambda_k)$~\footnote{Предполагается, что  
параметры $\lambda_k$ точно известны в рамках основной гипотезы.} и пусть мы наблюдаем избыток 
событий в каждой из $N$ независимых величин. Тогда мы можем определить вероятности 

\begin{equation}
P_k(n \ge n_{k,obs}|\lambda_k)=\displaystyle \sum_{n=n_{k,obs}}^{\infty}{P(n|\lambda_k)}, 
\label{eq:228}
\end{equation}

\noindent 
так называемые ``p-value''.


В методе Стоуфера используется тот факт, что сумма процессов Пуассона является процессом Пуассона 
и, как следствие, формула комбинирования вероятностей имеет вид 

\begin{equation}
P(P_1,\dots,P_n)=\displaystyle \sum^{\infty}_{n=\sum_k{n_{k,obs}}}{P(n|\sum_k{\lambda_k})}.
\label{eq:229}
\end{equation}

В случае  $\lambda_k\ll n_{k,obs}$ сумма~(\ref{eq:229}) 
определяется в основном первым членом в сумме  

\begin{equation}
P_k(n \ge n_{k,obs}|\lambda_k)\approx\displaystyle 
\frac{1}{n_{k,obs}!}\lambda_k^{n_{k,obs}}e^{-\lambda_k}
\label{eq:230}
\end{equation}

\noindent
и, следовательно, 

\begin{equation}
P(P_1,\dots,P_n)=\displaystyle 
\frac{1}{(\sum_{k=1}^N{n_{k,obs}})!}
(\sum_{k=1}^N{\lambda_k})^{\sum_{k=1}^N{n_{k,obs}}}e^{-\sum_{k=1}^N{\lambda_k}}.
\label{eq:231}
\end{equation}

\noindent
Для $n=2$ сравнение правил комбинирования~(\ref{eq:227}) и~(\ref{eq:231}) показывает, что 
численные результаты могут сильно различаться. 

В качестве примера рассмотрим комбинирование двух распределений Пуассона с 
$n_{1,obs} = n_{2,obs}=1$ и
$\lambda_1,~\lambda_2 \ll 1$. В этом случае формулы~(\ref{eq:230}) и~(\ref{eq:231}) 
примут вид $P_1\approx \lambda_1$, $P_2 \approx \lambda_2$, 

\begin{equation}
P(P_1,P_2) = \displaystyle \frac{(\lambda_1+\lambda_2)^2}{2} =
\frac{(P_1+P_2)^2}{2}.
\label{eq:232}
\end{equation}

\noindent
Как видим, различие между формулой~(\ref{eq:232}) и формулой Фишера~(\ref{eq:227}) 
весьма существенно. 

\subsection{``Look elsewhere'' эффект}\label{sec:LEE}

Предположим, что мы ищем бозон Хиггса посредством реакции 
$pp\rightarrow H\rightarrow \gamma\gamma + \dots$, то есть мы ищем узкий пик 
в распределении дифотонных событий по инвариантной массе $m(\gamma \gamma)$. 
Предположим, что мы ищем эффект в интервале масс $m_{-} < m(\gamma \gamma) \le m_{+}$. 
Стандартный путь -- разбиваем интервал $[m_-,m_+]$ на $k$ бинов, каждый с шириной 
$\displaystyle \Delta=\frac{m_+-m_-}{k}$   и ищем усиление сигнала в одном из 
$k$ бинов. Предположим для простоты, что распределение фона плоское, то есть фон не зависит 
от $m$ при $m_-\le m\le m_+$. 

Оценка фона в каждом из бинов в силу независимости распределения от $m$ получается равной 
$\lambda =\displaystyle \frac{1}{k}\sum{N_l}$ и для большого количества $k$ 
и $N$ параметр фона $\lambda$ определяется с хорошей точностью из экспериментальных данных.

Мы ищем превышение сигнала над фоном в каждом бине. Вероятность детектирования  
$N_k$ событий в бине определяется формулой
Пуассона~(\ref{eq:43}) и является функцией от наблюдаемого 
числа частиц в бине и среднего $\lambda$~\footnote{Мы здесь рассматриваем случай, когда фон 
плоский и параметр $\lambda$, описывающий фон в распределении Пуассона, известен.}.

``Look elsewhere'' эффект состоит в том, что, поскольку мы не знаем точное расположение 
сигнала, вероятность реализации событий вследствие флуктуации фона увеличивается 
на фактор $k$ по сравнению с наименее вероятным 
значением $P_{min}=\displaystyle min_k{P(N_{k,obs}|\lambda)}$, где 

$P(N_{k,obs}|\lambda)=\displaystyle \sum_{n=N_{k,obs}}^{\infty}{P(n|\lambda)}$.  

\noindent
Иными словами ``look elsewhere'' эффект приводит к изменению вероятности 
$P_{min} \rightarrow k~P_{min}$.

\subsection{Тест Колмогорова-Смирнова}

Предположим мы имеем набор измерений ${x_1,x_2, \dots, x_n}$ и мы хотим проверить 
насколько совместим этот набор с функцией распределения $f(x)$. 

В методе Колмогорова-Смирнова это достигается путем сравнения теоретической кумулятивной 
функции распределения 

\begin{equation}
F(x)=\displaystyle \int_{-\infty}^x{f(x')dx'} 
\label{eq:233}
\end{equation}

\noindent
и ``экспериментальной'' кумулятивной функции распределения  

\begin{equation}
F_n(x)=\displaystyle \frac{1}{n}\sum_{i=1}^n{\theta(x-x_i)}, 
\label{eq:234}
\end{equation}

\noindent
где $\theta(x)=\cases{1,~~~x > 0, \cr 0,~~~x \le 0.}$

Расстояние между теоретической кумулятивной функцией распределения и 
экспериментальной определяется как 

\begin{equation}
D_n = \displaystyle {sup}_x|F_n(x)-F(x)|. 
\label{eq:235}
\end{equation}

Для больших $n$ функция $D_n\rightarrow 0$. Параметр  $k=\sqrt{n}D_n$ имеет распределение 
$P(k\le x)=\displaystyle 1 - 2\sum_{l=1}^{\infty}{(-1)^{l-1}e^{-l^2x^2}}.$

\noindent
Подробное рассмотрение и доказательства можно найти в монографии~\cite{James}.  

С помощью теста Колмогорова-Смирнова можно решить вопрос следуют ли выборки 
${x_1, \dots, x_n}$ и ${y_1, \dots, y_N}$ одному и тому же распределению. 
Для этого расстояние между двумя выборками определяется как 

\begin{equation}
D_{n,m} = \displaystyle {sup}_x|F_n(x)-F_m(x)|. 
\label{eq:236}
\end{equation}

\noindent
Можно показать, что переменная $\displaystyle \sqrt{\frac{nm}{n+m}}D_{n,m}$ 
асимптотически удовлетворяет распределению Колмогорова.

%
%
%
%
%
%


\newpage

\section{Статистическое программное обеспечение в задачах физики высоких энергий}

\subsection{Обзор основных пакетов}

Базовым инструментарием для решения задач физики высоких энергий является система 
программ ROOT~\cite{ROOT}. ROOT -- пакет объектно-ориентированных программ 
и библиотек, разработанный в ЦЕРН. Проект базируется на свободном программном 
обеспечении. Наряду со специальными средствами программирования и стандартных 
математических вычислений, ROOT обеспечивает пользователя средствами для построения и 
анализа гистограмм и графиков функций, средствами фитирования и подбора теоретических 
и экспериментальных зависимостей, инструментарием для проведения статистического 
(в том числе многофакторного) анализа данных~\cite{ROOTstat}.

Большинство программных наработок, по возможности, либо встраиваются в ROOT, либо являются 
надстройкой над ROOT, то есть базируются на ROOTовских библиотеках.
Проводятся работы по совмещению и/или по созданию интерфейса между
пакетом ROOT и астрофизическим языком и оболочкой для статистических
расчетов и построения графиков R~\cite{R}. 

Надстройкой над ROOT является пакет программ RooFit~\cite{RooFit}. 
RooFit это инструментарий для моделирования ожидаемых распределений событий
в физическом анализе. Пакет был изначально ориентирован на эксперимент BaBar.
Пакет удобен для быстрого Монте Карло розыгрыша событий 
и статистической обработки полученых распределений. Со временем он стал универсальным и 
базовым для пакетов, расширяющих его возможности, например, 
RooStats~\cite{RooStats}.  
 
Байесовская парадигма реализована в пакете программ BAT - 
The Bayesian Analysis Toolkit~\cite{BAT}. 
Анализ базируется на теореме Байеса и использует Монте Карло
моделирование Марковских цепей. Это позволяет строить апостериорное 
распределение вероятностей, производить оценивание параметров, строить 
доверительные интервалы, осуществлять перенос неопределенностей. 

Много внимания Статистические группы экспериментов уделяют статистическим 
пакетам многомерного анализа. Было проведено несколько мини-совещаний по 
данной тематике, например,  Miniworkshop on Statistical Tools (2008) 
DESY~\cite{MWST}, CMS tutorials on Multivariate Analysis 
Methods (2007) CERN~\cite{CMSMVA}. Здесь можно 
выделить пакет TMVA - Toolkit for MultiVariate Data Analysis~\cite{TMVA}. 
Данный инструментарий также встроен в среду ROOT и ориентирован на использование 
многомерных классификационных алгоритмов для решения широкого спектра задач.  
Для решения задач физики высоких энергий и астрофизики также разрабатывается пакет 
StatPatternRecognition (SPR)~\cite{SPR}. Оба пакета имеют как перекрывающиеся, 
так и дополняющие друг друга возможности. 
Интересной новой разработкой является система поддержки принятия
решений при выборе переменных в многомерном анализе и уменьшения 
размерности задачи PARADIGM~\cite{PARADIGM}.

\subsection{Проект RooStats}

Для анализа экспериментальных данных коллабораций CMS и ATLAS разрабатывается  
проект RooStats~\cite{RooStats}, основанный на комплексе программ 
ROOT~\cite{ROOT}. Главные цели проекта  
\begin{itemize}
\item предоставить пользователю компьютерные программы с наиболее распространенными 
статистическими методами, которые применяются при анализе данных в физике высоких энергий, 
\item стандартизовать используемые методы для легкого сравнения результатов полученных 
разными группами и разными экспериментами. 
\end{itemize}

RooStats использует три наиболее распространенных подхода в статистике: 

\begin{enumerate}
\item частотный подход, 
\item метод максимального правдоподобия, 
\item Байесовский подход. 
\end{enumerate}

Заметим, что программа $RooStats$ постоянно развивается, совершенствуется и,   
на сегодняшний день, она содержит программы, позволяющие решать следующие задачи:

\begin{enumerate}
\item Точечная оценка для определения наилучшего в некотором смысле (например оценки с 
минимальной дисперсией или наиболее вероятные значения) значения параметра. 
\item Определение доверительного интервала: областей параметров функции распределения, 
не противоречащих наблюдаемым данным. 
\item Проверка гипотез: оценка значения вероятности $p$ для одной или нескольких 
гипотез (значимость). 
\item Оценка качества фита -- количественное определение насколько хорошо модель описывает 
данные.
\end{enumerate}

Программа RooStats написана на языке $C++$ и содержит следующие классы, 
позволяющие решать эти задачи. 

\begin{itemize}
\item[-] ProfileLikelihoodCalculator вычисляет значимость сигнала и определяет 
наилучшее значение сигнала на основе метода максимального правдоподобия. 
Возможность учета систематических эффектов также включена в калькулятор. 

\item {\it ProfileLikelihood} возможно использовать для оценки интервалов доверия. 
Возможно вычисление верхнего и нижнего пределов, а также центрального доверительного 
интервала.

\item {\it BayesianCalculator} позволяет решать задачи на основе 
метода Байеса. Причем Байесовское интегрирование может производиться численно, 
аналитически и методом Монте Карло с помощью Марковских цепей 
(MCMCCalculator - Monte Carlo Markov Chain Calculator). Здесь, конечно, очень 
важно -- выбор функции приора $\pi(\lambda)$, а также выбор интервала -- центрального  
интервала, интервала минимальной длины или одностороннего интервала. 

\item {\it HybridCalculator} вычисляет частотную вероятность событий. Учет систематики проводится 
с помощью метода Кузинса-Хайлэнда. В частности вычисляет ``п''-значения (p-value). 
Вычисления производятся с помощью Монте Карло розыгрыша псевдоэкспериментов. Также 
возможно осуществить построение Неймана (NeymanConstruction) для определения интервалов 
доверия частотным способом. Предоставляется возможность использовать несколько правил 
конструирования интервалов (интервал минимальной длины, центральный интервал, 
метод Фельдмана-Кузинса). 
Также при определении значения верхних пределов на сигнал можно использовать чисто 
$P_{sl}$ частотный подход и $CL_s=\displaystyle \frac{P_{sl}}{1-p_B}$ -- модифицированный  
частотный подход. 

\item {\it HypoTestInverter}  преобразует результат по проверке гипотез (HybridCalculator) 
в доверительный интервал (или в предел доверия) для параметра. 

\item {\it HistFactory}  обеспечивает использование статистического инструментария пакета программ 
RooStats без необходимости использовать язык моделирования данных для пакета RooFit.

\item {\it BATCalculator}. Сам пакет Байесовских вычислений с 
помощью Монте Карло цепей Маркова (BAT) является внешним к RooStats, но как класс 
в пакете программ RooStats полезен.

\end{itemize}

\subsection{Проект BAT}

BAT (the Bayesian Analysis Toolkit)~\cite{BATCPC} -- инструментарий для статистического анализа 
 также как и RooStats возник недавно. Статистический анализ 
данных в пакете программ BAT основывается на теореме Байеса и реализуется с помощью 
метода Монте Карло с использованием цепей Маркова~\cite{Markov}. Это позволяет строить апостериорные 
распределения параметров и, соответственно, проводить оценку параметров, устанавливать 
доверительные пределы и интервалы, а также осуществлять непосредственно перенос 
неопределенностей на уровне апостериорных распределений.  

Одна из главных целей анализа данных -- сравнить модельные предсказания с экспериментальными 
данными и либо сделать заключение о корректности модели по отношению к данным, либо построить 
области доверия с той или иной точностью для параметров заданной модели.

\newpage

\begin{center} {\large \bf Заключение}
\end{center}

{Основной задачей Большого адронного коллайдера является поиск бозона Хиггса -- 
последней неоткрытой частицы, предсказываемой Стандартной Моделью. 
Так же весьма важен поиск проявлений новой физики вне рамок Стандартной Модели, 
включая в первую очередь суперсимметрию. При обработке экспериментальных данных 
и получении ограничений на новые частицы и взаимодействия используются частотный 
подход, метод максимального правдоподобия и Байесовский подход. Как правило эти 
подходы приводят к численно близким результатам. При извлечении ограничений на 
новую физику из данных БАКа очень важно правильно оценивать влияние систематических 
эффектов, что во многих случаях весьма и весьма нетривиально. 

Авторы благодарны академику В.А. Матвееву по инициативе которого был написан этот  
обзор. Работа поддержана грантом РФФИ 10-02-00468-а.
}

\newpage




\end{document}